\documentclass[preprint,review,12pt]{elsarticle}
\pdfoutput=1
\journal{Computers and Fluids}

\usepackage{mathtools}
\usepackage[export]{adjustbox}
\usepackage{BOONDOX-calo}   
\usepackage{algpseudocode}
\usepackage{algorithm}
\usepackage{alphalph}
\usepackage{amsmath}
\usepackage{amssymb}
\usepackage{amsthm}
\usepackage{amsbsy}
\usepackage{array}
\usepackage{appendix}
\usepackage{calc}
\usepackage{changepage}
\usepackage{dcolumn}        
\usepackage{enumitem}
\usepackage{epsfig}
\usepackage{esvect}
\usepackage[pscoord]{eso-pic}
\usepackage{etoolbox}
\usepackage{fancybox}
\usepackage{fancyhdr}
\usepackage{float}
\usepackage{hyperref}
\usepackage[T1]{fontenc}
\usepackage[hang,bottom]{footmisc}
\usepackage[nomessages]{fp}
\usepackage{graphicx}
\usepackage{grffile}        
\usepackage{ifthen}
\usepackage[utf8]{inputenc}
\usepackage{lastpage}
\usepackage{letltxmacro}
\usepackage{mathptmx}
\usepackage{marvosym}
\usepackage{multirow}
\usepackage{natbib}
\usepackage{parskip}
\usepackage{pgffor}
\usepackage{pgfplots}
\usepackage{pifont}
\usepackage{ragged2e}
\usepackage{scrextend}
\usepackage{setspace}
\usepackage{siunitx}
\usepackage{soul}
\usepackage{subfig}
\usepackage{tabularx}
\usepackage{titlesec}
\usepackage{textcomp}
\usepackage{txfonts}
\usepackage[normalem]{ulem}
\usepackage{url}
\usepackage{varwidth}
\usepackage{verbatim}
\usepackage{wasysym}
\usepackage{xcolor}
\usepackage{xstring}
\usepackage[
bindingoffset=0in,
top    = 0.8in,
bottom = 0.8in,
left   = 0.75in,
right  = 0.75in]{geometry}
\usepackage{pdflscape}
\usepackage{listings}
\usepackage{multicol}
\usepackage{color}
\usepackage{booktabs}
\usepackage[section]{placeins}
\usepackage{lineno}
\usepackage{graphics}
\usepackage{wrapfig}
\usepackage{epstopdf}
\usepackage{tikz}
\usetikzlibrary{arrows,matrix,positioning,fit}
\usetikzlibrary{shapes,positioning}
\usetikzlibrary{backgrounds}
\usepgfplotslibrary{colorbrewer}
\usepgfplotslibrary{patchplots}
\usepgfplotslibrary[colorbrewer]
\usetikzlibrary{pgfplots.colorbrewer}
\usetikzlibrary[pgfplots.colorbrewer]
\usepgfplotslibrary{units}
\usetikzlibrary{spy}
\usepackage{pgfplotstable}
\graphicspath{ {./figures/} }
\newsavebox\CBox
\newcommand\hcancel[2][0.3pt]{%
    \ifmmode\sbox\CBox{$#2$}\else\sbox\CBox{#2}\fi%
    \makebox[0pt][l]{\usebox\CBox}%
    \rule[0.3\ht\CBox-#1/2]{\wd\CBox}{#1}}
\definecolor{green}{rgb}{0,0.79,0}

\hypersetup{
    linktoc     = all,
    colorlinks  = true,
    allcolors = [rgb]{0,0,0.75}, 
}
\hypersetup{final}

\usetikzlibrary{arrows,chains,matrix,positioning,scopes}

\makeatletter
\def\resetMathstrut@{%
    \setbox\z@\hbox{%
        \mathchardef\@tempa\mathcode`\\relax
        \def\@tempb##1"##2##3{\the\textfont"##3\char"}%
        \expandafter\@tempb\meaning\@tempa \relax
    }%
    \ht\Mathstrutbox@1.2\ht\z@ \dp\Mathstrutbox@1.2\dp\z@
}
\makeatother

\catcode`@=11
\def\caseswithdelim#1#2{\left#1\,\vcenter{\normalbaselines\m@th
        \ialign{\strut$##\hfil$&\quad##\hfil\crcr#2\crcr}}\right.}
\catcode`@=12

\def\breakloop{\fi\iffalse}

\newcommand{\xmark}{\ding{55}}

\newcommand\bm[1]{\boldsymbol{#1}} 
\newcommand\p{\mathcal{p}}

\newcommand\drmb[1]{\mathrm{d} \bm{\mathrm{#1}}}
\newcommand\drm[1]{\mathrm{d} #1}
\newcommand\SD{\mathrm{SD}}
\newcommand\FR{\mathrm{FR}}
\newcommand\FRSD{\mathrm{FR_{SD}}}
\newcommand\FRDG{\mathrm{FR_{DG}}}
\newcommand\DFR{\mathrm{DFR}}

\newcommand{\fsize}{}

\newcommand{\forloop}[5][1]%
{%
    \setcounter{#2}{#3}%
    \ifthenelse{#4}%
    {%
        #5%
        \addtocounter{#2}{#1}%
        \forloop[#1]{#2}{\value{#2}}{#4}{#5}%
    }%
    {%
    }%
}%

\newcommand*\overbar[1]{%
    \vbox{%
        \hrule height 0.9pt%
        \kern0.35ex%
        \hbox{%
            \kern-0.1em%
            \ifmmode#1\else\ensuremath{#1}\fi%
            \kern0.0em%
        }
    }
}

\renewcommand{\algorithmiccomment}[1]{\bgroup\hfill!#1\egroup}

\definecolor{bluesteel}{rgb}{0.392, 0.392, 0.498}
\definecolor{lightgray}{gray}{1.0}
\newlength\myheight
\newlength\mydepth
\settototalheight\myheight{Xygp}
\newcommand*\inlinegraphics[1]{%
    \settototalheight\myheight{Xygp}%
    \settodepth\mydepth{Xygp}%
    \raisebox{-\mydepth}{\includegraphics[height=\myheight]{#1}}%
}
\newcommand\orcid[1]{\href{https://orcid.org/#1}{\inlinegraphics{orcid_16x16.png}}}

\makeatletter
\def\BState{\State\hskip-\ALG@thistlm}
\makeatother

\newtheorem{theorem}{Theorem}[section]

\newcommand{\etal}{et~al.}

\newcommand{\hfd}{\hat{f}^{\delta}}

\newcommand{\hfI}{\hat{f}^{\delta I}}
\newcommand{\hfC}{\hat{f}^{\delta C}}

\newcommand{\hud}{\hat{u}^{\delta}}

\newcommand\bd[1]{\bm{#1}^\delta}
\newcommand\hb[1]{\hat{\bm{#1}}}
\newcommand\hbd[1]{\hb{#1}^\delta}
\newcommand\hbI[1]{\hb{#1}^{\delta I}}
\newcommand\hbC[1]{\hb{#1}^{\delta C}}

\newcommand\px[2]{\frac{\partial #1}{\partial {#2}}}
\newcommand\pxi[3]{\frac{\partial^{#1}#2}{\partial {#3}^{#1}}}
\newcommand\dx[2]{\frac{\mathrm{d} #1}{\mathrm{d} #2}}
\newcommand\dxi[3]{\frac{\mathrm{d}^{#1}#2}{\mathrm{d} {#3}^{#1}}}
\newcommand\rint[2]{\int^{1}_{-1}{#1} \mathrm{d}{#2}}

\newcommand\pxvar[2]{\partial_{#2} #1}

\newcommand{\half}{\frac{1}{2}}
\newcommand{\hnabla}{\hat{\nabla}}

\begin{document}

\begin{frontmatter}
    \title{Accuracy, Stability, and Performance Comparison between the Spectral Difference and Flux Reconstruction Schemes}
    
    \author{C. Cox\fnref{label1}\corref{cor1}}
    \ead{coxc@tamu.edu}
    
    \author{W. Trojak\fnref{label2}}
    \ead{wt247@tamu.edu}
    
    \author{T. Dzanic\fnref{label2}}    
    \ead{tdzanic@tamu.edu}

    \author{F. D. Witherden\fnref{label2}}
    \ead{fdw@tamu.edu}
    
    \author{A. Jameson\fnref{label1,label2}}
    \ead{antony.jameson@tamu.edu}
    
    \cortext[cor1]{Corresponding author: C. Cox}
    \address[label1]{Department of Aerospace Engineering, Texas A\&M University, College Station, TX 77843, USA\vspace{-0.3cm}}
    \address[label2]{Department of Ocean Engineering, Texas A\&M University, College Station, TX 77843, USA}
    \date{}
    
    \begin{abstract}
        We report the development of a discontinuous spectral element flow solver that includes the implementation of both spectral difference and flux reconstruction formulations. With this high order framework, we have constructed a foundation upon which to provide a fair and accurate assessment of these two schemes in terms of accuracy, stability, and performance with special attention to the true spectral difference scheme and the modified spectral difference scheme recovered via the flux reconstruction formulation. Building on previous analysis of the spectral difference and flux reconstruction schemes, we provide a novel nonlinear stability analysis of the spectral difference scheme. Through various numerical experiments, we demonstrate the additional stability afforded by the true, baseline spectral difference scheme without explicit filtering or de-aliasing due to its inherent feature of staggered flux points. This arrangement leads to favorable suppression of aliasing errors and improves stability needed for under-resolved simulations of turbulent flows.
    \end{abstract}

    \begin{keyword}
        {discontinuous spectral element \sep spectral difference \sep flux reconstruction \sep implicit large eddy simulation}
        \begin{MSC}[2010]
            46E39 \sep 46N35 \sep 65M70 \sep 76N15
        \end{MSC}
    \end{keyword}
\end{frontmatter}


\section{Introduction}\label{s:intro}

Computational fluid dynamics presents practitioners with many challenges, chief among which is resolving the often wide range of length scales while keeping computational cost sufficiently low. This is crucial if such simulations are to meaningfully impact engineering design cycles. Reynolds-Averaged Navier-Stokes (RANS) methods, the prevailing mode of choice in the industry, have exhibited significant shortcomings in simulating complex turbulent flows, and as such, there is considerable interest in the development of high-fidelity scale-resolving simulations. Although far superior in terms of accuracy, these scale-resolving simulations can be orders of magnitude more computationally expensive than their RANS counterparts which makes them intractable for many practical engineering purposes. To address this challenge, various families of methods have emerged over several decades, one of which is the spectral element method (SEM), a set of high-order techniques that has been successfully used for many applications. These methods developed out of discontinuous techniques, such as that of Reed and Hill~\cite{reed-hill:1973}, which forwent some solution continuity in favor of localizing the calculation to sub-domains. This sub-domain structure---with reduced inter-element communication---can increase the computational efficiency through structured compute regions that are well suited to modern massively parallel computer architectures such as graphic processing units (GPU).

Discontinuous SEM offers geometric flexibility and reduced dissipation/dispersion errors for high-fidelity computations; however, application of these schemes to turbulent flow problems can be problematic due to numerical instability issues. As the cost of resolving the finest physical length scales grows prohibitively large with increasing Reynolds number, scale-resolving simulations are typically restricted to resolving only the statistically significant length scales. For a sufficiently high-order scheme, this lack of resolution can cause aliasing errors to occur and produce unstable simulations~\cite{Chapelier2012}. These errors originate from the high-order of the flux function and/or the geometry and limit the space in which the approximate solution can reside~\cite{sherwin-karniadakis:2005}. To ameliorate these errors and achieve stability, various techniques have been introduced, such as spectral vanishing viscosity methods (SVV)~\cite{Tadmor1990,Tadmor1993,Karamanos2000,Maday1993}, modal filtering~\cite{Gassner2012,hesthaven-warburton:2008,Gottlieb1997,Gottlieb2001}, and split skew-symmetric methods~\cite{Kravchenko1997,Blaisdell1996,Gassner2016,winters:2018}. However, these techniques do come with a notable computational cost and, in some cases, tunable parameters, and it has become commonplace to perform simulations without explicit filtering or de-aliasing applied to the solution. One such approach in the context of solving turbulent flows with discontinuous SEM is implicit large eddy simulation (ILES)~\cite{uranga:2011,beck:2014,cartondewiart:2012,vermeire:2016,moura:2017}, from which high-fidelity solutions can be obtained without any added modeling or filtering traditionally used to account for sub-grid length scales by utilizing the inherent numerical dissipation of the scheme. However, this dissipation may be insufficient when using high-order discretizations for high Reynolds number turbulent flows, and it is not yet evident which method is best suited for robustly achieving stable and accurate simulations for these flows. There is speculation that certain methods may have more favorable de-aliasing properties which can result in improvements in stability, although it has not been thoroughly explored. 


In this paper, we investigate two nodal discontinuous spectral element methods with several similarities. The first method is the flux reconstruction (FR) method of Huynh~\cite{huynh:2007} and Vincent~\etal~\cite{vincent-castonguay-jameson:2011}. This method uses a local polynomial approximation of the solution to form an approximation to the flux such that continuity is enforced through inter-element communication and correction functions. This method has been adapted for several element topologies~\cite{witherden:2015} and has been applied to various equation sets including the Euler equations~\cite{Williams2011,Castonguay2011}, Navier--Stokes equations~\cite{Williams2011,Castonguay2013}, and their incompressible counterparts~\cite{cox-liang-plesniak:2012,Loppi2018}. Several implementations of FR are available that have demonstrated the possibility to achieve high computational efficiency and scalability on large problems~\cite{witherden:2014,Vincent2016}. The second method is the spectral difference (SD) method originally put forth by Kopriva~\etal~\cite{Kopriva1996,Kopriva1996b}, where a staggered arrangement of points is used within each element, with one set of points for the solution and another for the flux and its gradient. The formal stability of this method for linear problems was explored by Jameson~\cite{jameson:2010}, who found a Lobatto-type distribution for the flux points to be important. Furthermore, Huynh~\cite{huynh:2007} found that the accuracy of the scheme is independent of the solution point locations for linear problems. Similar to FR, this method has been successfully applied to non-linear equations~\cite{wang-liu-may-jameson:2007,sun-wang-liu:2007,yu-wang-hu:2011} as well as in the simulation of complex physics~\cite{chan:2011,Lodato2016,lodato:2019,wang:2016}. 

The SD method is of interest as the approximation of the flux function, which is projected into the solution space through differentiation, is one degree higher than the solution. It is conjectured that this increased order of the flux equips SD with a favorable amount of de-aliasing in comparison to FR. In the body of SD and FR literature, there has been little comparative study between these related methods and the effect that different techniques for the flux function approximation will have on the stability and accuracy of the methods. We investigate the differences and similarities for these schemes when used in ILES, and show the effects of the higher degree of the flux approximation on the stability of the method. To this end, this work is structured with the formulation of SD and FR schemes on hyper-cube elements in Section~\ref{s:formulation}. Non-linear analysis of the SD method is presented in Section~\ref{s:sd_stability}, where the instability mechanics are considered as well as scaling arguments for the error. Section~\ref{s:governing} sets forth the formulation used for the Navier--Stokes equations and Section~\ref{s:numerical_results} details results from numerical experiments for a series of test cases. Finally, conclusions are drawn in Section~\ref{s:conclusions}.

\section{Discontinuous Spectral Element Formulations on Hexahedral Elements}\label{s:formulation}

For the sake of completeness, we briefly describe in the following sections the SD and FR schemes on tensor product hexahedral elements such that a self-contained comparison of the different formulations can be made.

\subsection{Element mapping}
We will begin by prescribing the shared definitions for partitioning the domain, reference domain, and how transformation from the reference domain and physical domain are constructed. The arbitrary connected solution domain $\mathrm{\Omega}\subset\mathbb{R}^3$ is partitioned into $N_e$ non-overlapping, conforming, hexahedral elements, each denoted by $\mathrm{\Omega}_e$, such that
\begin{align}
    \mathrm{\Omega} = \bigcup_{e=1}^{N_e} \mathrm{\Omega}_e,
    \hspace{0.4in}
    \bigcap_{e=1}^{N_e} \mathrm{\Omega}_e = \emptyset.
\end{align}
Each three-dimensional physical element $\mathrm{\Omega}_e$ is mapped to a reference element $\mathrm{\Omega}_r = \left\{ \xi,\eta,\beta~|~-1 \leqslant \xi,\eta,\beta \leqslant 1 \right\}$ through a mapping of the form
\begin{align}
	\bm{x}(\xi,\eta,\beta) &= \sum^K_{k=1} \bm{x}_k \phi_k(\xi,\eta,\beta),
	\label{e:xyz_transform}
\end{align}
%
where $K$ is the number of nodes per element $\mathrm{\Omega}_e$, $\bm{x}_k=(x_k,y_k,z_k)$ are nodal Cartesian coordinates, and $\phi_{k} \left( \xi,\eta,\beta \right)$ are the nodal shape functions.
After transformation into the computational domain, the governing equations in Eq.~(\ref{e:strongconservation_3d}) can be re-written in the form
\begin{align}
    \px{\hb{U}}{t} + \px{\hb{f}}{\xi}  + \px{\hb{g}}{\eta} + \px{\hb{h}}{\beta} = 0	
    \label{e:strongconservation_3d_hat}
\end{align}
where the relationship between physical and reference quantities for a stationary mesh is given by
\begin{align}
    \hb{U} = |\bm{J}| \bm{U},
    \hspace{0.4in}
    \begin{bmatrix}
        \hb{f} \\ \hb{g} \\ \hb{h}
    \end{bmatrix}
    =
    |\bm{J}| \bm{J}^{-1}
    \begin{bmatrix}
        \bm{f} \\ \bm{g} \\ \bm{h}
    \end{bmatrix}.
\end{align}
For stationary grids, the Jacobian is defined as $\bm{J}=\partial(x,y,z)/\partial(\xi,\eta,\beta)$. This information is needed at both the solution and flux points within each reference element in accordance with the spectral difference and flux reconstruction methodologies described in Sections~\ref{s:sd} and \ref{s:fr}.

\subsection{Spectral difference}\label{s:sd}

\begin{figure}[t]
        \renewcommand{\fsize}{55mm}
        \centering
        \subfloat[]{\includegraphics[width=\fsize,keepaspectratio]
            {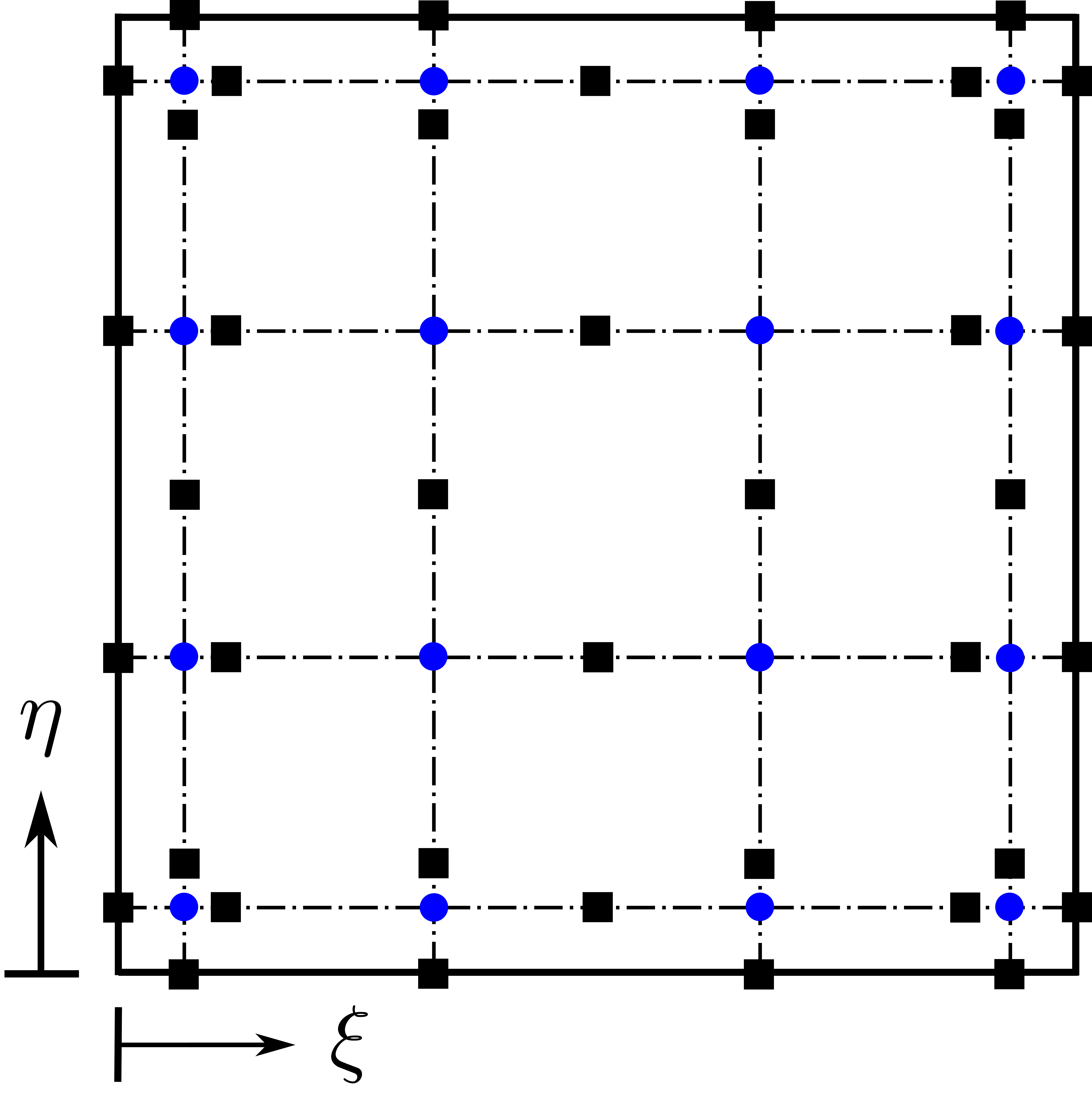}
            \label{f:SD_p3}
        }
        \subfloat[]{\includegraphics[width=\fsize,keepaspectratio]
            {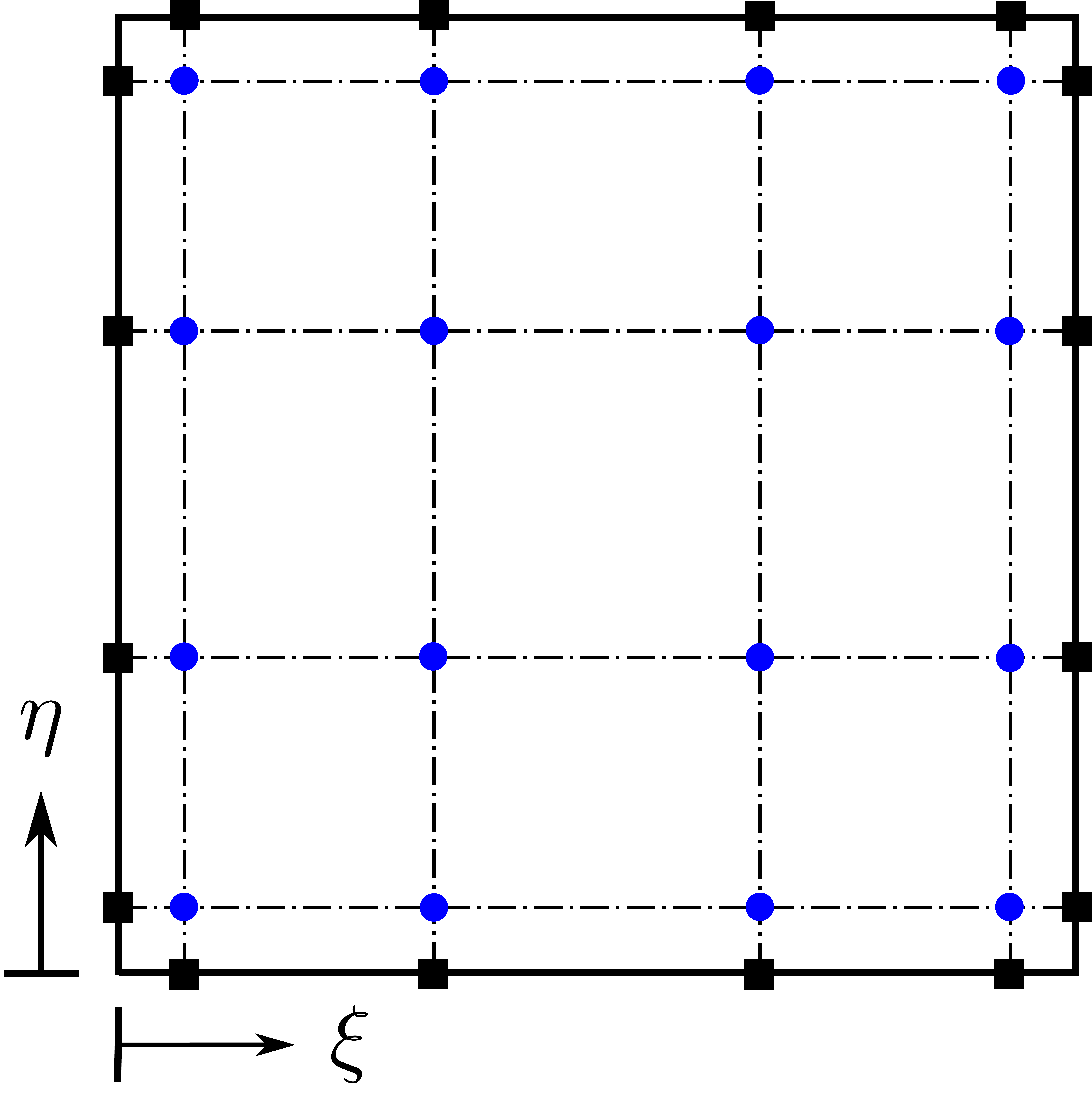}
            \label{f:FR_p3}
        }
        
        \caption{Distribution of solution points (SP {\color{blue} {\Large$\bullet$}}) and flux points (FP $\blacksquare$) with $\p=3$ for (a) spectral difference method and (b) flux reconstruction method inside a unit reference element $\mathrm{\Omega}_r$.}
        \label{f:FR_SD_p3}
    \end{figure}

    Following the original work of Kopriva and Kolias~\cite{Kopriva1996,Kopriva1996b} and Lui~\etal~\cite{liu-vinokur-wang:2006}, we briefly describe here the three-dimensional spectral difference formulation for which the distribution of solution points in a reference cube can be interpreted from the distribution of points in the reference square shown in Fig.~\ref{f:SD_p3}. In this two-dimensional representation, the number of solution points (blue circles) along each direction is four---these points, representing a polynomial of order $\p=3$, are located at Gauss--Legendre quadrature points. The number of flux points (black squares) along each direction is one higher than the number of solution points---these points are also located at Gauss--Legendre quadrature points in the interior plus the two end points at -1 and 1. Using the $\p+1$ solution points and the $\p+2$ flux points, two sets of Lagrange interpolating polynomials---of degree $\p$ and $\p+1$---along the $\xi$ direction can be built using
    \begin{subequations}
        \begin{align}
            \mathcal{l}_i(\xi) &= \prod_{\substack{s=1\\s\neq i}}^{\p+1} \bigg(\frac{\xi-\xi_s}{\xi_i-\xi_s}\bigg) \quad \forall~i\in \{1, \dots, \p+1\},\label{e:lagrange_l}\\
            \mathcal{h}_{i+\half}(\xi) &= \prod_{\substack{s=0\\s\neq i}}^{\p+1}\bigg(\frac{\xi-\xi_{s+\half}}{\xi_{i+\half} - \xi_{s+\half}}\bigg) \quad \forall~i\in \{0, \dots, \p+1\}\label{e:lagrange_h},
        \end{align}
    \end{subequations}
    with analogous definitions made for the $\eta$ and $\beta$ directions. Here it can be observed that $\mathcal{l}_i(\xi_s)=\delta_{is}$, and the complete polynomial approximation can be obtained within $\mathrm{\Omega}_r$ through tensor products of the three $\p$ degree one-dimensional Lagrange polynomials by
    \begin{align}
        \bd{U}_r(\xi,\eta,\beta) &=
            \sum_{k=1}^{\p+1} \sum_{j=1}^{\p+1} \sum_{i=1}^{\p+1}
            \frac{\hbd{U}_{r|i,j,k}}{|\bm{J}_{r|i,j,k}|}
            ~\mathcal{l}_i(\xi) ~ \mathcal{l}_j(\eta) ~ \mathcal{l}_k(\beta)
            \label{e:reconstructedU_3d_SD}
    \end{align}
    where $\hbd{U}_{r|i,j,k}=\hbd{U}_r(\xi_i,\eta_j,\beta_k)$ are the nodal coefficients of the solution in $\mathrm{\Omega}_r$ that represent the value of the approximate solution polynomial $\hbd{U}_r$ evaluated at the set of solution points. The values of the flux vectors can be obtained in a similar manner, but instead by using the three $\p+1$ degree one-dimensional polynomials $\mathcal{h}_{i+\frac{1}{2}}$, $\mathcal{h}_{j+\frac{1}{2}}$ and $\mathcal{h}_{k+\frac{1}{2}}$ by
    \begin{subequations}
    \begin{align}
        \hbd{f}_r(\xi,\eta,\beta) &= \sum_{k=1}^{\p+1}\sum_{j=1}^{\p+1}\sum_{i=0}^{\p+1}
            \hbd{f}_{r|i+\half,j,k}~\mathcal{h}_{i+\half}(\xi)~\mathcal{l}_j(\eta)~\mathcal{l}_k(\beta)\label{e:reconstructedF_3d_SD},\\
        \hbd{g}_r(\xi,\eta,\beta) &= \sum_{k=1}^{\p+1}\sum_{j=0}^{\p+1}\sum_{i=1}^{\p+1}
            \hbd{g}_{r|i,j+\half,k}~\mathcal{l}_i(\xi)~\mathcal{h}_{j+\half}(\eta)~\mathcal{l}_k(\beta)\label{e:reconstructedG_3d_SD},\\
        \hbd{h}_r(\xi,\eta,\beta) &= \sum_{k=0}^{\p+1}\sum_{j=1}^{\p+1}\sum_{i=1}^{\p+1}
            \hbd{h}_{r|i,j,k+\half}~\mathcal{l}_i(\xi)~\mathcal{l}_j(\eta)~\mathcal{h}_{k+\half}(\beta)\label{e:reconstructedH_3d_SD}.
    \end{align}
    \end{subequations}
    The nodal coefficients, $\hbd{f}_{r|i+1/2,j,k}=\hbd{f}_r(\xi_{i+1/2},\eta_j,\beta_k)$, of the approximate discontinuous fluxes, $\hbd{f}_r$, are computed from the solution at the flux points $\hat{\bm{U}}_{r|i+1/2,j,k}$ obtained by Eq.~(\ref{e:reconstructedU_3d_SD}). Similar expressions can be defined for $\hbd{g}$ and $\hbd{h}$. The gradients at the solution points are computed using the solution at the flux points with the derivative of the Lagrange polynomial approach (see Sun~\etal~\cite{sun-wang-liu:2007}). The gradients can then be interpolated from the solution points to the flux points using a similar Lagrange interpolation approach given in Eq.~(\ref{e:reconstructedU_3d_SD}) to obtain the terms $\hnabla\hbd{U}_{r|i+1/2,j,k}$, $\hnabla\hbd{U}_{r|i,j+1/2,k}$, and $\hnabla\hbd{U}_{r|i,j,k+1/2}$. These gradients are needed only for evaluation of the viscous fluxes.
    
    The common inviscid flux $\hbI{f}_e(\bd{U}_L,\bd{U}_R)$ at an interface between elements in the reference space can be computed using any suitable approximate or exact Riemann solver, where the subscripts $L$ and $R$ denote left and right states of an \emph{interface}. Similar expressions can be defined for $\hbI{g}_e$ and $\hbI{h}_e$. In the SD implementation, the common viscous fluxes such as $\hbI{f}_v(\bd{U}_L,\nabla\bd{U}_L,\bd{U}_R,\nabla\bd{U}_R)$ are computed using an approach analogous to inviscid Riemann solvers. In this work, we use the simple averaging approach from Bassi and Rebay (BR1)~\cite{bassi-rebay:1997}. Note that the fluxes in Eqs.~(\ref{e:reconstructedF_3d_SD})-(\ref{e:reconstructedH_3d_SD}) are continuous within each element, but discontinuous across element interfaces. Globally continuous fluxes can be achieved in SD by replacing the interpolated values of the fluxes at element interfaces (denoted by a $1/2$ or $\p+3/2$ index) with the common fluxes such that derivatives of the continuous fluxes can then be written as
    \begin{subequations}
    \begin{align}
        \px{\hbC{f}_r}{\xi} &= \sum_{k=1}^{\p+1}\sum_{j=1}^{\p+1}
            \Bigg[\hbI{f}_{r|\half,j,k} \dx{\mathcal{h}_{\half}(\xi)}{\xi} +
                   \hbI{f}_{r|\p+\frac{3}{2},j,k}\dx{\mathcal{h}_{\p+\frac{3}{2}}(\xi)}{\xi} + 
            \sum_{i=1}^{\p}\hbd{f}_{r|i+\half,j,k}~\dx{\mathcal{h}_{i+\half}(\xi)}{\xi}\Bigg]~\mathcal{l}_j(\eta)~
            \mathcal{l}_k(\beta),\label{e:dfhat_3D_SD}\\
        \px{\hbC{g}_r}{\eta} &= \sum_{k=1}^{\p+1}\sum_{i=1}^{\p+1}
            \Bigg[\hbI{g}_{r|i,\half,k} \dx{\mathcal{h}_{\half}(\eta)}{\eta} +
                   \hbI{g}_{r|i,\p+\frac{3}{2},k} \dx{\mathcal{h}_{\p+\frac{3}{2}}(\eta)}{\eta} + 
            \sum_{j=1}^{\p}\hbd{g}_{r|i,j+\half,k}~\dx{\mathcal{h}_{j+\half}(\eta)}{\eta}\Bigg]~\mathcal{l}_i(\xi)~
            \mathcal{l}_k(\beta),\label{e:dghat_3D_SD}\\
        \px{\hbC{h}_r}{\beta} &= \sum_{j=1}^{\p+1}\sum_{i=1}^{\p+1}
            \Bigg[\hbI{h}_{r|i,j,\half} \dx{\mathcal{h}_{\half}(\beta)}{\beta} +
                  \hbI{h}_{r|i,j,\p+\frac{3}{2}} \dx{\mathcal{h}_{\p+\frac{3}{2}}(\beta)}{\beta} + 
            \sum_{k=1}^{\p}\hbd{h}_{r|i,j,k+\half}~\dx{\mathcal{h}_{k+\half}(\beta)}{\beta} \Bigg]~\mathcal{l}_i(\xi)~
            \mathcal{l}_j(\eta)\label{e:dhhat_3D_SD}.
    \end{align}
    \end{subequations}

\subsection{Flux reconstruction}\label{s:fr}
    Following the original work by Huynh~\cite{huynh:2007,huynh:2009}, we briefly describe here the three-dimensional flux reconstruction formulation for which the distribution of solution points in a reference cube can be interpreted from the distribution of points in the reference square shown in Fig.~\ref{f:FR_p3}. In this 2D representation, the number of solution points (blue circles) along each direction is four---these points, representing a polynomial of order $\p=3$, are located at Gauss--Legendre quadrature points. The flux points (black squares) along each direction are located at the two end points at -1 and 1. Using the solution at the $\p+1$ solution points, a $\p$ degree Lagrange interpolating polynomial along each $\xi$, $\eta$, and $\beta$ direction can be constructed using Eq.~(\ref{e:lagrange_l}). Tensor products may once again be applied on the one dimensional Lagrange polynomial to obtain a complete approximation of the solution and the fluxes by
    \begin{subequations}
    \begin{align}
        \bd{U}_r(\xi,\eta,\beta) &= \sum_{k=1}^{\p+1}\sum_{j=1}^{\p+1}\sum_{i=1}^{\p+1}
            \frac{\hbd{U}_{r|i,j,k}}{|\bm{J}_{r|i,j,k}|}~\mathcal{l}_i(\xi)~\mathcal{l}_j(\eta)~\mathcal{l}_k(\beta)\label{e:reconstructedU_3d},\\
        \hbd{f}_r(\xi,\eta,\beta) &= \sum_{k=1}^{\p+1}\sum_{j=1}^{\p+1}\sum_{i=1}^{\p+1}
            \hbd{f}_{r|i,j,k}~\mathcal{l}_i(\xi)~\mathcal{l}_j(\eta)~\mathcal{l}_k(\beta)\label{e:reconstructedF_3d},\\
        \hbd{g}_r(\xi,\eta,\beta) &= \sum_{k=1}^{\p+1}\sum_{j=1}^{\p+1}\sum_{i=1}^{\p+1}
            \hbd{g}_{r|i,j,k}~\mathcal{l}_i(\xi)~\mathcal{l}_j(\eta)~\mathcal{l}_k(\beta)\label{e:reconstructedG_3d},\\
        \hbd{h}_r(\xi,\eta,\beta) &= \sum_{k=1}^{\p+1}\sum_{j=1}^{\p+1}\sum_{i=1}^{\p+1}
            \hbd{h}_{r|i,j,k}~\mathcal{l}_i(\xi)~\mathcal{l}_j(\eta)~\mathcal{l}_k(\beta)\label{e:reconstructedH_3d}.
    \end{align}
    \end{subequations}
    In FR, as in SD, the nodal coefficients, $\hbd{f}_{r|i,j,k}=\hbd{f}_r(\xi_i,\eta_j,\beta_k)$, of the approximate discontinuous fluxes $\hbd{f}_r$ are computed from $\hbd{U}_{r|i,j,k}$ and $\hnabla\hbd{U}_{r|i,j,k}$, where the latter term is only required for the viscous fluxes. Similar expressions can be defined for $\hbd{g}$ and $\hbd{h}$. In accordance with the methodology of the flux reconstruction scheme, the continuous flux functions defined along $\xi$, $\eta$ and $\beta$ directions can be written as
    \begin{subequations}
    \begin{align}
        \hbC{f}_r &= \hbd{f}_r + \big[\hbI{f}_{r-\half} - \hbd{f}_r(-1,\eta,\beta)\big]g_{LB}(\xi) +
            \big[\hbI{f}_{r+\half} - \hbd{f}_r(1,\eta,\beta)\big]g_{RB}(\xi),\label{e:fhat_FR}\\
        \hbC{g}_r &= \hbd{g}_r + \big[\hbI{g}_{r-\half} - \hbd{g}_r(\xi,-1,\beta)\big]g_{LB}(\eta) +
            \big[\hbI{g}_{r+\half} - \hbd{g}_r(\xi,1,\beta)\big]g_{RB}(\eta),\label{e:ghat_FR}\\
        \hbC{h}_r &= \hbd{h}_r + \big[\hbI{h}_{r-\half} - \hbd{h}_r(\xi,\eta,-1)\big]g_{LB}(\beta) +
            \big[\hbI{h}_{r+\half} - \hbd{h}_r(\xi,\eta,1)\big]g_{RB}(\beta)\label{e:hhat_FR}
    \end{align}
    \end{subequations}
    where $g_{LB}$ and $g_{RB}$ represent left boundary (LB) and right boundary (RB) correction functions in the reference element, respectively. A stable correction function as defined by Huynh~\cite{huynh:2007} and Vincent~\etal~\cite{vincent-castonguay-jameson:2011} can be generalised for the left boundary as
    \begin{equation}
        g_{LB}(\xi) = \alpha \mathcal{R}_{R,\mathcal{p}+1}(\xi) + (1-\alpha)\mathcal{R}_{R,\mathcal{p}}(\xi)
        \label{e:general_correction_function}
    \end{equation}
    where $\mathcal{R}_{R,(\cdot)}(\xi)$ represents the right Radau polynomial~\cite{Radau1880}. The expression for a correction to the right boundary is obtained simply by reflection of $g_{LB}(\xi)$ such that $g_{RB}(\xi) = g_{LB}(-\xi)$ on the interval $\mathrm{\Omega}_r=\{\xi~|~-1 \leqslant \xi \leqslant 1\}$. Choosing $\alpha=1$ for the correction function in Eq.~(\ref{e:general_correction_function}) recovers the collocation based nodal DG method. Alternatively, choosing $\alpha=(\mathcal{p}+1) / (2\mathcal{p}+1)$ recovers a modified SD method---in the current work, it is this scheme to which we directly compare true SD. Another type of scheme can be obtained by setting $\alpha=\mathcal{p} / (2\mathcal{p}+1)$, which leads to the lumped Lobatto $g_2$ scheme identified by Huynh~\cite{huynh:2007} that collocates solution points with the Lobatto points. These three schemes are referred to herein as $\FRDG$, $\FRSD$, and $\mathrm{FR_2}$, respectively. Lastly, Romero~\etal~\cite{romero:2016} provided a simplified formulation of the FR scheme that substitutes a Lagrange interpolation operation for the correction functions. They offered a proof of equivalence of their scheme to $\FRDG$, provided that solution points are placed at the corresponding Gauss--Legendre points. This method is referred to as direct FR ($\DFR$)~\cite{romero:2016,huynh:2020}.
    
    From Eqs.~(\ref{e:fhat_FR})-(\ref{e:hhat_FR}), we can obtain the derivatives of the continuous flux functions
    \begin{subequations}
        \begin{align}
            \begin{split}
                \px{\hbC{f}_r}{\xi} = \sum_{k=1}^{\p+1}\sum_{j=1}^{\p+1}\sum_{i=1}^{\p+1}
                    \hbd{f}_{r|i,j,k}~\dx{\mathcal{l}_i(\xi)}{\xi}~\mathcal{l}_j(\eta)~\mathcal{l}_k(\beta)
                    &+ \big[\hbI{f}_{r-\half} - \hbd{f}_r(-1,\eta,\beta) \big]\dx{g_{LB}(\xi)}{\xi} \\
                    &+ \big[\hbI{f}_{r+\half} - \hbd{f}_r(1,\eta,\beta) \big]\dx{g_{RB}(\xi)}{\xi}\label{e:dfhat_3D_FR},
            \end{split}
        \end{align}
        \begin{align}
            \begin{split}
                \px{\hbC{g}_r}{\eta} = \sum_{k=1}^{\p+1}\sum_{j=1}^{\p+1}\sum_{i=1}^{\p+1}
                    \hbd{g}_{r|i,j,k}~\mathcal{l}_i(\xi)~\dx{\mathcal{l}_j(\eta)}{\eta}~\mathcal{l}_k(\beta)
                    &+ \big[\hbI{g}_{r-\half} - \hbd{g}_r(\xi,-1,\beta) \big]\dx{g_{LB}(\eta)}{\eta} \\
                    &+ \big[\hbI{g}_{r+\half} - \hbd{g}_r(\xi,1,\beta) \big]\dx{g_{RB}(\eta)}{\eta}\label{e:dghat_3D_FR},
            \end{split}
        \end{align}
        \begin{align}
            \begin{split}
                \px{\hbC{h}_r}{\beta} = \sum_{k=1}^{\p+1}\sum_{j=1}^{\p+1}\sum_{i=1}^{\p+1}
                    \hbd{h}_{r|i,j,k}~\mathcal{l}_i(\xi)~\mathcal{l}_j(\eta)~\dx{\mathcal{l}_k(\beta)}{\beta} 
                    &+ \big[\hbI{h}_{r-\half} - \hbd{h}_r(\xi,\eta,-1) \big]\dx{g_{LB}(\beta)}{\beta} \\
                    &+ \big[\hbI{h}_{r+\half} - \hbd{h}_r(\xi,\eta,1) \big]\dx{g_{RB}(\beta)}{\beta}\label{e:dhhat_3D_FR}.
            \end{split}
        \end{align}
    \end{subequations}
    In the FR implementation, the common viscous fluxes are computed using a BR2-type, second procedure of Bassi and Rebay~\cite{bassi-rebay:2000} written using the flux reconstruction methodology~\cite{huynh:2009} to achieve compactness of the stencil in multiple dimensions.
    
    Once the divergence of the continuous flux is obtained by Eqs.~(\ref{e:dfhat_3D_SD})-(\ref{e:dhhat_3D_SD}) for the spectral difference scheme or Eqs.~(\ref{e:dfhat_3D_FR})-(\ref{e:dhhat_3D_FR}) for the flux reconstruction scheme, an appropriate time stepping technique can be applied to march the solution forward in time. The implementation of both schemes is done within a single coding framework such that fair and proper comparisons of the two methodologies can be made in terms of stability, accuracy, and performance.
\section{Nonlinear Stability of Spectral Difference}\label{s:sd_stability}
	
	In the work of Jameson~\etal~\cite{jameson-castonguay-vincent:2012}, the non-linear stability of the flux reconstruction method was investigated, and it was found that the solution decay could be decomposed into a stable component and a non-linear component which can cause instabilities. As this analysis was useful in understanding the mechanism by which non-linearities affect stability and how de-aliasing methods can mitigate this, we will perform a similar analysis for the spectral difference method in order to highlight the differences that arise between these two techniques. 
	
	Consider a scalar conservation law in one dimension
	\begin{equation}
		\px{u}{t} + \px{f}{x} = 0,
	\end{equation}
	where the lower-case terms denote a scalar quantity. This may be cast in the reference domain as
	\begin{equation}
		\px{\hud}{t} + \px{\hfd}{\xi} = 0.
	\end{equation}
	As was introduced in the previous section, the approximate flux $\hfd$ in the FR and SD methodologies is replaced by a corrected flux $\hfC$ that enforces $C^0$ continuity in the flux between elements. A similar expression for the corrected flux used in the flux reconstruction method can be written for the spectral difference method as 
	\begin{equation*}
		\hfC = \hfd + \left(\hfI_{L} - \hfd_{L}\right)\mathcal{h}_{\half}  + \left(\hfI_{R} - \hfd_{R}\right)\mathcal{h}_{\p+\frac{3}{2}}.
	\end{equation*}
	For brevity, we temporarily drop the subscript $r$ and refer to left and right interfaces of a given element with the subscripts $L$ and $R$. Therefore, 
	\begin{equation}\label{eq:sd}
		\px{\hud}{t} = -\px{\hfd}{\xi} \;\;-\;\; \left(\hfI_{L} - \hfd_{L}\right)\dx{\mathcal{h}_{\half}}{\xi} 
		\;\;-\;\; \left(\hfI_{R} - \hfd_{R}\right)\dx{\mathcal{h}_{\p+\frac{3}{2}}}{\xi}.
	\end{equation}
	To analyze stability in a norm that induces a Sobolev space, namely
	\begin{equation}\label{eq:sobolev_norm}
		\|\hat{v}\|^2_{W_{2,\p,\iota}} = \rint{\hat{v}^2 + \frac{\iota}{2}\big(\partial^\p\hat{v}\big)^2}{\xi},
	\end{equation}
	we investigate the behavior of
	\begin{equation}
		\dx{}{t} \|\hat{v}\|^2_{W_{2,\p,\iota}} =  \dx{}{t}\rint{\hat{v}^2 + \frac{\iota}{2}\big(\partial^\p\hat{v}\big)^2}{\xi}.
	\end{equation}	 
	
	By taking Eq.~(\ref{eq:sd}) and, following the work of Jameson~\etal~\cite{jameson-castonguay-vincent:2012}, multiplying it by $\hud$ and integrating, we obtain
	\begin{equation}
		\half\dx{}{t}\rint{(\hud)^2}{\xi} = -\rint{\hud\px{\hfd}{\xi}}{\xi} 
		\;\;-\;\; \left(\hfI_{L} - \hfd_{L}\right)\rint{\hud\dx{\mathcal{h}_{\half}}{\xi}}{\xi}
		\;\;-\;\; \left(\hfI_{R} - \hfd_{R}\right)\rint{\hud\dx{\mathcal{h}_{\p+\frac{3}{2}}}{\xi}}{\xi}.
	\end{equation}
	Upon using the product rule, this may be rewritten as
	\begin{equation}
	    \begin{split}
		    \half\dx{}{t}\rint{(\hud)^2}{\xi} = \rint{\hfd\px{\hud}{\xi}}{\xi}
		    &\;\;+\;\; \left(\hfI_{L} - \hfd_{L}\right)\rint{\mathcal{h}_\half\dx{\hud}{\xi}}{\xi} \\
		    &\;\;+\;\; \left(\hfI_{R} - \hfd_{R}\right)\rint{\mathcal{h}_{\p+\frac{3}{2}}\dx{\hud}{\xi}}{\xi} \;\;+\;\; \left(\hfI_{L}\hud_{L} - \hfI_{R}\hud_{R}\right).
	    \end{split}
	\end{equation}
	Furthermore, taking Eq.~(\ref{eq:sd}) and differentiating it $\p$ times gives
	\begin{equation}\label{eq:sd_p1}
		\px{}{t}\bigg(\pxi{\p}{\hud}{\xi}\bigg) = -\pxi{\p+1}{\hfd}{\xi} 
		\;\;-\;\; \left(\hfI_{L} - \hfd_{L}\right)\dxi{\p+1}{\mathcal{h}_{\half}}{\xi} 
		\;\;-\;\; \left(\hfI_{R} - \hfd_{R}\right)\dxi{\p+1}{\mathcal{h}_{\p+\frac{3}{2}}}{\xi}.
	\end{equation}
	A key difference between this derivation and that for the flux reconstruction method is that $\hfd$ is a polynomial of degree $\p+1$, and therefore the first term on the right-hand side is not zero, but a constant. If Eq.~(\ref{eq:sd_p1}) is then multiplied by the $\p^\mathrm{th}$ derivative of $\hud$ and integrated, the following is obtained
	\begin{equation}
	    \begin{split}
		    \half\dx{}{t}\rint{\bigg(\pxi{\p}{\hud}{\xi}\bigg)^2}{\xi} = -\rint{\pxi{\p+1}{\hfd}{\xi}\pxi{\p}{\hud}{\xi}}{\xi}
		    &-\;\; \left(\hfI_{L} - \hfd_{L}\right)\rint{\pxi{\p}{\hud}{\xi}\dxi{\p+1}{\mathcal{h}_{\half}}{\xi}}{\xi} \\
		    &-\;\; \left(\hfI_{R} - \hfd_{R}\right)\rint{\pxi{\p}{\hud}{\xi}\dxi{\p+1}{\mathcal{h}_{\p+\frac{3}{2}}}{\xi}}{\xi},
	    \end{split}
	\end{equation}	 
	which, in turn, may be written as	
	\begin{equation}\label{eq:sd_int1}
	    \half\dx{}{t}\rint{\bigg(\pxi{\p}{\hud}{\xi}\bigg)^2}{\xi} = -2\pxi{\p+1}{\hfd}{\xi}\pxi{\p}{\hud}{\xi} 
	    \;-\; 2\left(\hfI_{L} - \hfd_{L}\right)\pxi{\p}{\hud}{\xi}\dxi{\p+1}{\mathcal{h}_{\half}}{\xi}
		\;-\; 2\left(\hfI_{R} - \hfd_{R}\right)\pxi{\p}{\hud}{\xi}\dxi{\p+1}{\mathcal{h}_{\p+\frac{3}{2}}}{\xi}.
	\end{equation}	
	By combining Eqs.~(\ref{eq:sd_p1})~and~(\ref{eq:sd_int1}) and taking the norm as given by Eq.~(\ref{eq:sobolev_norm}), we then obtain
	\begin{equation}\label{eq:sd_sobolev_0}
	    \begin{split}
    		\half\dx{}{t}\|\hud\|^2_{W_{2,\p,\iota}} = &\rint{\hfd\px{\hud}{\xi}}{\xi}
    		\;+\; \left(\hfI_{L} - \hfd_{L}\right)\rint{\mathcal{h}_\half\dx{\hud}{\xi}}{\xi} 
    		\;+\; \left(\hfI_{R} - \hfd_{R}\right)\rint{\mathcal{h}_{\p+\frac{3}{2}}\dx{\hud}{\xi}}{\xi} \\
		    &-\;\; \iota\pxi{\p+1}{\hfd}{\xi}\pxi{\p}{\hud}{\xi}
		    \;-\; \iota\left(\hfI_{L} - \hfd_{L}\right)\pxi{\p}{\hud}{\xi}\dxi{\p+1}{\mathcal{h}_\half}{\xi}
		    \;-\; \iota\left(\hfI_{R} - \hfd_{R}\right)\pxi{\p}{\hud}{\xi}\dxi{\p+1}{\mathcal{h}_{\p+\frac{3}{2}}}{\xi}\\
    		&+\;\; \left(\hfI_{L}\hud_{L} - \hfI_{R}\hud_{R}\right).
        \end{split}
	\end{equation}
	
	To proceed further, we refer to the work of Huynh~\cite{huynh:2007} who showed that for the linear case, SD could be recovered from FR for a given correction function. This correction function, which we denote as $g$, recovered SD when the SD flux points were formed from the $\p$ degree Gauss--Legendre quadrature points with points added at $-1$ and $1$. This is the logical choice as Jameson~\etal~\cite{jameson:2010} showed that these points resulted in the only SD scheme with provable linear stability. The connection between the SD and FR formulations is given by 
	\begin{equation}
		\mathcal{h}_\half = g_{L} \quad \mathrm{and} \quad \mathcal{h}_{\p+\frac{3}{2}} = g_{R}.
	\end{equation}
	Therefore, we may write Eq.~(\ref{eq:sd_sobolev_0}) as 
	\begin{equation}\label{eq:sd_sobolev_1}
	    \begin{split}
	        \half\dx{}{t}\|\hud\|^2_{W_{2,\p,\iota}} = &\rint{\hfd\px{\hud}{\xi}}{\xi} 
	        \;\;+\;\; \left(\hfI_{L} - \hfd_{L}\right)\rint{g_{L}\dx{\hud}{\xi}}{\xi}
		    \;\;+\;\; \left(\hfI_{R} - \hfd_{R}\right)\rint{g_{R}\dx{\hud}{\xi}}{\xi}\\
		    &-\;\;\iota\pxi{\p+1}{\hfd}{\xi}\pxi{\p}{\hud}{\xi}
		    \;\;-\;\; \iota\left(\hfI_{L} - \hfd_{L}\right)\pxi{\p}{\hud}{\xi}\dxi{\p+1}{g_{L}}{\xi} 
		    \;\;-\;\; \iota\left(\hfI_{R} - \hfd_{R}\right)\pxi{\p}{\hud}{\xi}\dxi{\p+1}{g_{R}}{\xi}\\
		    &+\;\; (\hfI_{L}\hud_{L} - \hfI_{R}\hud_{R}),
	    \end{split}
	\end{equation}
	and from Vincent~\etal~\cite{vincent-castonguay-jameson:2011}, we use
	\begin{subequations}
		\begin{align}
			\rint{g_{L}\px{\hud}{\xi}}{\xi} - \iota\pxi{\p}{\hud}{\xi}\dxi{\p+1}{g_{L}}{\xi} &= 0, \\
			\rint{g_{R}\px{\hud}{\xi}}{\xi} - \iota\pxi{\p}{\hud}{\xi}\dxi{\p+1}{g_{R}}{\xi} &= 0,
		\end{align}
	\end{subequations}
	which reduces Eq.~(\ref{eq:sd_sobolev_1}) to
	\begin{equation}
		\half\dx{}{t}\|\hud\|^2_{W_{2,\p,\iota}} = \rint{\hfd\px{\hud}{\xi}}{\xi} 
		+ \left(\hfI_{L}\hud_{L} - \hfI_{R}\hud_{R}\right) - \iota\pxi{\p+1}{\hfd}{\xi}\pxi{\p}{\hud}{\xi}.
	\end{equation}
	If the broken norm is then constructed from this for $N$ elements on a periodic domain, we obtain
	\begin{equation}
		\half\dx{}{t}\|u^\delta\|^2_{W_{2,\p,\iota}} = \Theta + \sum_{i=0}^{N-1}\epsilon_i - \iota\sum^{N-1}_{i=0}\pxi{\p+1}{f^\delta}{x}\pxi{\p}{u^\delta}{x},
	\end{equation}
	where 
	\begin{equation}
	    \epsilon_i = \int_{\mathrm{\Omega}_i}(f^\delta - f)\px{u^\delta}{x} \mathrm{d}x.
	\end{equation}
	Here, the term $\Theta$ is the interface contribution to the stability for which a full derivation can be found in \cite{jameson-castonguay-vincent:2012}, and the reader is referred to that work for a more complete derivation. If the common interface values are set such that they form an E-flux~\cite{Lax1957,Osher1984}, then $\Theta \leqslant 0$ and therefore the stability is controlled by the latter two terms. In contrast, the last term is not present in FR and it is possible that this term could have a stabilizing effect for SD. 
	
    To illustrate more clearly the effect that the difference between the schemes has on the approximation of the flux gradient, we will now examine the error scaling. Using theorems and corollaries presented by Bernardi and Maday~\cite{Bernardi1997}, we further analyze the behavior of the error in the flux evaluated in the $L^2$ norm
    \[
    \bigg\|\px{f}{x}-\px{f^{\delta C}}{x}\bigg\|_{L^2}.
    \]
    Throughout, we adopt a similar notation to Bernardi and Maday~\cite{Bernardi1997}, where we define the Sobolev space
    \begin{equation*}
        H^k(X) = \Big\{v\in L^2(X)\;\; |\;\; \forall\: m\in \mathbb{N},\; m\leqslant k,\; \partial^mv\in L^2(X) \Big\},
    \end{equation*}
    where $X$ is an open, bounded, Lipschitz-continuous set of $\mathbb{R}$, and the norm induced on the space $H^k$ is
    \begin{equation}
        \|u\|_{H^k} = \sqrt{\int_X \sum^k_{m=0}\big(\partial^mu\big)^2\mathrm{d}x}.
    \end{equation}
    To analyze the behavior of the flux error, we establish two necessary theorems. 
    \begin{theorem}\label{thm:u} (See Bernardi and Maday~\cite{Bernardi1997}, Thm.~13.2.)
        For some function $u\in H^k$ with $k>1/2$ and the Lagrange interpolation operator $I_\p g\in \mathbb{P}_\p$ such that $I_\p g(\zeta_i)=g(\zeta_i)$ for some points set of points $\{\zeta_i\}_{i\leqslant \p+1}$, the following estimate holds
        \begin{equation}
            \|u-I_\p u\|_{L^2} \leqslant C(k)(\p+1)^{-k}\|u\|_{H^k},
        \end{equation}
        for some constant $C$ that is only dependent on $k$.
	\end{theorem}
	\begin{theorem}\label{thm:du} (See Bernardi and Maday~\cite{Bernardi1997}, Thm.~13.4.)
	    For some function $u\in H^k$ with real numbers $k$ and $r$ such that $k\geqslant 1$ and $r<k$ and the Lagrange interpolation operator $I_\p g$ defined in Thm. \ref{thm:u}, the following estimates hold
	    \begin{equation}
	        \|u - I_\p u\|_{H^r} \leqslant
	            \begin{cases}
	            C(\p+1)^{3r/2-k}\|u\|_{H^k} &\mbox{if } r\leqslant1,\\
	            C(\p+1)^{2r-1/2-k}\|u\|_{H^k} &\mbox{if } r\geqslant1.
	            \end{cases}
	   \end{equation}
	\end{theorem}
	With these theorems established, we look to determine the bound on 
	\begin{equation}
        \bigg\|\px{f}{x}-\px{f^{\delta C}}{x}\bigg\|_{L^2} = \bigg\|\px{f}{x} - \bigg[\px{f^\delta}{x} + \left(f^{\delta I}_{L} - f^\delta_{L}\right)\dx{g_{L}}{x}+ \left(f^{\delta I}_{R} - f^\delta_{R}\right)\dx{g_{R}}{x}\bigg]\bigg\|_{L^2}.
	\end{equation}
	From the triangle inequality, this may be rewritten as 
	\begin{equation}
	    \bigg\|\px{f}{x}-\px{f^{\delta C}}{x}\bigg\|_{L^2} \leqslant \bigg\|\px{f}{x}-\px{f^\delta}{x}\bigg\|_{L^2} + \big|f^{\delta I}_{L}-f^\delta_{L}\big|\bigg\|\dx{g_{L}}{x}\bigg\|_{L^2} + \big|f^{\delta I}_{R}-f^\delta_{R}\big|\bigg\|\dx{g_{R}}{x}\bigg\|_{L^2}.
	\end{equation}
	We impose that the interface values take the form
	\begin{equation}
	    f^{\delta I}_{r|L} = \kappa_{L}f^\delta_{r-1|R} + (1-\kappa)f^\delta_{r|L}, \quad \mathrm{for} \quad \kappa_{L} \in[0,1],
	\end{equation}
    where $r$ denotes the element index, and impose similar behavior at the opposite interface. We may then write 
	\begin{equation}\label{eq:sd_ineq}
	    \bigg\|\px{f}{x}-\px{f^{\delta C}}{x}\bigg\|_{L^2} \leqslant \bigg\|\px{f}{x}-\px{f^\delta}{x}\bigg\|_{L^2} 
	    + \kappa_{L}\big|f^\delta_{r-1|R} - f^\delta_{r|L}\big|\bigg\|\dx{g_{L}}{x}\bigg\|_{L^2} 
	    + \kappa_{R}\big|f^\delta_{r|R} - f^\delta_{r+1|L}\big|\bigg\|\dx{g_{R}}{x}\bigg\|_{L^2}.
	\end{equation}
	As the true interface term is the same for both sides of the interface, we may write
	\begin{equation}
    	\kappa_{L}\left(f^\delta_{r-1|R} - f^\delta_{r|L}\right) = \kappa_{L}\left(f^\delta_{r-1|R} - f_{r-1|R} + f_{r|L} - f^\delta_{r|L}\right),
    \end{equation}
    which can be generalized for the other interface. Under the assumption that $f$ is a high-order function of $u$ such that if $u\in H^k$ then $f\in H^{mk}$ for $m \geqslant 1$, the interface correction will scale with the interpolation error of $f$
    \begin{equation}
        \kappa_{L}\big|f^\delta_{r-1|R} - f^\delta_{r|L}\big| \leqslant \kappa_{L}C(\p+2)^{-mk}\|f\|_{H^{mk}}.
    \end{equation}
    It is then straightforward to prove the following bound for a Lagrange polynomial 
    \begin{equation*}
        \bigg\|\dx{h_{i+\half}}{x}\bigg\|_{L^2} \leqslant C(\p+2),
    \end{equation*}
    and as the correction function for SD is a Lagrange polynomial, we may use this to give
    \begin{equation}
        \kappa_{L}\big|f^\delta_{r-1|R} - f^\delta_{r|L}\big|\bigg\|\dx{g_L}{x}\bigg\|_{L^2} \leqslant \kappa_{L}C(\p+2)^{1-mk}\|f\|_{H^{mk}}.
    \end{equation}
    Considering the first term on the right-hand-side of Eq.~(\ref{eq:sd_ineq}), we can modify Thm.~\ref{thm:du} to yield
    \begin{equation*}
        \bigg\|\px{f}{x}-\px{f^{\delta}}{x}\bigg\|_{L^2} \leqslant \big\|f-f^\delta\big\|_{H^1} \leqslant C(\p+2)^{3/2-mk}\|f\|_{H^{mk}}.
    \end{equation*}
    Combining these results and taking into account that $\kappa\in[0,1]$, the upper bound is found to be 
    \begin{equation}\label{eq:sd_error}
        \bigg\|\px{f}{x}-\px{f^{\delta C}}{x}\bigg\|_{L^2} \leqslant C(\p+2)^{3/2-mk}\|f\|_{H^{mk}}.
    \end{equation}
	Repeating these steps for FR, we find the similar and expected relation that 
    \begin{equation}\label{eq:fr_error}
        \bigg\|\px{f}{x}-\px{f^{\delta C}}{x}\bigg\|_{L^2} \leqslant C(\p+1)^{3/2-mk}\|f\|_{H^{mk}}.
    \end{equation}
    As a result, the error of SD can be lower due to the different scaling, the difference being most evident when the ratio $(\p+2)/(\p+1)$ is largest and $k$ is large compared to $\p$ (i.e. in under-resolved cases). We remark that this result is separate from arguments concerning the study of the scheme's asymptotic rate of convergence with respect to grid spacing. In that case, it is known that DG-type FR schemes can obtain super-convergence one degree higher than SD and other FR variants~\cite{vincent-castonguay-jameson:2011b,Asthana2017}.
\section{Governing Equations}\label{s:governing}
Consider the full three-dimensional compressible Navier--Stokes equations written in strong conservation form for a Cartesian coordinate system $\left(x,y,z\right)$
\begin{align}
    \px{\bm{U}}{t} + \px{\bm{f}}{x} + \px{\bm{g}}{y} + \px{\bm{h}}{z} = 0	
    \label{e:strongconservation_3d}.
\end{align}
The vector of state variables, $\bm{U}(x,y,z,t)$, is defined for $[x,y,z]\in \mathrm{\Omega}\subset \mathbb{R}^3$ and $t \in \mathbb{R}^+$, with $\bm{U}=[\rho~\rho u~\rho v~\rho w~\rho E]^T$ and the flux vectors $\bm{f}$, $\bm{g}$ and $\bm{h}$ contain both inviscid terms, denoted by $(\cdot)_e$, and viscous terms, denoted by $(\cdot)_v$, where
\begin{align}
    \bm{f} &= \bm{f}_e-\bm{f}_v, &
    \bm{g} &= \bm{g}_e-\bm{g}_v, &
    \bm{h} &= \bm{h}_e-\bm{h}_v.
\end{align}
The inviscid flux vectors can be written as
\begin{align}
    \bm{f}_e &= \begin{bmatrix}
                    \rho u\\ \rho u^{2}+p\\ \rho uv\\ \rho uw\\ (\rho E + p)u
                \end{bmatrix}, &
    \bm{g}_e &= \begin{bmatrix}
                    \rho v\\ \rho vu \\ \rho v^{2}+p\\ \rho vw\\ (\rho E + p)v
                \end{bmatrix}, &
    \bm{h}_e &= \begin{bmatrix}
                    \rho w\\ \rho wu \\ \rho wv\\ \rho w^{2}+p\\ (\rho E + p)w
                \end{bmatrix}
\end{align}
and the viscous flux vectors can be written as
\begin{align}
    \bm{f}_v &= \begin{bmatrix}
                    0\\ \tau_{xx}\\ \tau_{xy}\\ \tau_{xz}\\
                    \kappa\frac{\partial T}{\partial x} + u\tau_{xx} + v\tau_{xy} + w\tau_{xz}
                \end{bmatrix}, &
    \bm{g}_v &= \begin{bmatrix}
                    0\\ \tau_{yx}\\ \tau_{yy}\\ \tau_{yz}\\
                    \kappa\frac{\partial T}{\partial y} + u\tau_{yx} + v\tau_{yy} + w\tau_{yz}
                \end{bmatrix}, &
    \bm{h}_v &= \begin{bmatrix}
                    0\\ \tau_{zx}\\ \tau_{zy}\\ \tau_{zz}\\
                    \kappa\frac{\partial T}{\partial z} + u\tau_{zx} + v\tau_{zy} + w\tau_{zz}
                \end{bmatrix}.
\end{align}
The total energy is $E=p/[\rho(\gamma - 1)] + (u^2 + v^2 + w^2)/2$ and the thermal conductivity is $\kappa=(\mu c_p)/Pr$. Under Stokes' hypothesis, the bulk viscosity is assigned a value of zero, leading to the second coefficient of viscosity taking the value $\lambda = -2/3\mu$; therefore, we can write
\begin{subequations}
\begin{align}
    \tau_{xx} &= 2\mu \frac{\partial u}{\partial x} + \lambda \nabla \cdot \bm{u}, &
    \tau_{yy} &= 2\mu \frac{\partial v}{\partial y} + \lambda \nabla \cdot \bm{u}, &
    \tau_{zz} &= 2\mu \frac{\partial w}{\partial z} + \lambda \nabla \cdot \bm{u},
    \\
    \tau_{xy} &= \tau_{yx} = \mu \left( \frac{\partial v}{\partial x} + \frac{\partial u}{\partial y} \right), &
    \tau_{xz} &= \tau_{zx} = \mu \left( \frac{\partial w}{\partial x} + \frac{\partial u}{\partial z} \right), &
    \tau_{yz} &= \tau_{zy} = \mu \left( \frac{\partial w}{\partial y} + \frac{\partial v}{\partial z} \right).
\end{align}
\end{subequations}
In the formulation above, $u$, $v$, and $w$ are the
components of velocity in the $x$, $y$, and $z$ directions, respectively, and
$\rho$ represents the density, $p$ the pressure, $\mu$ the dynamic viscosity, $\nu$ the kinematic viscosity, $Pr$ the Prandtl number, $\gamma$ the specific heat ratio, and $c_p$ the specific heat at constant pressure. Unless stated otherwise, the Prandtl number and specific heat ratio are set constant at $Pr = 0.72$ and $\gamma = 1.4$ for all simulations. 

\section{Numerical Experiments}\label{s:numerical_results}
    The results from a series of numerical experiments performed comparing $\SD$ and $\FRSD$ will now be presented. 
\subsection{Heterogeneous Linear Advection Equation}\label{s:heterogeneous_1D}
We will begin with a 1D linear test case that can be modified such at aliasing is introduced. Given a linear advection equation with variable propagation speed, an equivalent scalar conservation form can be derived
\begin{equation}
	\px{u}{t} + (2-\sin{x})\px{u}{x} = 0 \;\;\Leftrightarrow\;\; \px{u}{t} + \px{(2-\cos{x})u}{x} = u\sin{x}.
\end{equation}
In the latter form, the equation introduces aliasing errors in numerical calculations, and thus is a suitable candidate for identifying de-aliasing properties of numerical schemes without the presence of non-linearities. Furthermore, when this equation is applied to a periodic domain $\mathrm{\Omega}=[0,2\pi]$, the solution is shown to analytically have a time period of $T = 4\pi/\sqrt{3}$, allowing for exact calculations of the error~\cite{Trojak2018a}.

The initial condition for this test was chosen to be a reconstruction of the energy spectra
\begin{equation}
	E(k,t=0) = \frac{Ck^4}{k_0^5}\exp{\bigg(-\frac{k^2}{k_0^2}\bigg)}, \quad \mathrm{where} \quad C = \frac{2}{3\sqrt{\pi}}, \quad \mathrm{and} \quad k_0 = 10,
\end{equation}
which is similar to the condition used by Alhawwary~\etal~\cite{Alhawwary2018} and San~\cite{San2016}. A 1D scalar field was then reconstructed from the spectra as
\begin{equation}
	u(x,t=0) = \sum^{k_\mathrm{max}}_{k=0}\sqrt{2E(k,0)}\cos{(kx + \Psi(k))},
\end{equation}
where $k_\mathrm{max} = 2048$ is some maximum wavenumber and $\Psi(k)\in(0,2\pi]$ is a random phase angle for wavenumber $k$. With this initial condition, multiple modes are excited while $E(k)\rightarrow0$ as $k\rightarrow\infty$, which makes differences in aliasing evident.

\begin{figure}{t}
	\centering
    \subfloat[$120$ DoF.]{
		\centering
			\resizebox{0.5\linewidth}{!}{	\begin{tikzpicture}
		\begin{loglogaxis}[name=plot1,
			xlabel={$k$},
		    xtick={1,10,100},
    		xmin=1,xmax=60,
    		ylabel={$E(k)$},
    		ymin=1e-7,ymax=1e4,
    		grid=both,
    		legend style={at={(0.025,0.35)},anchor=south west,font=\scriptsize},
    		legend cell align={right},
    		style={font=\small}]
			\addplot[color=black, style={thick}]
				table[x=f,y=t05,col sep=comma,unbounded coords=jump]{./figures/data/hetero_120_c.csv};
			\addlegendentry{$t_0, \p=5$}
			\addplot[color=black, dashed, style={thick}] 
				table[x=f,y=t04,col sep=comma,unbounded coords=jump]{./figures/data/hetero_120_c.csv};
			\addlegendentry{$t_0, \p=4$}
			
			\addplot[color={RdBu-C}, style={thick}]
				table[x=f,y=frp5c,col sep=comma,unbounded coords=jump]{./figures/data/hetero_120_c.csv};
			\addlegendentry{$\FRSD, \p=5$}
			\addplot[color={RdBu-F}, style={thick}]
				table[x=f,y=frp4c,col sep=comma,unbounded coords=jump]{./figures/data/hetero_120_c.csv};
			\addlegendentry{$\FRSD, \p=4$}
			\addplot[color={RdBu-J}, style={thick}]
				table[x=f,y=sdp4c,col sep=comma,unbounded coords=jump]{./figures/data/hetero_120_c.csv};
			\addlegendentry{$\SD, \p=4$}
			\addplot[color={RdBu-M}, style={thick}]
				table[x=f,y=sdp5c,col sep=comma,unbounded coords=jump]{./figures/data/hetero_120_c.csv};
			\addlegendentry{$\SD, \p=5$}
		\end{loglogaxis} 		
	\end{tikzpicture}}
		\label{fig:hetero_120}}
    \subfloat[$600$ DoF.]{
		\centering
			\resizebox{0.51\linewidth}{!}{	\begin{tikzpicture}
		\begin{loglogaxis}[name=plot1,
			xlabel={$k$},
    		xmin=1,xmax=300,
    		ylabel={$E(k)$},
    		ymin=1e-14,ymax=1e0,
    		grid=both,
    		legend style={at={(0.05,0.05)},anchor=south west,font=\scriptsize},
    		legend cell align={right},
    		style={font=\small}]
			\addplot[color=black, style={thick}]
				table[x=f,y=t05,col sep=comma,unbounded coords=jump]{./figures/data/hetero_600_c.csv};
			\addlegendentry{$t_0, \p=5$}
			\addplot[color=black, dashed, style={thick}] 
				table[x=f,y=t04,col sep=comma,unbounded coords=jump]{./figures/data/hetero_600_c.csv};
			\addlegendentry{$t_0, \p=4$}
			
			\addplot[color={RdBu-C}, style={thick}]
				table[x=f,y=frp5c,col sep=comma,unbounded coords=jump]{./figures/data/hetero_600_c.csv};
			\addlegendentry{$\FRSD, \p=5$}
			\addplot[color={RdBu-F}, style={thick}]
				table[x=f,y=frp4c,col sep=comma,unbounded coords=jump]{./figures/data/hetero_600_c.csv};
			\addlegendentry{$\FRSD, \p=4$}
			\addplot[color={RdBu-J}, style={thick}]
				table[x=f,y=sdp4c,col sep=comma,unbounded coords=jump]{./figures/data/hetero_600_c.csv};
			\addlegendentry{$\SD, \p=4$}
			\addplot[color={RdBu-M}, style={thick}]
				table[x=f,y=sdp5c,col sep=comma,unbounded coords=jump]{./figures/data/hetero_600_c.csv};
			\addlegendentry{$\SD, \p=5$}
		\end{loglogaxis} 		
	\end{tikzpicture}}
		\label{fig:hetero_600}}
	\caption{Energy spectrum comparison of $\FRSD$ and $\SD$ with centrally-differenced interfaces after one time period averaged over \num{1e3} initial conditions.}
	\label{fig:hetero_comp}
\end{figure}
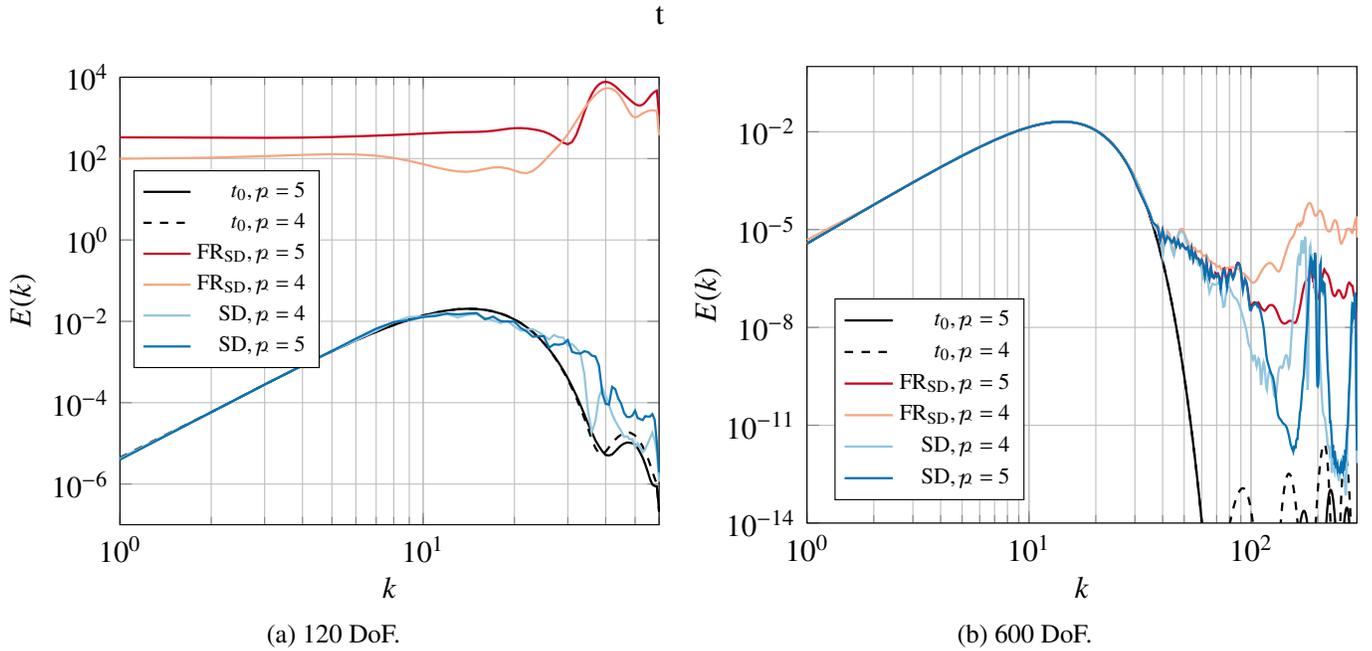

A comparison of the numerical results is shown in Fig.~\ref{fig:hetero_comp} for various polynomial orders and grid resolutions after one time period. For this case, only the average spectra results of the experiments using centrally-differenced interfaces are shown as negligible differences between $\SD$ and $\FRSD$ were observed when upwinding was used. This effect can be attributed to the numerical dissipation of upwinded schemes at high frequencies which can be sufficient to dampen aliasing errors in this case. When using central-differencing on the coarse grid, instabilities were evident in the spectra of the calculations using the FR method, whereas the SD method was stable. As the grid was refined, the FR method became stable but had notably more energy at high wavenumbers than the SD method. As it can be shown analytically for this equation that aliasing will be introduced at the highest wavenumbers and propagated to the lower wavenumbers, it is evident that pure SD is more stable due to less aliasing error.

\subsection{Isentropic Euler Vortex}\label{s:ev}

To demonstrate and compare super-convergence of the flux reconstruction and spectral difference schemes~\cite{Cockburn2000,Adjerid2002} for the Euler equations within the current implementation, we solve the isentropic Euler vortex~\cite{shu:1998} in a free-stream flow for which there exists an exact analytical solution. Super-convergence for these types of schemes is said to be achieved once the observed order of accuracy is greater than $\p+1$. The vortex is initially prescribed a size $r_c$ and strength $\epsilon$, positioned in the domain at $(x_o,y_o)$, and here we will consider a vortex advecting purely in the $y$-direction. The analytical solution at $(x,y,t)$ for this test case is given by
\begin{subequations}
\begin{align}
    \rho(x,y,t) &= \rho_{\infty} \left( 1 - \frac{(\gamma-1) \epsilon^2 M_{\infty}^2}{8 \pi^2} \exp{(2f)} \right)^{\frac{1}{\gamma - 1}}\label{e:ev:rho},\\
    u(x,y,t) &= U_{\infty} \left(\frac{\epsilon (y-y_o-U_\infty t)}{2 \pi r_c} \exp{(f)} \right)\label{e:ev:u},\\
    v(x,y,t) &= U_{\infty} \left( 1 - \frac{\epsilon (x-x_o)}{2 \pi r_c} \exp{(f)} \right)\label{e:ev:v},\\
    p(x,y,t) &= p_{\infty}\left(\frac{\rho}{\rho_\infty}\right)^\gamma\label{e:ev:p}.
\end{align}
\end{subequations}
where $f = (1 - (x-x_o)^2 - (y-y_o-U_\infty t)^2) / 2r_c^2$. To match the conditions of Vincent~\etal~\cite{vincent-castonguay-jameson:2011b} and Witherden~\etal~\cite{witherden:2014}, we set the free-stream conditions to $\rho_\infty=1$, $U_\infty=1$, and $p_\infty= (\rho_\infty U_\infty^2) / (\gamma  M_\infty^2)$, where the free-stream Mach number is $M_\infty=0.4$. We prescribe the size and strength of the vortex to be $r_c=1.5$ and $\epsilon=13.5$, respectively, and initially position the vortex at the center of the domain located at $(x_o,y_o)=(20,20)$.

The computational domain $\mathrm{\Omega}=\{x,y\in\mathbb{R}~|~0 \leqslant x,y \leqslant 40\}$ is partitioned using four different meshes of $120^2$, $140^2$, $160^2$, and $180^2$ elements. The upper and lower boundaries are treated as periodic while the left and right boundaries are prescribed free-stream conditions. These conditions result in modeling an infinite array of coupled vortices; however, the impact of the vortex on the free-stream at the boundaries is negligible since the vortex size $r_c$ is small compared to the length $L=40$ of the domain and the vortex strength exponentially decays from its origin~\cite{vincent-castonguay-jameson:2011b}. Therefore, we are effectively modeling a vortex propagating through an infinite domain. We consider a polynomial order $\p=3$, which gives $480^2$, $560^2$, $640^2$, and $720^2$ DoF for the various meshes. We use Davis' form of the Rusanov approximate Riemann solver~\cite{davis:1988} to compute inviscid numerical fluxes at the interfaces between elements, and we use the low-storage, five-stage, fourth-order accurate Runge--Kutta scheme of Carpenter and Kennedy~\cite{carpenter-kennedy:1994b} with a time step of $\Delta t = \num{1.25e-3}$ to explicitly march the solution through time. This time step is small enough such that all truncation errors are dominated by the spatial discretization.

\begin{figure}[H]
    \renewcommand{\fsize}{70mm}
    \centering
    \subfloat[]{
        \includegraphics[height=\fsize,keepaspectratio]
        {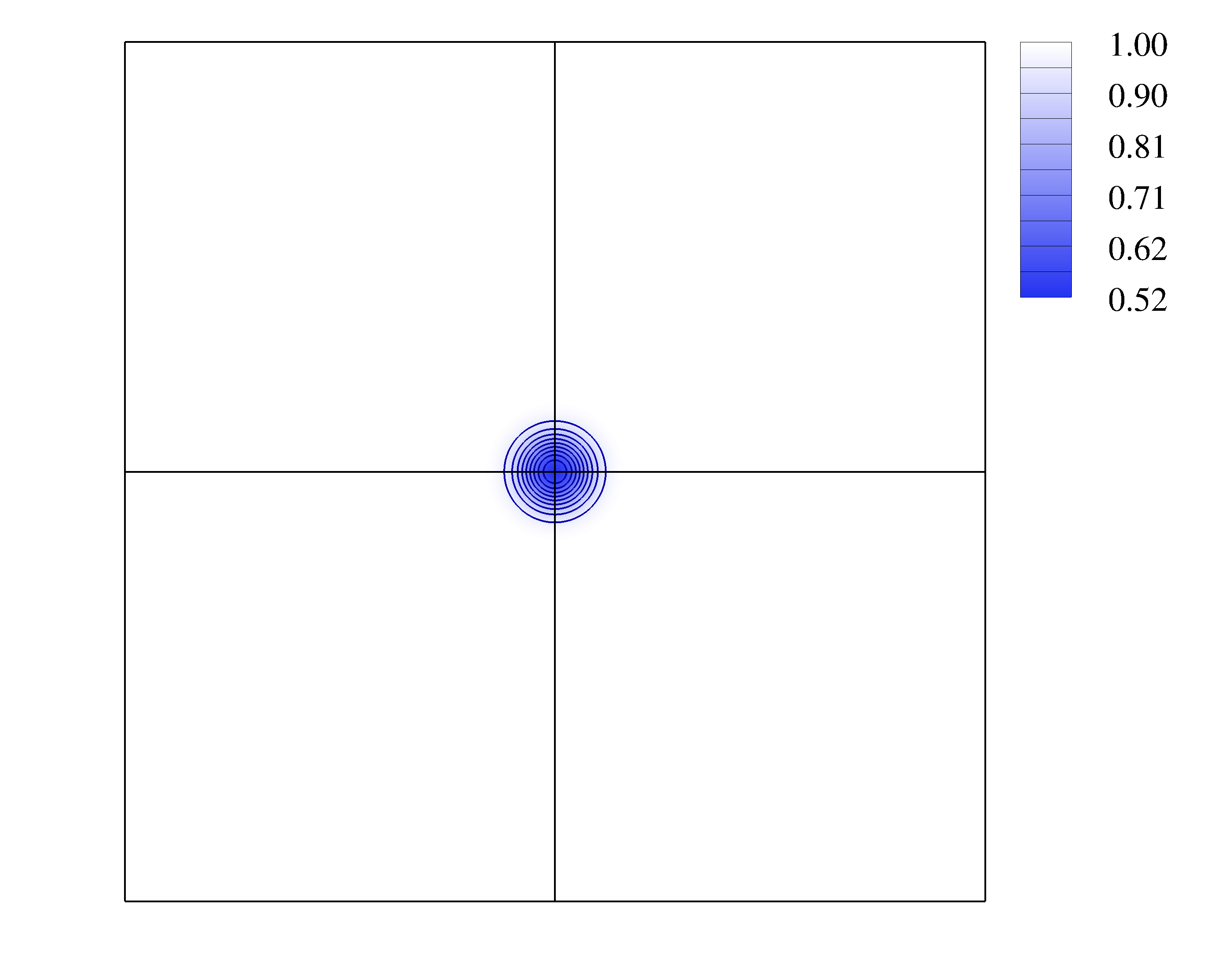}
    }
    \subfloat[]{
        \includegraphics[height=72.5mm,keepaspectratio]
        {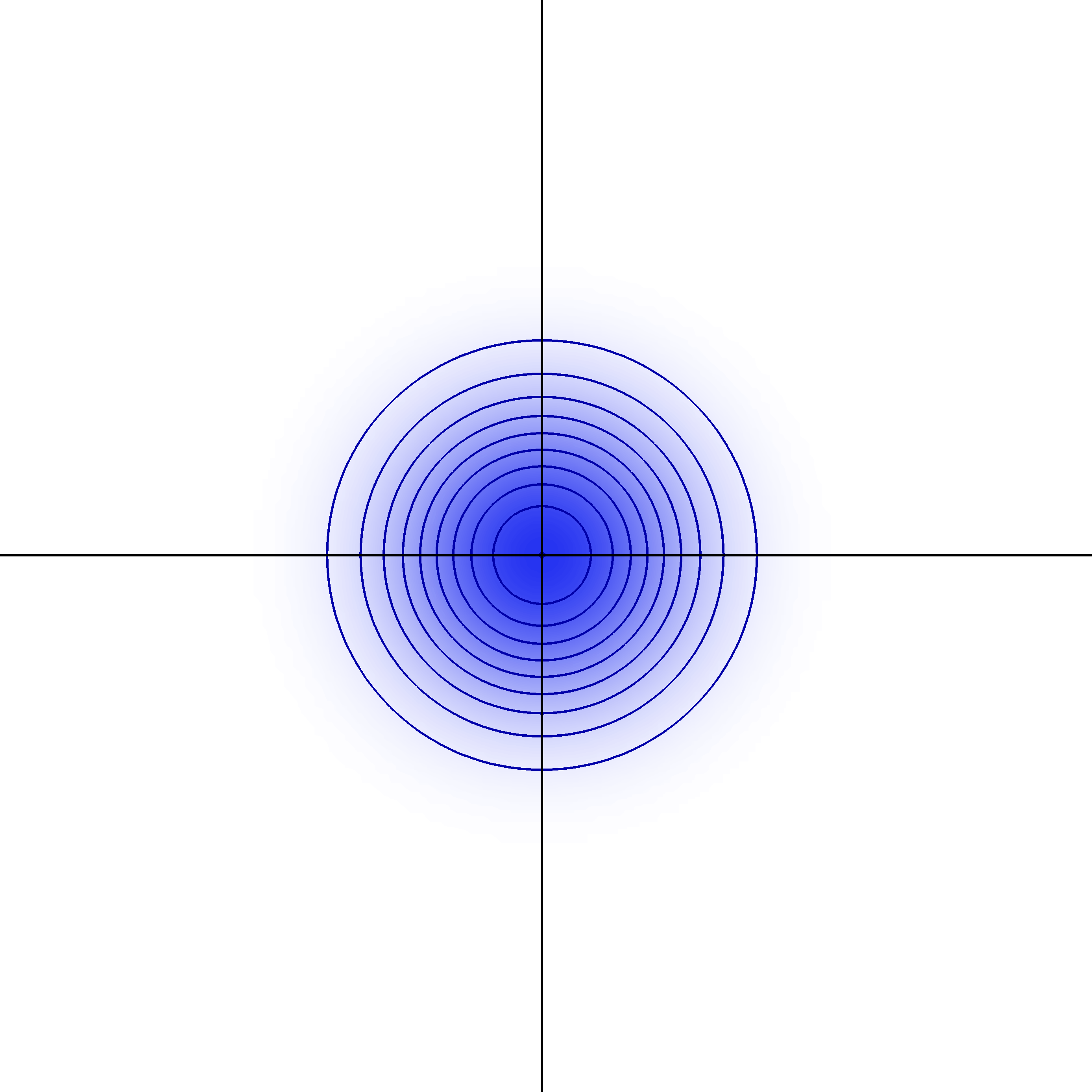}
    }
    
    \caption{Isentropic Euler Vortex: (a) $\p=3$ solution of density $\rho$ after 45 advective flow cycles ($t=\SI{1800}{\second}$) in the domain $\mathrm{\Omega}=\{x,y\in\mathbb{R}~|~0 \leqslant x,y \leqslant 40\}$ which is partitioned into $120^2$ elements, (b) enlarged image of the vortex centered at the origin $(x,y)=(20,20)$ at $t=\SI{1800}{\second}$.}
    \label{f:euler_vortex:density}
\end{figure}

\begin{figure}[H]
    \renewcommand{\fsize}{70mm}
    \centering
    \includegraphics[height=\fsize,keepaspectratio]
    {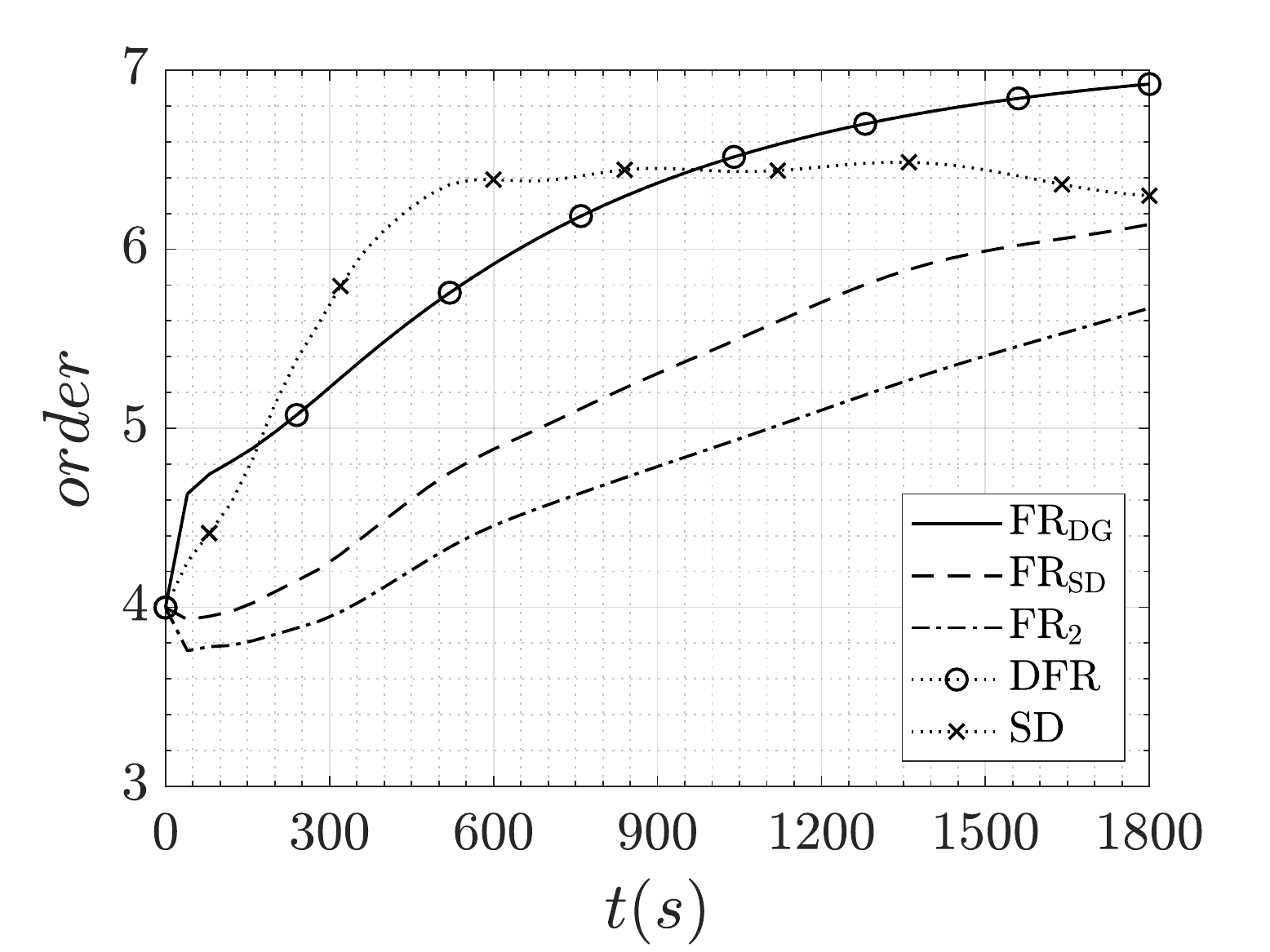}
    
    \caption{Isentropic Euler Vortex: super accuracy with $\p=3$ for various discontinuous spectral element schemes---$\FRDG$, $\FRSD$, $\mathrm{FR_{2}}$, $\DFR$ and $\SD$.}
    \label{f:euler_vortex:accuracy}
\end{figure}

To assess the order of accuracy, we compute the $L^2$-norm of the density error $\|e\|_2$ inside an integration window $\mathrm{\Omega}_I=\{x,y\in\mathbb{R}~|~-2 \leqslant x-x_o \leqslant 2,\; -2 \leqslant y-y_o \leqslant 2 \}$ at each moment in time the vortex advects through the entire computational domain and returns to the origin, which occurs when $t=t^\star L / u_\infty$ for $t^\star\in\{1,2,\ldots,45\}$. The $L^2$-norm of the density error is defined as
\begin{align}
    \|e\|_2 = \sqrt{\int_{\mathrm{\Omega}_I}(\rho_n(x,y) - \rho_e(x,y))^2 \, \drmb{x}}
    \label{e:ev:L2-error}
\end{align}
where $\rho_n(x,y)$ is the numerical density and $\rho_e(x,y)$ is the exact analytical solution given in Eq.~(\ref{e:ev:rho}) at $t=0$. To approximate the integrals in Eq.~(\ref{e:ev:L2-error}), we apply a more than sufficient high-strength quadrature rule. To compare our results against those obtained by~\cite{vincent-castonguay-jameson:2011b} and~\cite{witherden:2014}, we plot the observed convergence of the $\FRSD$ and $\SD$ schemes in Fig.~\ref{f:euler_vortex:accuracy}, where the order of accuracy at any given time is determined by computing the slope of the line given by a least-squares fit of $\log(\|e\|_2)$ as a function of $\log(h)$. For the four different meshes, we use grid spacings $h\in\{1/3,2/7,1/4,2/9\}$. For comparison, we also plot results from other FR schemes built into the current solver in Fig.~\ref{f:euler_vortex:accuracy} including $\FRDG$, $\mathrm{FR_2}$ and $\DFR$. We observe an approximate $2\p+1$ level of accuracy under $\FRDG$ at $t=\SI{1800}{\second}$ and $2\p$ under $\FRSD$. We also confirm that the super accuracy of the $\DFR$ scheme is equivalent to that of $\FRDG$ since solution points are placed at corresponding Gauss--Legendre points.

\begin{figure}[H]
    \renewcommand{\fsize}{78mm}
    \centering
    \subfloat[]{
        \includegraphics[height=\fsize,keepaspectratio]
        {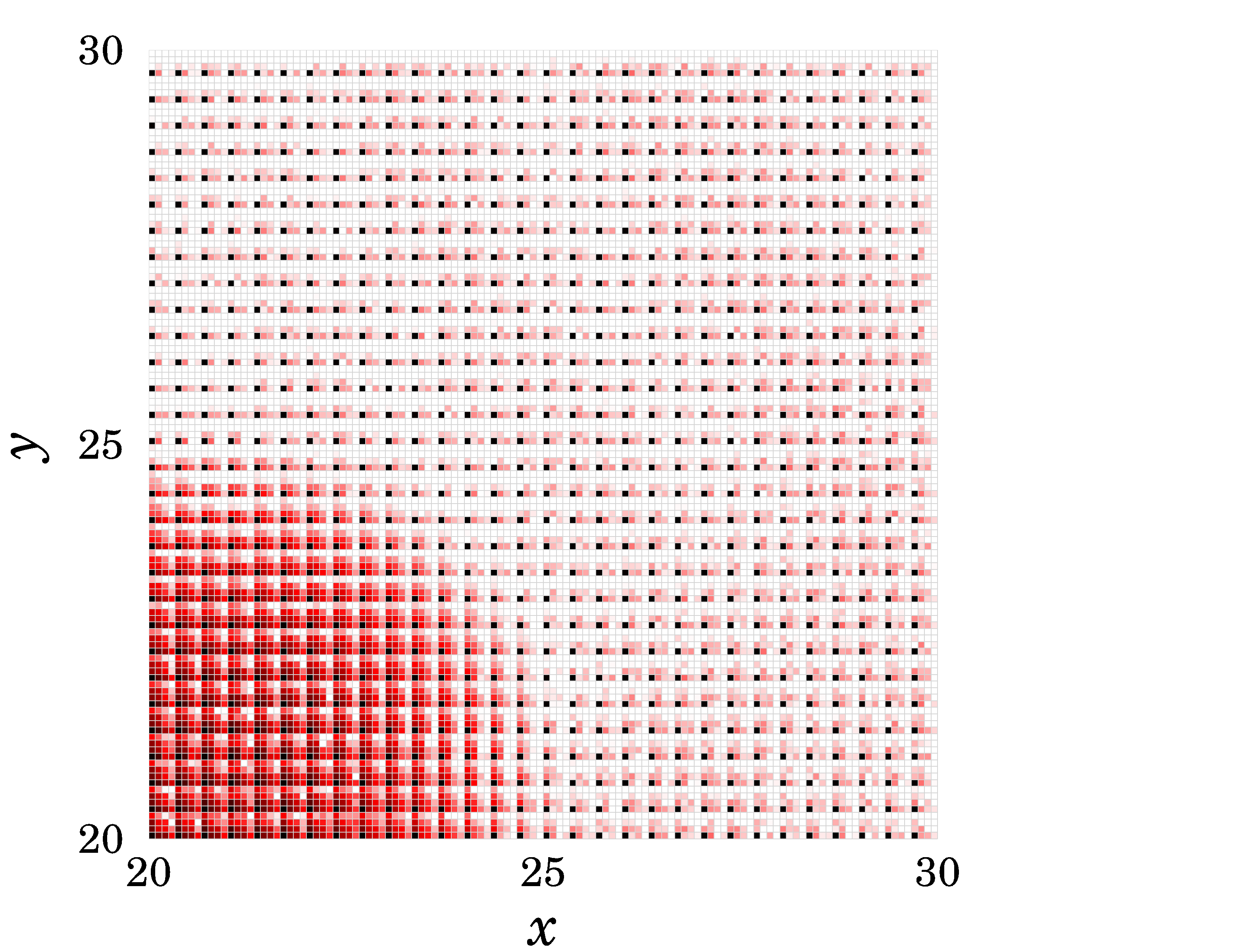}
        \label{f:ev:gSD:p3_120:40}
    }
    \subfloat[]{
        \hspace{-0.9in}
        \includegraphics[height=\fsize,keepaspectratio]
        {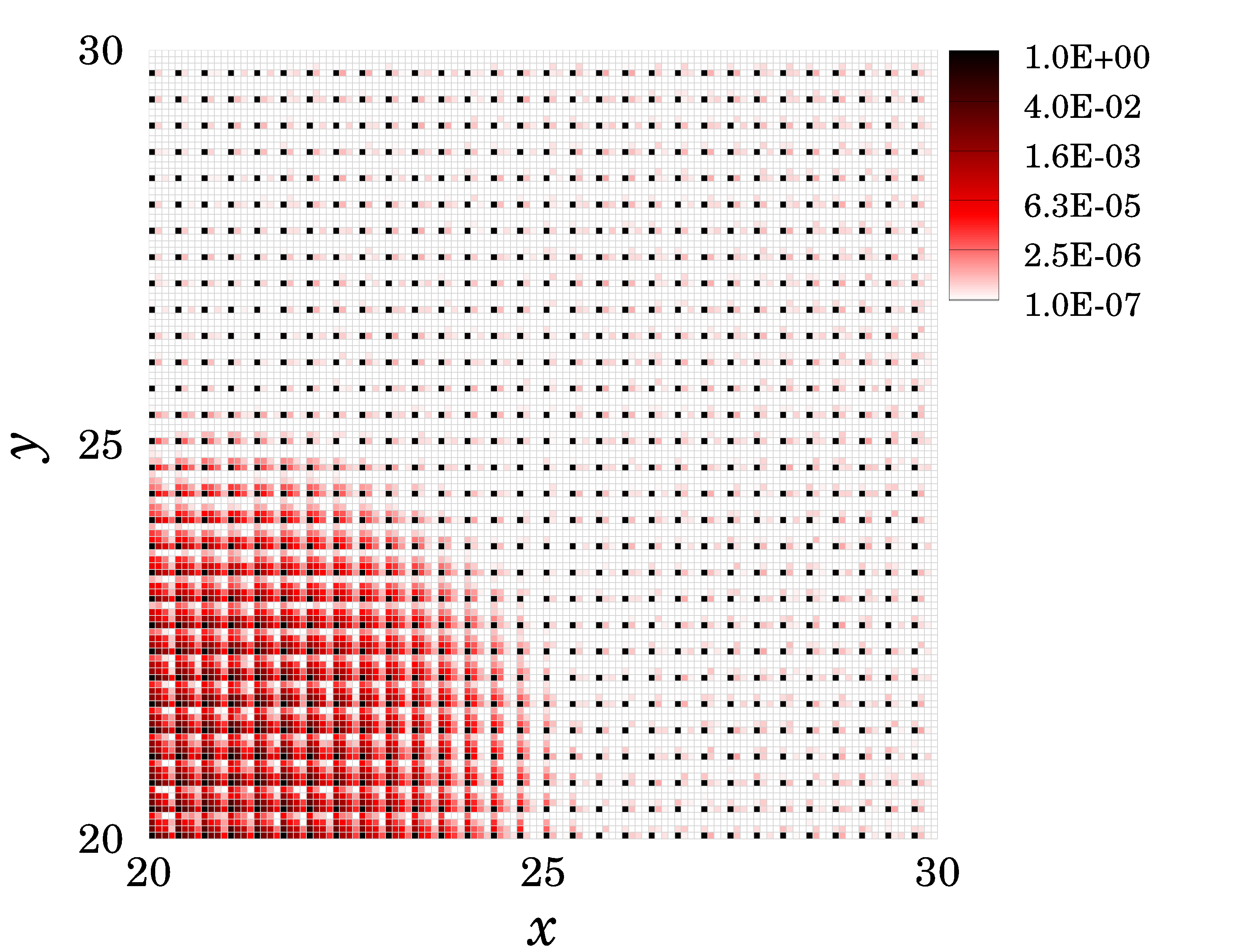}
        \label{f:ev:SD:p3_120:40}
    }
    
    \caption{Isentropic Euler Vortex: modal coefficients for $\p=3$ on the subdomain $\{x,y~|~20<x<30,20<y<30\}$ on a $120\times120$ grid after one advective flow cycle ($t=\SI{40}{\second}$). (a) $\FRSD$, (b) $\SD$.}
    \label{f:ev:120x120:t40}
\end{figure}

\begin{figure}[H]
    \renewcommand{\fsize}{65mm}
    \centering
    \subfloat[]{
        \includegraphics[height=\fsize,keepaspectratio]
        {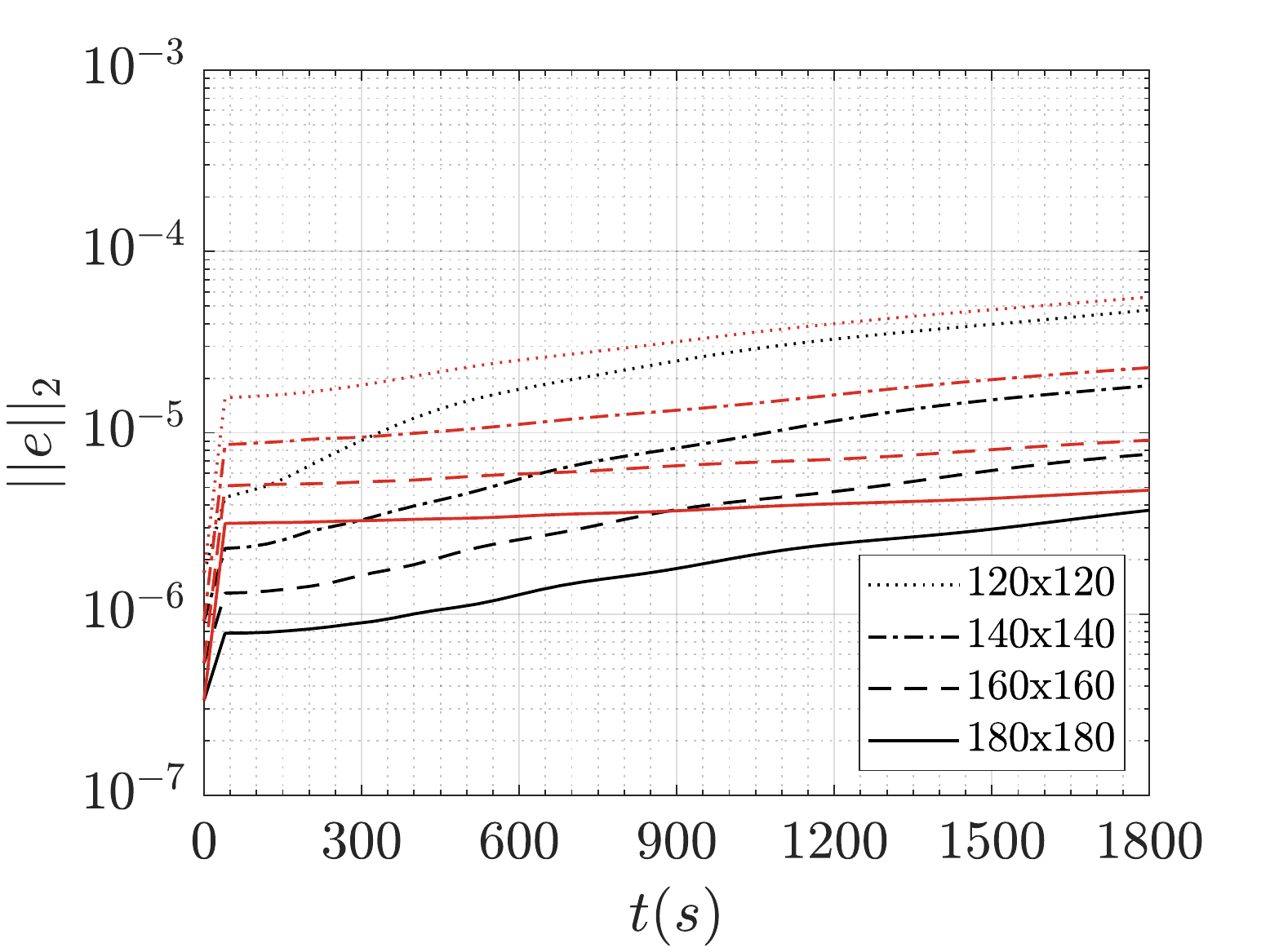}
        \label{f:euler_vortex:L2:gSD:SD}
    }
    \subfloat[]{
        \includegraphics[height=\fsize,keepaspectratio]
        {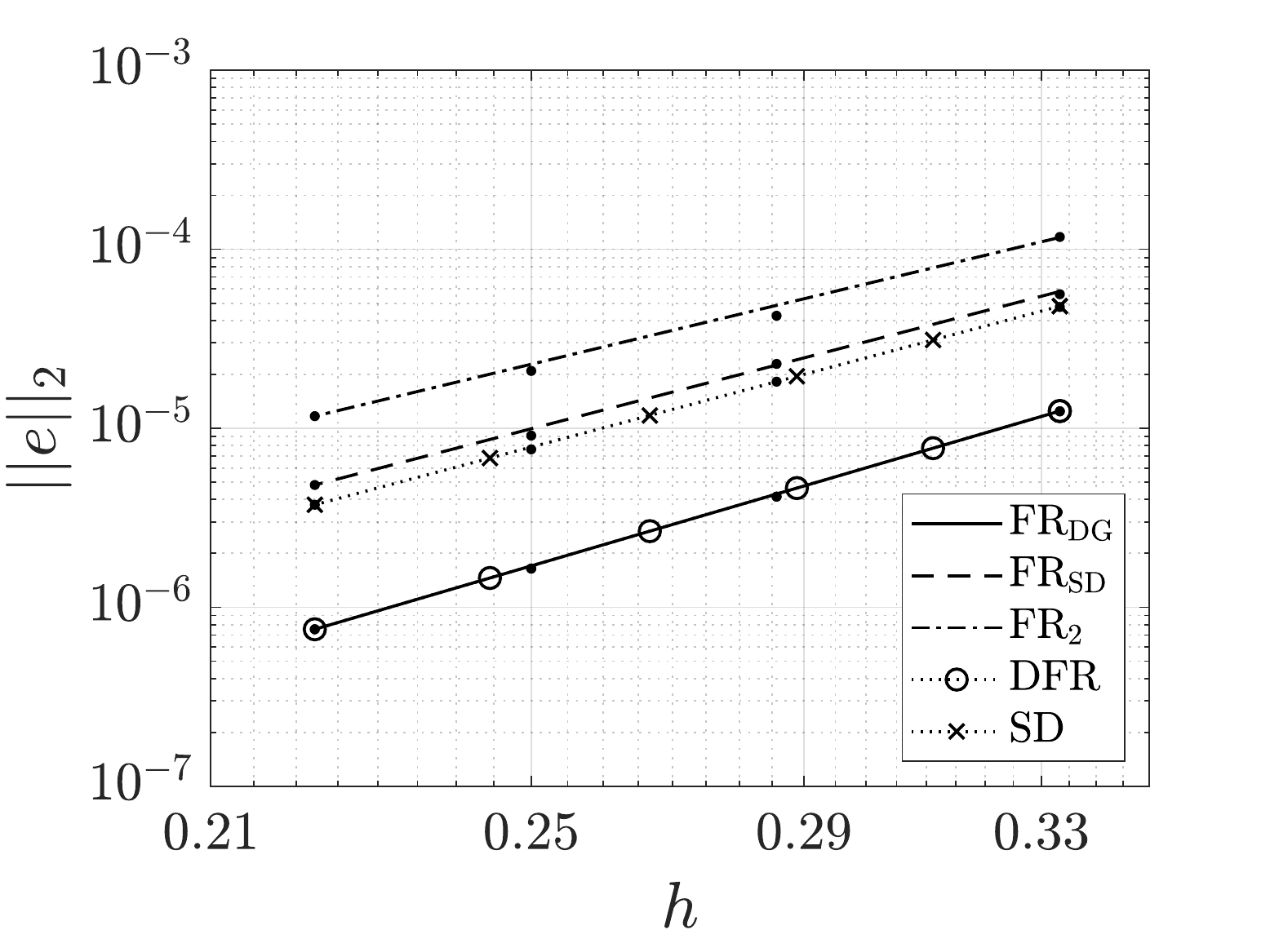}
        \label{f:euler_vortex:L2:h}
    }
    
    \caption{Isentropic Euler Vortex: $L^2$-norm of density error $||e||_2$ as a function of time for $\p=3$ (a) $\SD$ ({\it black}) and $\FRSD$ ({\it red}), (b) $L^2$-norm of density error as a function of grid spacing $h$ at $t=\SI{1800}{\second}$.}
    \label{f:euler_vortex:L2}
\end{figure}

We can recast the nodal form of the solution polynomial into its modal form by using a set of modal basis functions---orthogonal Legendre polynomials $\mathcal{L}_i(\xi)$, $\mathcal{L}_j(\eta)$---and their corresponding modal coefficients $c_{i,j}$~\cite{sherwin-karniadakis:2005}. Following the work of Spiegel~\etal~\cite{spiegel:2015}, we plot $|c_{i,j}|$ within each element (see Fig.~\ref{f:ev:120x120:t40}), normalizing by the mean mode and zeroing all modes less than $\num{1e-7}$. In these images, the values of $|c_{i,j}|$ in the lower left corner of each element correspond to the magnitude of the mean mode $c_{0,0}\mathcal{L}_0(\xi)\mathcal{L}_0(\eta)$. The values in the upper right corner of each element correspond to the magnitude of the highest Legendre mode $c_{\p,\p} \mathcal{L}_{\p}(\xi)\mathcal{L}_{\p}(\eta)$. From left to right and bottom to top, these modal coefficients correspond to the magnitude of the Legendre modes of increasing order with respect to $\xi$ and $\eta$, respectively, up to $\p$. Under $\FRSD$, we demonstrate in Fig.~\ref{f:ev:120x120:t40} that the higher frequency modes in regions away from the vortex are more energized in comparison to $\SD$. The larger magnitudes of the higher modes in $\FRSD$ can be attributed to aliasing errors. By comparison, the $\SD$ scheme is successful at suppressing this energy at the higher modes, with the dominant modes away from the vortex being the lowest order mean mode, which is consistent with analytic solution. In turn this produces a lower error in the solution, as demonstrated by the time history plot of the $L^2$-norm of density shown in Fig.~\ref{f:euler_vortex:L2:gSD:SD}. As a result, this causes rate of convergence history to initially increase sharply to a level above $2\p$ between $t=\SI{0}{\second}$ and $t=\SI{480}{\second}$, then level off for the remaining portion of the simulation. This rapid approach to an order greater than $2\p$ indicates favorable accuracy properties of the $\SD$ scheme, thereby reducing contamination of the solution from aliasing errors. This result is consistent with the analytical findings presented in Eqs.~(\ref{eq:sd_error}) and (\ref{eq:fr_error}). Ultimately, this offers improved stability when performing implicit large eddy simulations of turbulent flow problems such as those studied in Sec.~\ref{s:tgv} and Sec~\ref{s:sd7003}.

\subsection{Inviscid, subsonic flow over a cylinder}\label{s:inviscid_cylinder}

\begin{figure}[H]
    \renewcommand{\fsize}{50mm}
    \centering
    \subfloat[]{
        \includegraphics[height=\fsize,keepaspectratio]
        {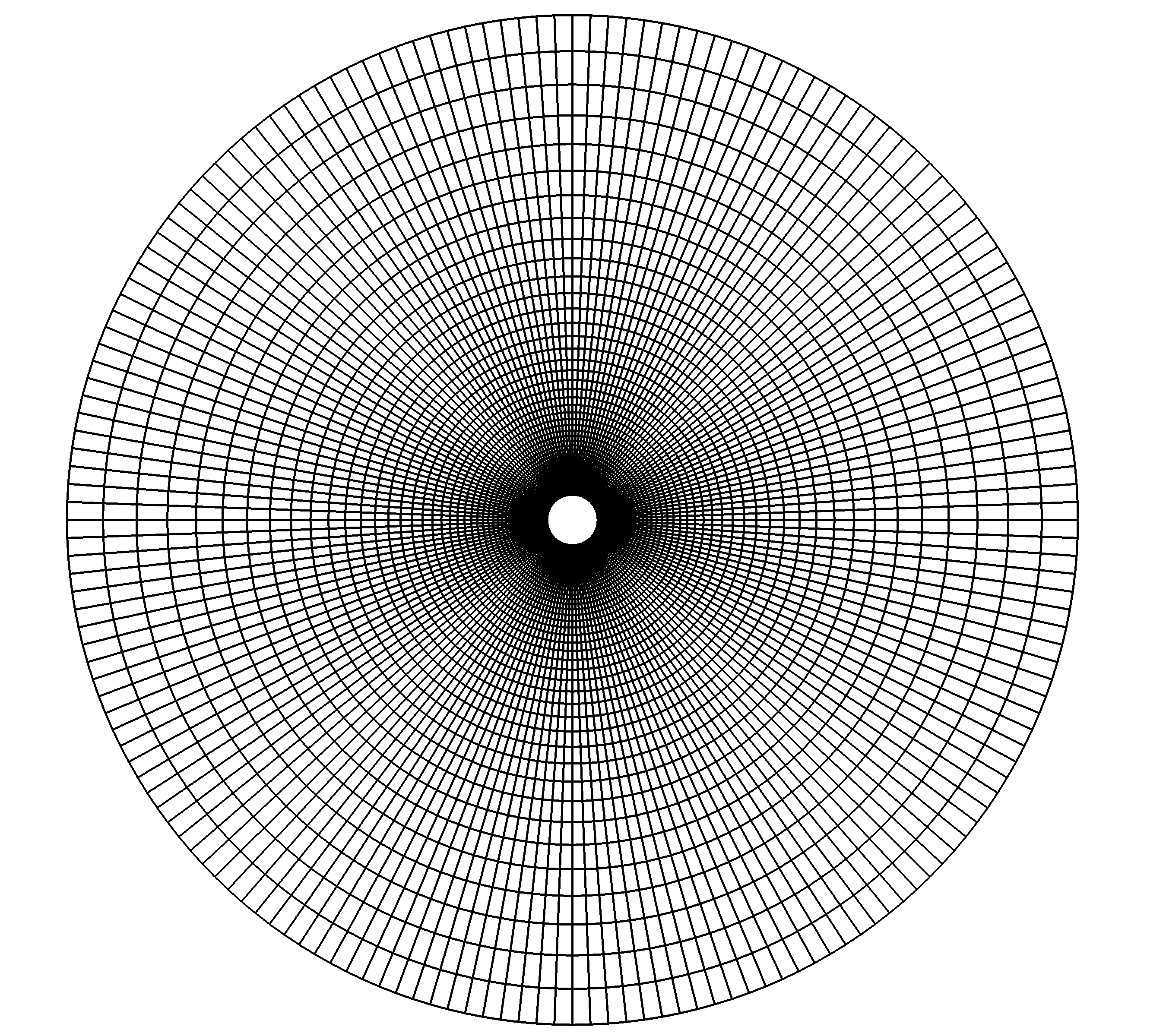}
        \label{f:cylinder:mesh}
    }
    \subfloat[]{
        \includegraphics[height=\fsize,keepaspectratio,frame]
        {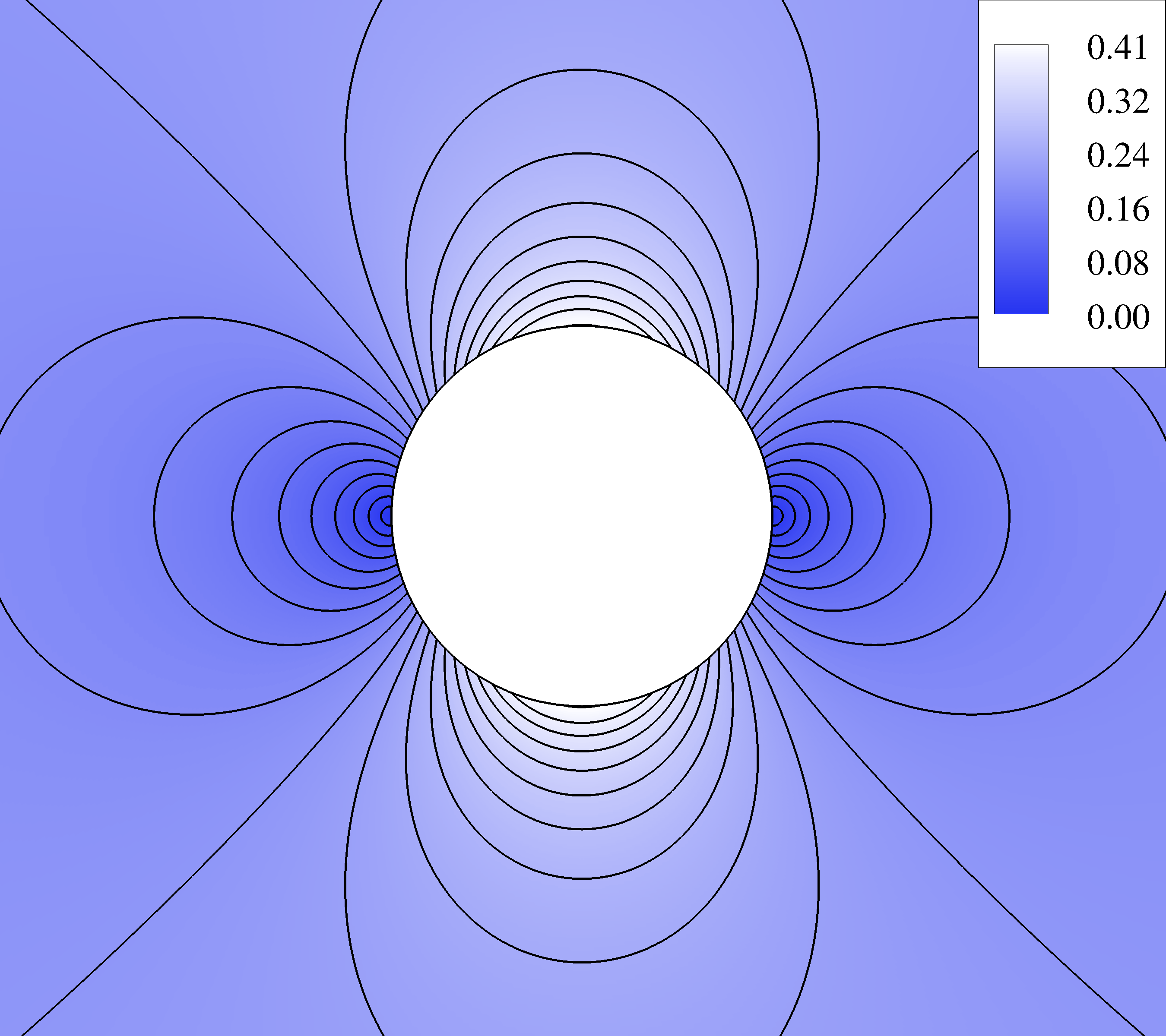}
        \label{f:cylinder:mach:gSD}
    }
    \subfloat[]{
        \includegraphics[height=\fsize,keepaspectratio,frame]
        {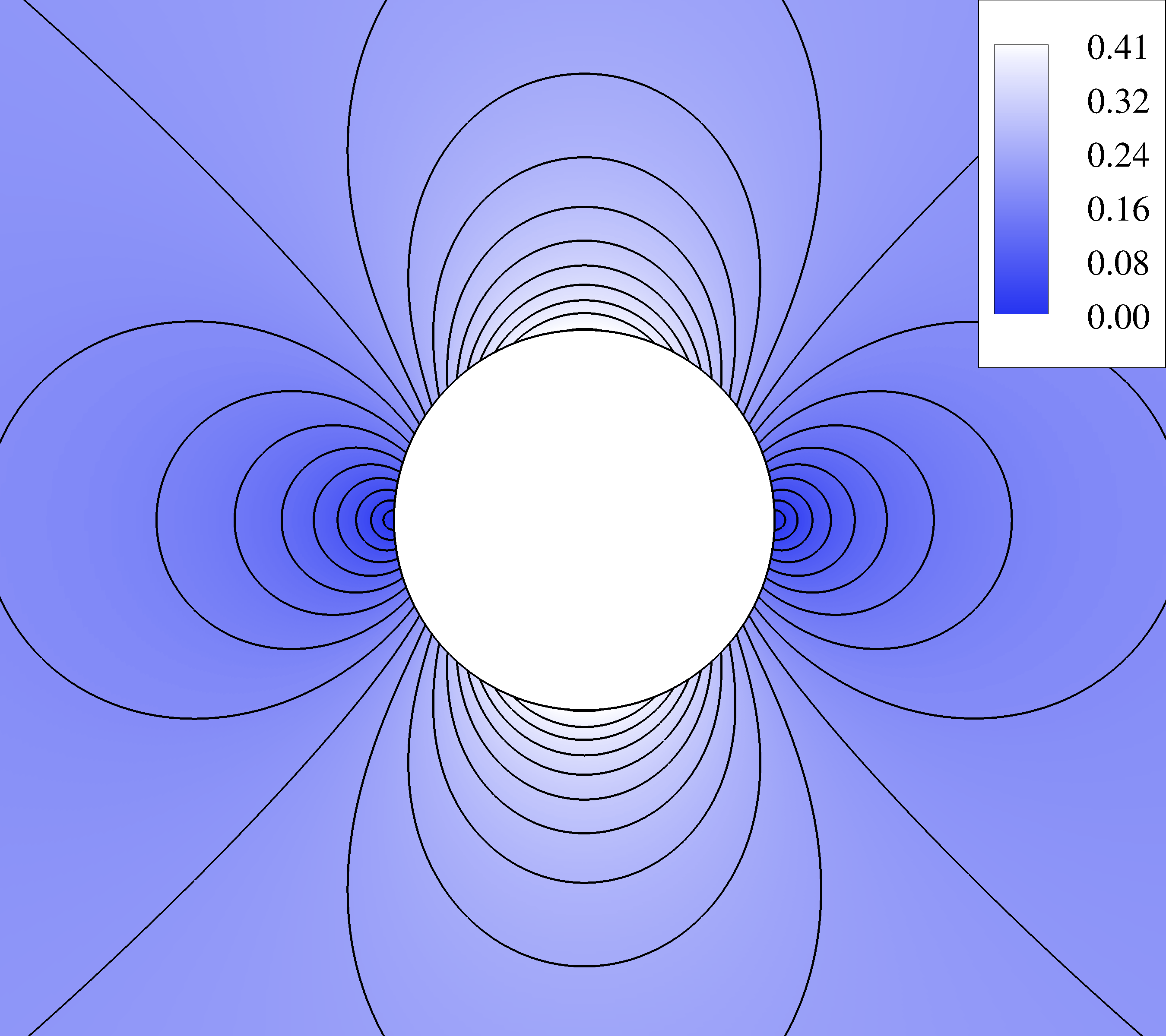}
        \label{f:cylinder:mach:SD}
    }
    
    \caption{Inviscid, subsonic flow over 2D cylinder: mach number (a) mesh; (b) $\FRSD$, $\p=6$; (c) $\SD$, $\p=6$.}
    \label{f:cylinder:mach}
\end{figure}

In this section, we simulate the steady, two-dimensional, inviscid, subsonic flow over a cylinder as governed by the compressible Euler equations. This test case is constructed to assess numerically-generated entropy and was used in Mengaldo~\etal~\cite{mengaldo:2015} to test the effectiveness of global de-aliasing for the $\FRDG$ scheme at different polynomial orders. Ideally, zero entropy should be generated for an inviscid, subsonic simulation, however aliasing in the numerical method introduces a mechanism allowing the build-up of entropy. To reduce numerical entropy generation due to the mesh representation of the cylinder wall, the curvature of the cylinder is represented with 176 quartic elements with 54 elements in the radial direction. The mesh, shown in Fig.~\ref{f:cylinder:mesh}, extends $10d$ into the farfield and contains a total of $176 \times 54 = 9\,504$ elements. The simulation was run at a freestream Mach number of $M_{\infty}=0.2$ with $\p=2$, $\p=4$ and $\p=6$ using the low-storage, four-stage, third-order embedded Runge--Kutta time integration scheme---abbreviated RK[4,3(2)]-2N---of Carpenter and Kennedy~\cite{carpenter-kennedy:1994a,kennedy-carpenter-lewis:2000} with adaptive time-stepping.

\begin{figure}[H]
    \renewcommand{\fsize}{50mm}
    \centering
    \subfloat[]{
        \includegraphics[height=\fsize,keepaspectratio,frame]
        {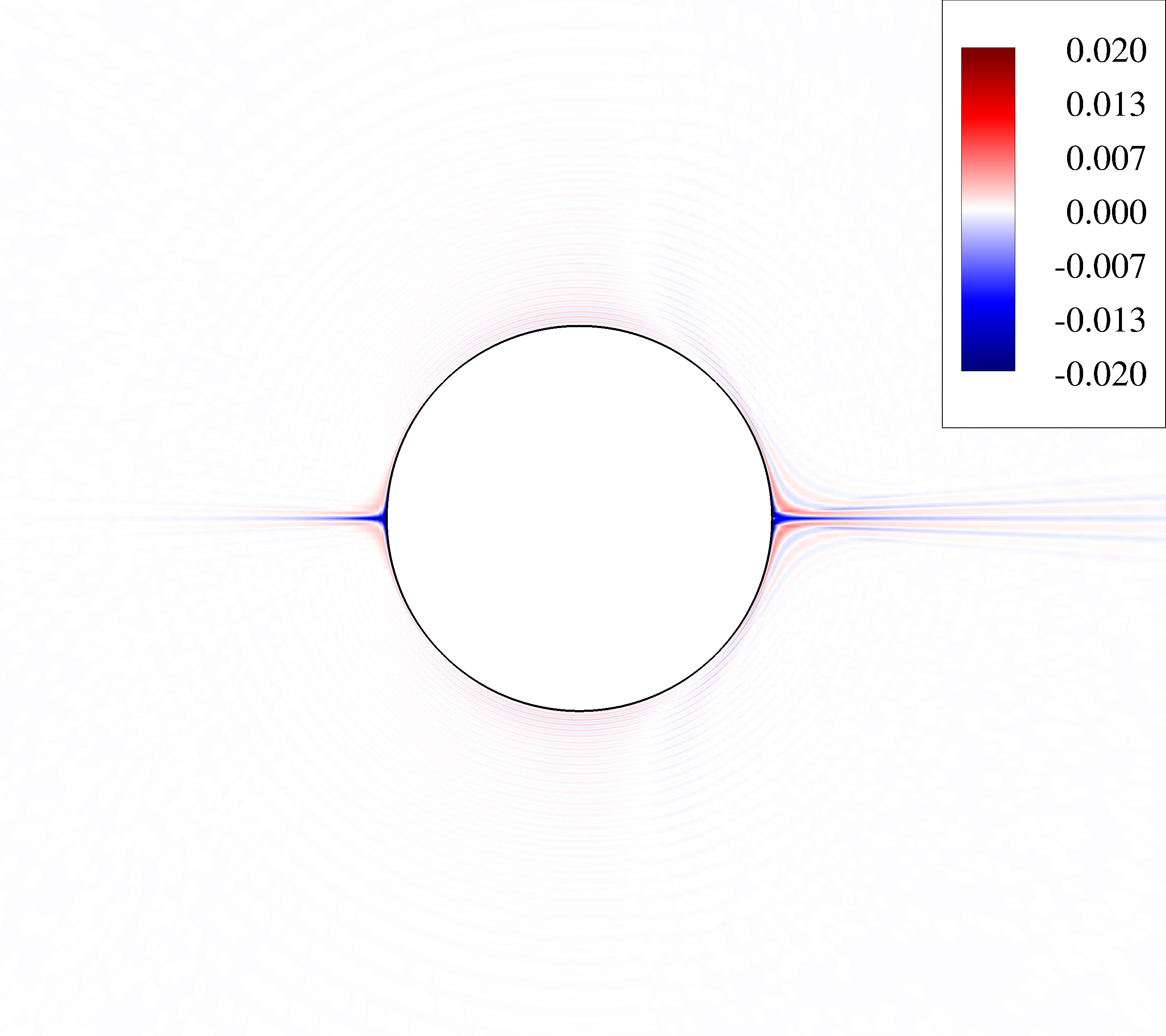}
        \label{f:entropy:gSD:p2}
    }
    \subfloat[]{
        \includegraphics[height=\fsize,keepaspectratio,frame]
        {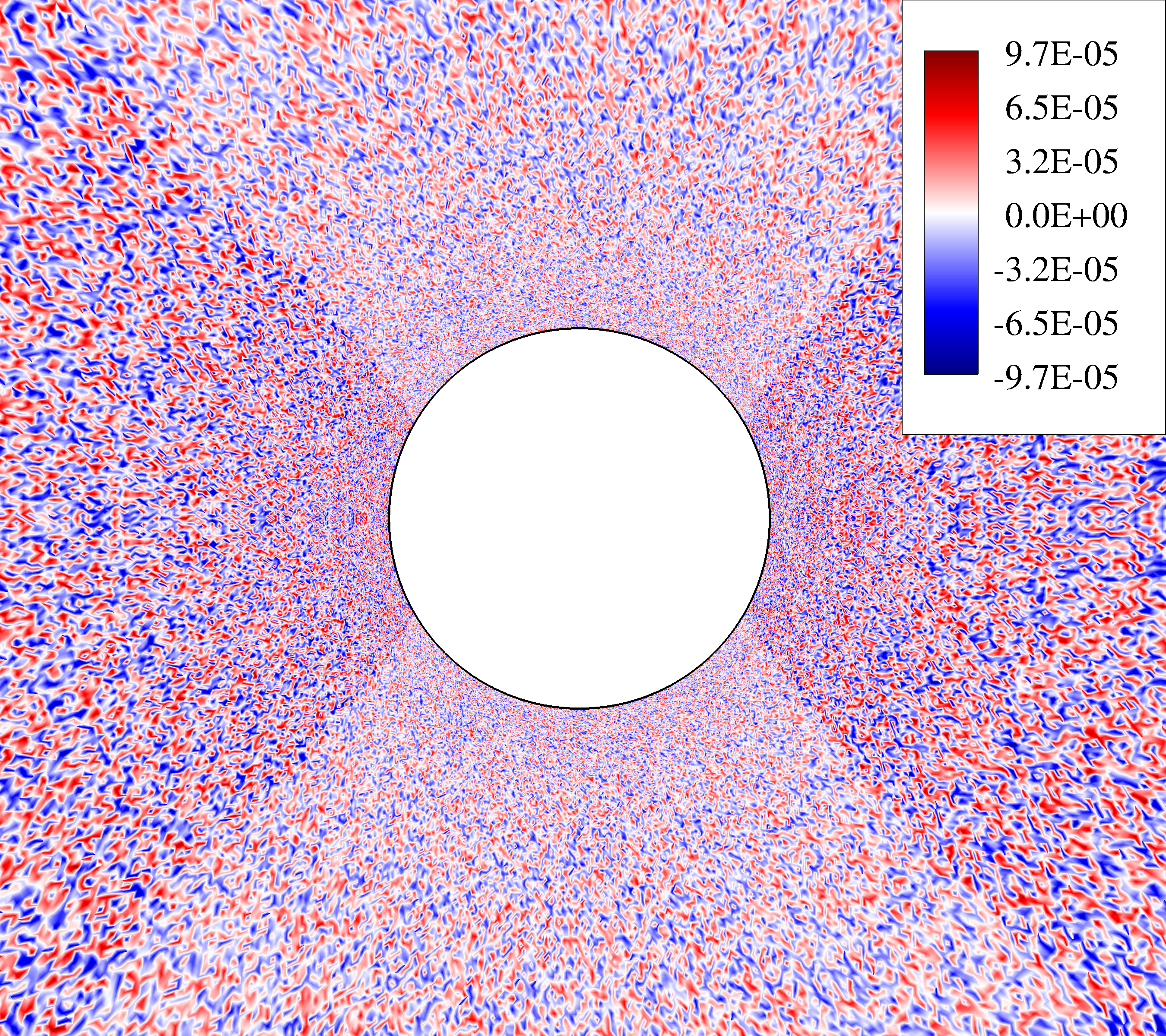}
        \label{f:entropy:gSD:p4}
    }
    \subfloat[]{
        \includegraphics[height=\fsize,keepaspectratio,frame]
        {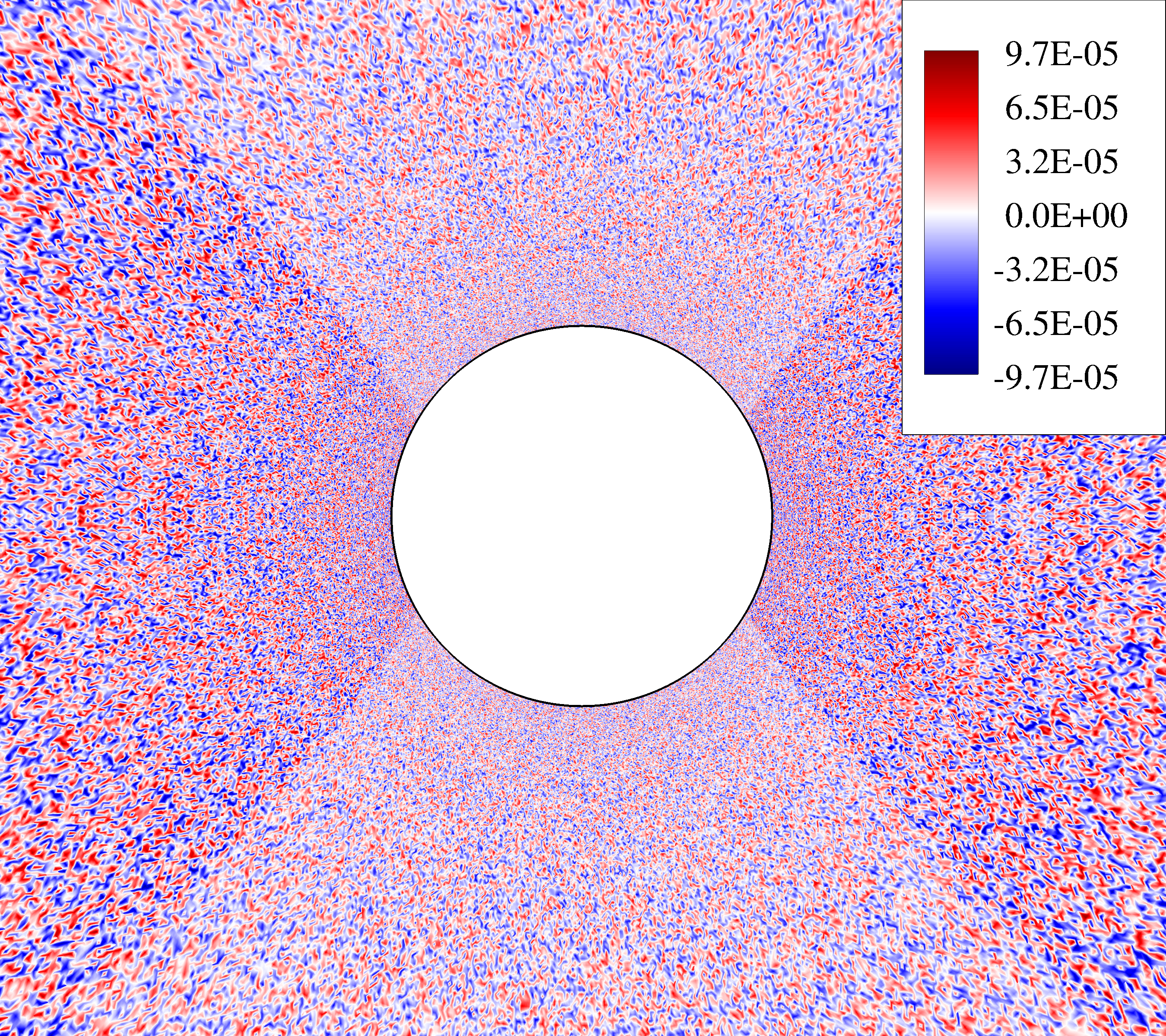}
        \label{f:entropy:gSD:p6}
    }
    \\
    \subfloat[]{
        \includegraphics[height=\fsize,keepaspectratio,frame]
        {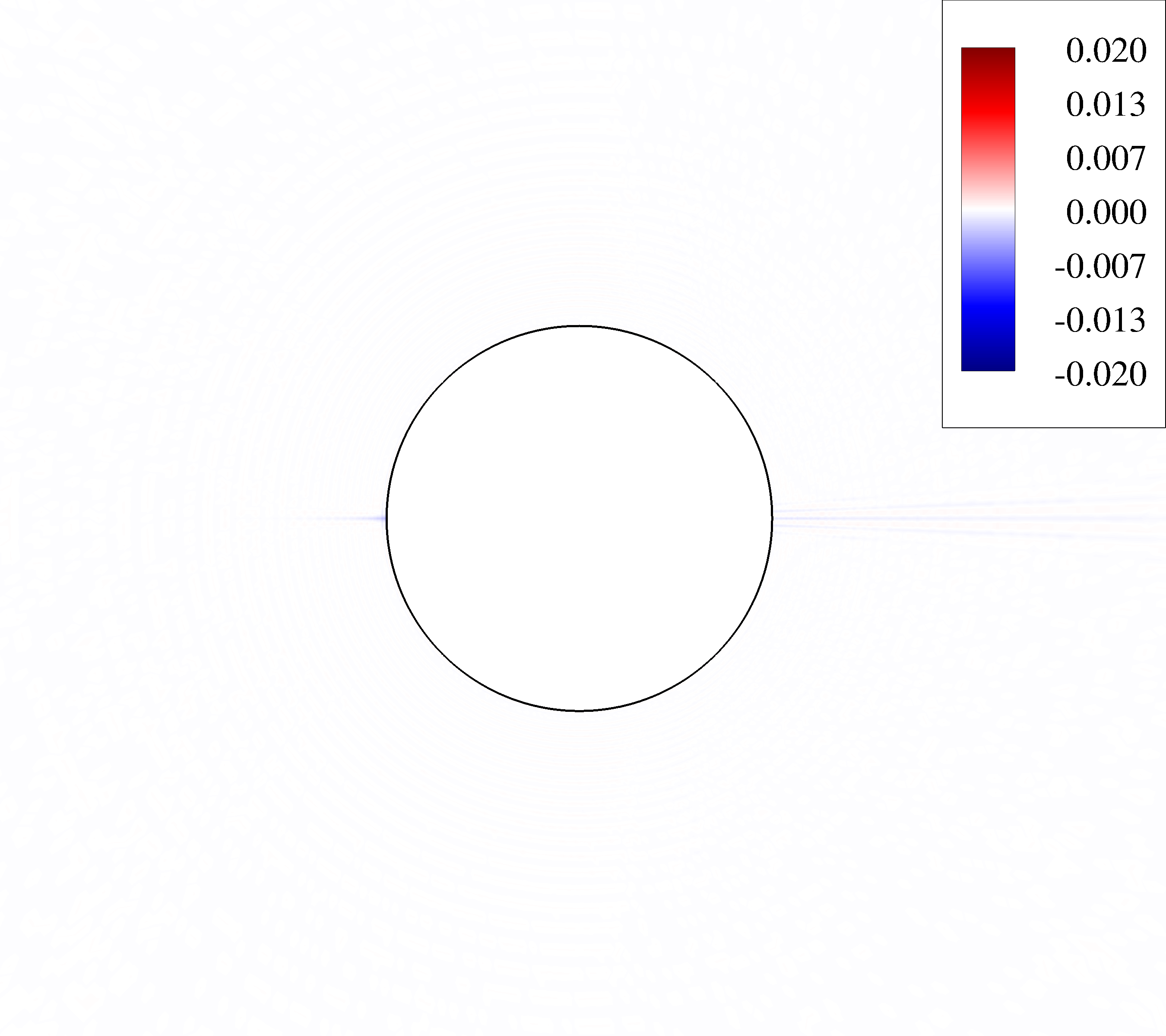}
        \label{f:entropy:SD:p2}
    }
    \subfloat[]{
        \includegraphics[height=\fsize,keepaspectratio,frame]
        {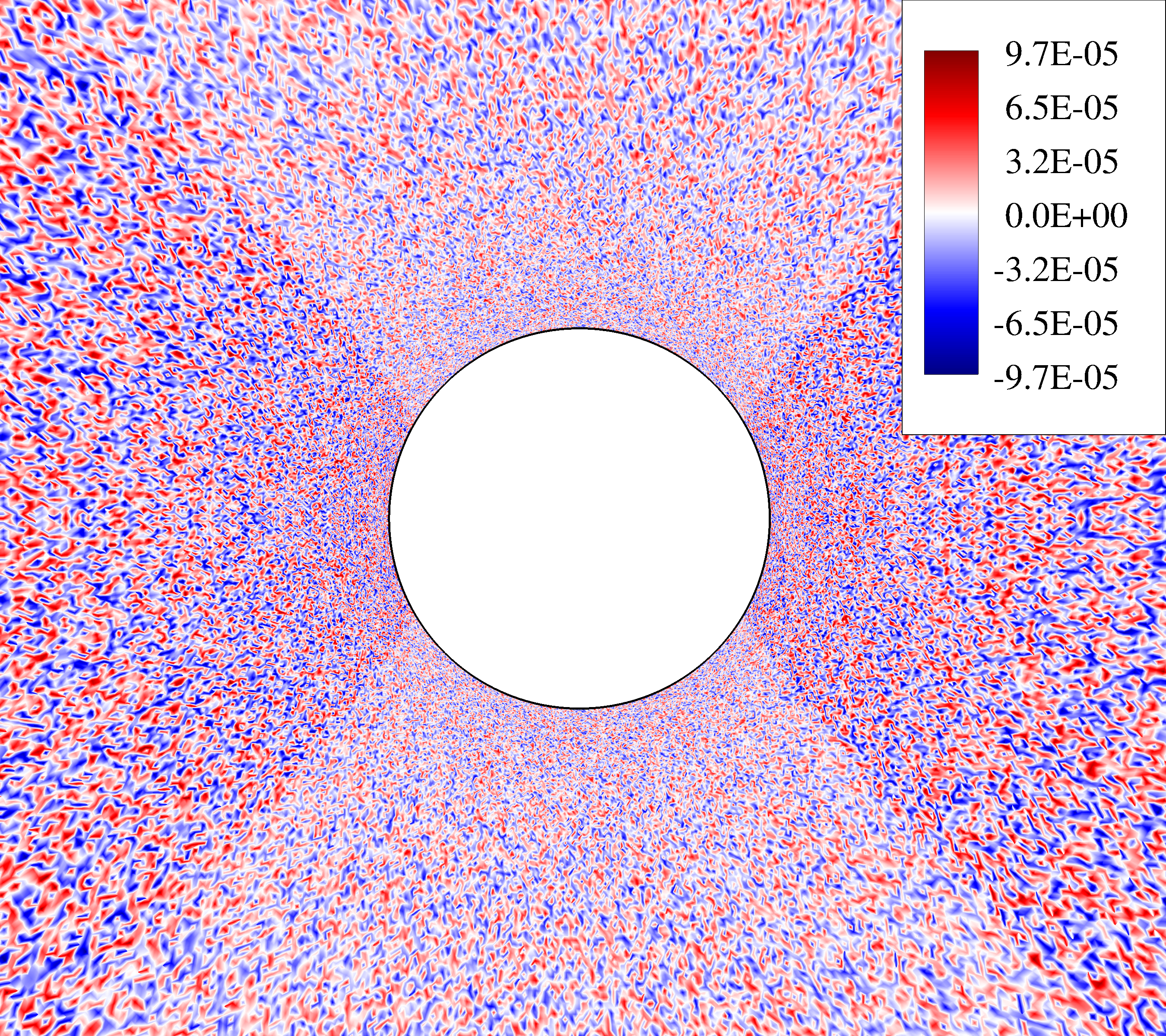}
        \label{f:entropy:SD:p4}
    }
    \subfloat[]{
        \includegraphics[height=\fsize,keepaspectratio,frame]
        {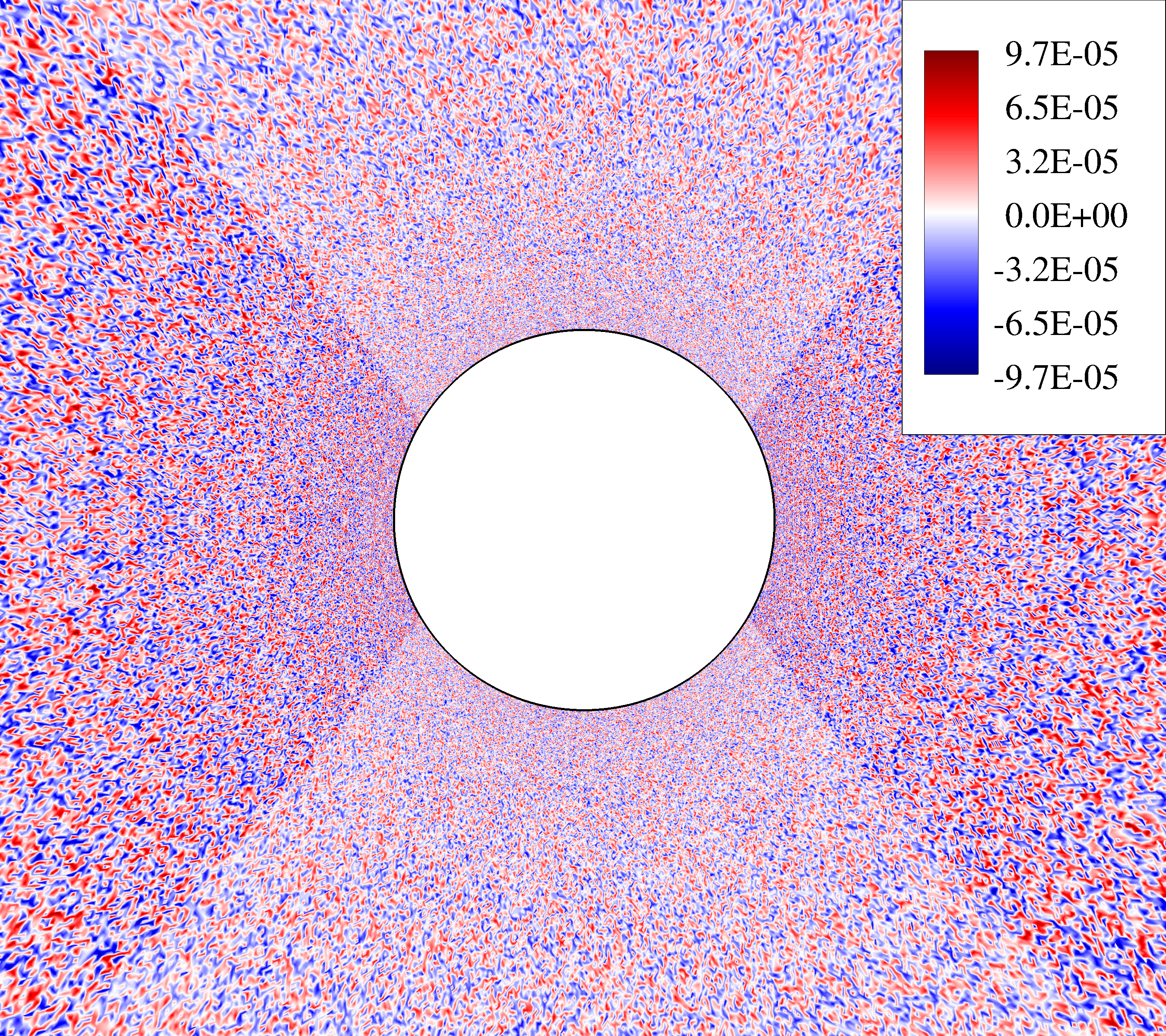}
        \label{f:entropy:SD:p6}
    }
    
    \caption{Inviscid, subsonic flow over 2D cylinder: entropy (a) $\FRSD$, $\p=2$; (b) $\FRSD$, $\p=4$; (c)  $\FRSD$, $\p=6$; (d) $\SD$, $\p=2$; (e) $\SD$, $\p=4$; (f) $\SD$, $\p=6$.}
    \label{f:cylinder:entropy}
\end{figure}

Mach number contours from the $\p=6$ solution for $\FRSD$ and $\SD$ can be seen in Figs.~\ref{f:cylinder:mach:gSD} and \ref{f:cylinder:mach:SD}, respectively, appearing qualitatively identical. Results of numerically-generated entropy (\SI{}{\joule\per\kilogram\per\kelvin})
for $\p\in\{2,4,6\}$ are shown in Fig.~\ref{f:cylinder:entropy} and tabulated in Tab.~\ref{t:cylinder}. For $\p=4$ and $\p=6$,  similar results for both $\FRSD$ and $\SD$ were observed, with entropy generation ranging between $\pm\num{9.79e-5}$ throughout the entire domain, with the difference in results between the two schemes being negligible at these polynomial orders. However, for the $\p=2$ case shown in Figs.~\ref{f:entropy:gSD:p2} and \ref{f:entropy:SD:p2}, the results demonstrate entropy build-up near the two stagnation points located on the windward side and leeward side of the cylinder, with a larger quantity of entropy build-up downstream. The minimum and maximum entropy values are approximately $\Delta s_{min}=\num{-1.95e-2}$ and $\Delta s_{max}=\num{1.83e-2}$ for $\FRSD$ and $\Delta s_{min}=\num{-6.72e-3}$ and $\Delta s_{max}=\num{4.93e-4}$ for $\SD$. These results demonstrate reduced numerical entropy generation under $\SD$ by a factor of approximately three, indicating more favorable results for this particular under-resolved case at $\p=2$ where the ratio of flux points to solution points $(\p+2) / (\p+1)$ is greatest for the $\SD$ scheme.

\begin{table}[H]
    \centering
    \setlength{\tabcolsep}{20pt}
    {\footnotesize
    \begin{tabular}{rccccc} \toprule
        \multirow{2}{0.5cm}{$\p$} & \multicolumn{2}{c}{$\FRSD$} & \multicolumn{2}{c}{$\SD$} &  \multirow{2}{0.8cm}{$\frac{\p+2}{\p+1}$} \\ \cmidrule(lr){2-3} \cmidrule(lr){4-5}
                & $\Delta s_{min}$ & $\Delta s_{max}$ & $\Delta s_{min}$ & $\Delta s_{max}$ &  \\ \midrule
         $2$ & \num{-1.95e-2} & \num{1.83e-2} & \num{-6.72e-3} & \num{4.93e-4} & $4/3=1.33$ \\
         $4$ & \num{-9.67e-5} & \num{9.68e-5} & \num{-9.69e-5} & \num{9.72e-5} & $6/5=1.20$ \\
         $6$ & \num{-9.79e-5} & \num{9.79e-5} & \num{-9.78e-5} & \num{9.69e-5} & $8/7=1.14$ \\ \bottomrule
    \end{tabular}
    }
    
    \caption{Inviscid, subsonic flow over 2D cylinder: numerically-generated entropy (\SI{}{\joule\per\kilogram\per\kelvin}) under $\FRSD$ and $\SD$.}
    \label{t:cylinder}
\end{table}

\subsection{Taylor--Green vortex at $Re=1\,600$}\label{s:tgv}

In this section, we simulate the Taylor--Green vortex (TGV)---a simple, canonical problem in fluid dynamics often used to study vortex dynamics and turbulent transition and decay~\cite{taylor-green:1937}. The problem consists of a cubic volume of fluid initially containing a smooth distribution of vorticity. As time evolves, the vortices roll-up, vortex lines stretch, and vorticity intensifies. The large-scale vortical structures break down and small-scale eddies are produced, ultimately resulting in the transition to turbulence~\cite{brachet:1983}. Eventually, the small-scale turbulent motion dissipates all the energy and the fluid comes to rest. This test case is consistently used to evaluate turbulent flow simulation methodologies by the International Workshop on High-order Methods in Computational Fluid Dynamics held at the American Institute of Aeronautics and Astronautics Aerospace Sciences Meeting~\cite{wang_ijnmf:2012}. Various authors have demonstrated success in using high-order schemes to predict this flow field, and the current work complements existing results in the literature from discontinuous spectral element methods~\cite{gassner:2013,cartondewiart:2014,bull:2015,vermeire:2016}. Specifically, we use the TGV to compare the accuracy and stability between the $\SD$ and $\FRSD$ schemes for under-resolved simulations of turbulent flow.

The initial conditions of velocity and pressure for the TGV are given by
\begin{subequations}
\begin{align}
    \rho(x,y,z,0) &= \frac{p}{RT_o}
    \label{e:tgv_rho}
    \\
    u(x,y,z,0) &= u_o \sin\left( \frac{x}{L} \right) \cos\left( \frac{y}{L} \right) \cos\left( \frac{z}{L} \right)
    \label{e:tgv_u}
    \\
    v(x,y,z,0) &= -u_o \cos\left( \frac{x}{L} \right) \sin\left( \frac{y}{L} \right) \cos\left( \frac{z}{L} \right)
    \label{e:tgv_v}
    \\
    w(x,y,z,0) &= 0
    \label{e:tgv_w}
    \\
    p(x,y,z,0) &= \rho_o u_o^2
                \left( \frac{1}{\gamma M_o^2}
                      + \frac{1}{16}\left[ \cos\left( \frac{2x}{L} \right)
                                         + \cos\left( \frac{2y}{L} \right) \right]
                                    \left[ \cos\left( \frac{2z}{L} \right) + 2 \right]
                \right)
    \label{e:tgv_p}
\end{align}
\end{subequations}
where the reference velocity, density, and Mach number are $u_o=1$, $\rho_o=1$, and $M_o=0.1$, respectively. The quantity $L$ defines a length scale for the problem; Reynolds number is defined as $Re=(\rho_o u_o L) / \mu$, and is set at $1\,600$. The fluid is modeled as a perfect gas with a specific heat ratio of $\gamma=1.4$ and Prandtl number of $Pr=0.71$. From the ideal gas law $RT_o=p_o/\rho_o$, and if we initialize the flow field with the assumption of isothermal flow, then $p/\rho = p_o/\rho_o$. This relationship allows the initial density field to be set according to Eq.~(\ref{e:tgv_rho}). The flow is computed inside a square domain $\mathrm{\Omega}=\{x,y,z~|~0 \leqslant x,y,z \leqslant 2\pi L\}$ with periodic boundaries. A characteristic convective time scale can be defined as $t_c = L / u_o$.
The non-dimensional integrated kinetic energy is
\begin{align}
    K = \frac{1}{\rho_o u_o^2 V} \int_\mathrm{\Omega} \frac{1}{2} \, \rho \, \bm{u}\cdot\bm{u} \, \drmb{x}
\end{align}
where $V$ is the total volume of the domain and $\drmb{x} = \drm x \drm y \drm z$. For this test case we choose $L=1$ such that the total volume is $V=8\pi^3$. The principal method of testing turbulent flow simulation methodologies using the TGV test case is to compute and track the dissipation rate of the kinetic energy through time. The dissipation rate based upon the kinetic energy is
\begin{align}
    \epsilon(K) = -\frac{\drm K}{\drm t^\star}
\end{align}
where $t^\star = t u_o / L$. The non-dimensional integrated enstrophy is
\begin{align}
    \zeta = \frac{t_c^2}{\rho_o V} \int_\mathrm{\Omega} \frac{1}{2} \, \rho \, \bm{\omega}\cdot\bm{\omega} \, \drmb{x}.
\end{align}
For strictly incompressible flow, the non-dimensional theoretical vorticity-based dissipation rate is proportional to $\zeta$ by
\begin{align}
    \epsilon(\zeta) = \frac{2\mu}{\rho_o u_o^2 t_c} \zeta.
\end{align}
In a compressible fluid, the non-dimensional theoretical dissipation rate is based upon the summation of the following three terms
\begin{subequations}
\begin{align}
    \epsilon(\bm{S}^d) = \frac{2 \mu t_c}{\rho_o u_o^2 V} \int_\mathrm{\Omega} \bm{S}^d:\bm{S}^d \, \drmb{x},
\end{align}
\begin{align}
    \epsilon(p) = -\frac{t_c}{\rho_o u_o^2 V} \int_\mathrm{\Omega} p \, \nabla \cdot \bm{u} \, \drmb{x},
\end{align}
\begin{align}
    \epsilon(\mu_b) = \frac{\mu_b t_c}{\rho_o u_o^2 V} \int_\mathrm{\Omega} (\nabla \cdot \bm{u})^2 \, \drmb{x}
\end{align}
\end{subequations}
where $\epsilon(\bm{S}^d)$ and $\epsilon(p)$ are the dissipation terms based upon the deviatoric strain-rate tensor $\bm{S}^d$ and pressure dilatation, respectively. Under Stokes' hypothesis, the bulk viscosity $\mu_b$ is assigned a value of zero, which leads to the second coefficient of viscosity taking the value $\lambda = -2/3\mu$; therefore, the dissipation due to the bulk viscosity is neglected. Furthermore, for low Mach number flows with negligible compressibility effects, the theoretical dissipation rate reasonably approximates the integrated enstrophy and can be estimated by $\epsilon(\bm{S}^d)$. In these simulations, we compute the theoretical dissipation rate as $\epsilon(\bm{S}^d)+\epsilon(p)$. All integrals are approximated with a sufficiently high-strength quadrature rule. The measured dissipation rate $\epsilon(K)$ is computed during post-processing using second-order finite differences to approximate the temporal derivative of the kinetic energy. A reference solution has been provided by van Rees~\etal~\cite{vanrees:2011}, which has to be scaled by a factor of $1/2~V$ to match the presentation of the current results. These authors performed a direct numerical simulation (DNS) at $Re=1\,600$ using a pseudo-spectral method on the incompressible Navier--Stokes equations with a resolution of $512^3$.

\begin{figure}[H]
    \renewcommand{\fsize}{60mm}
    \centering
    \subfloat[]{
        \hspace{-0.4in}
        \includegraphics[height=\fsize,keepaspectratio]
        {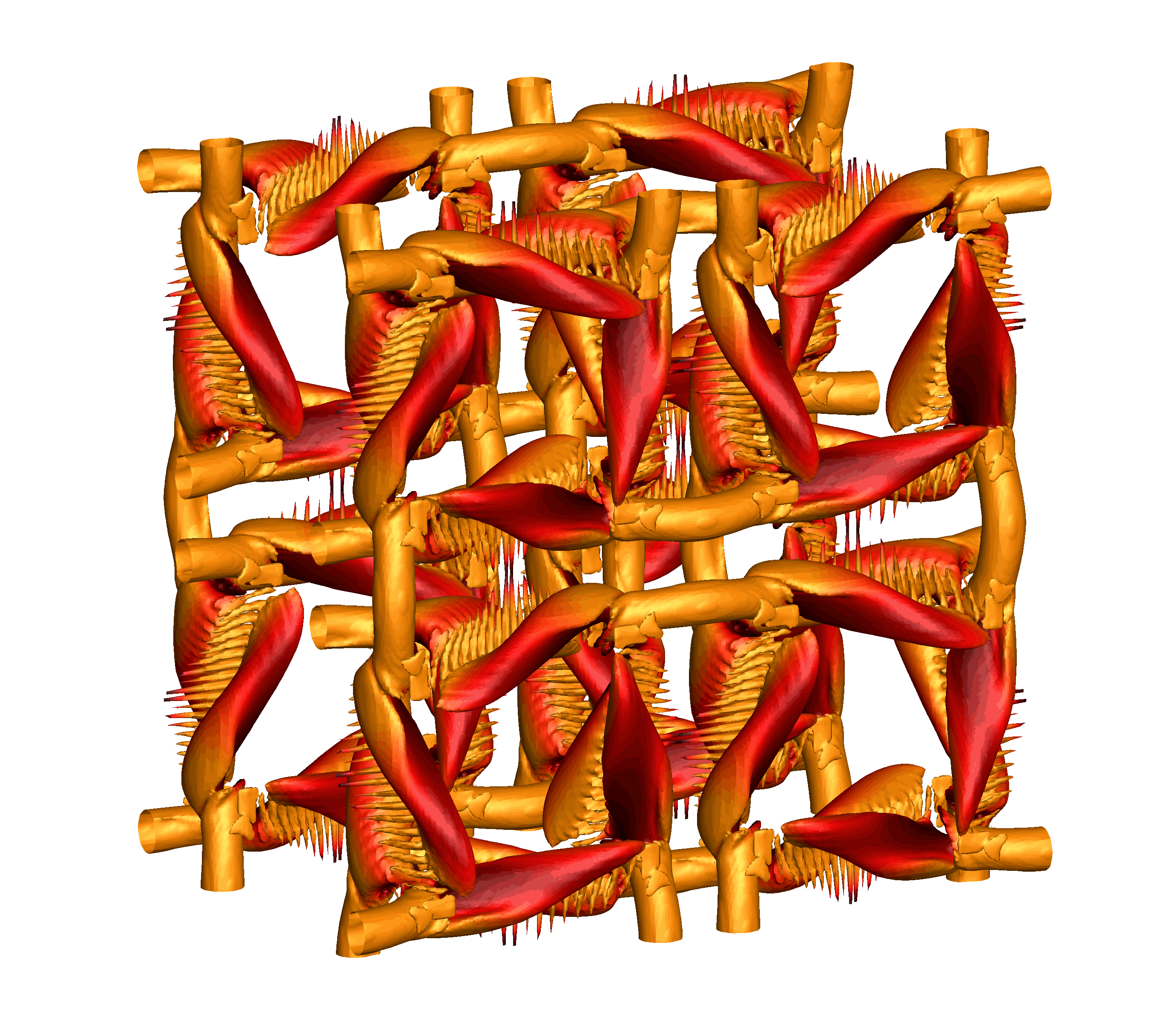}
        \label{f:tgv:Q:t5}
    }
    \subfloat[]{
        \hspace{-0.4in}
        \includegraphics[height=\fsize,keepaspectratio]
        {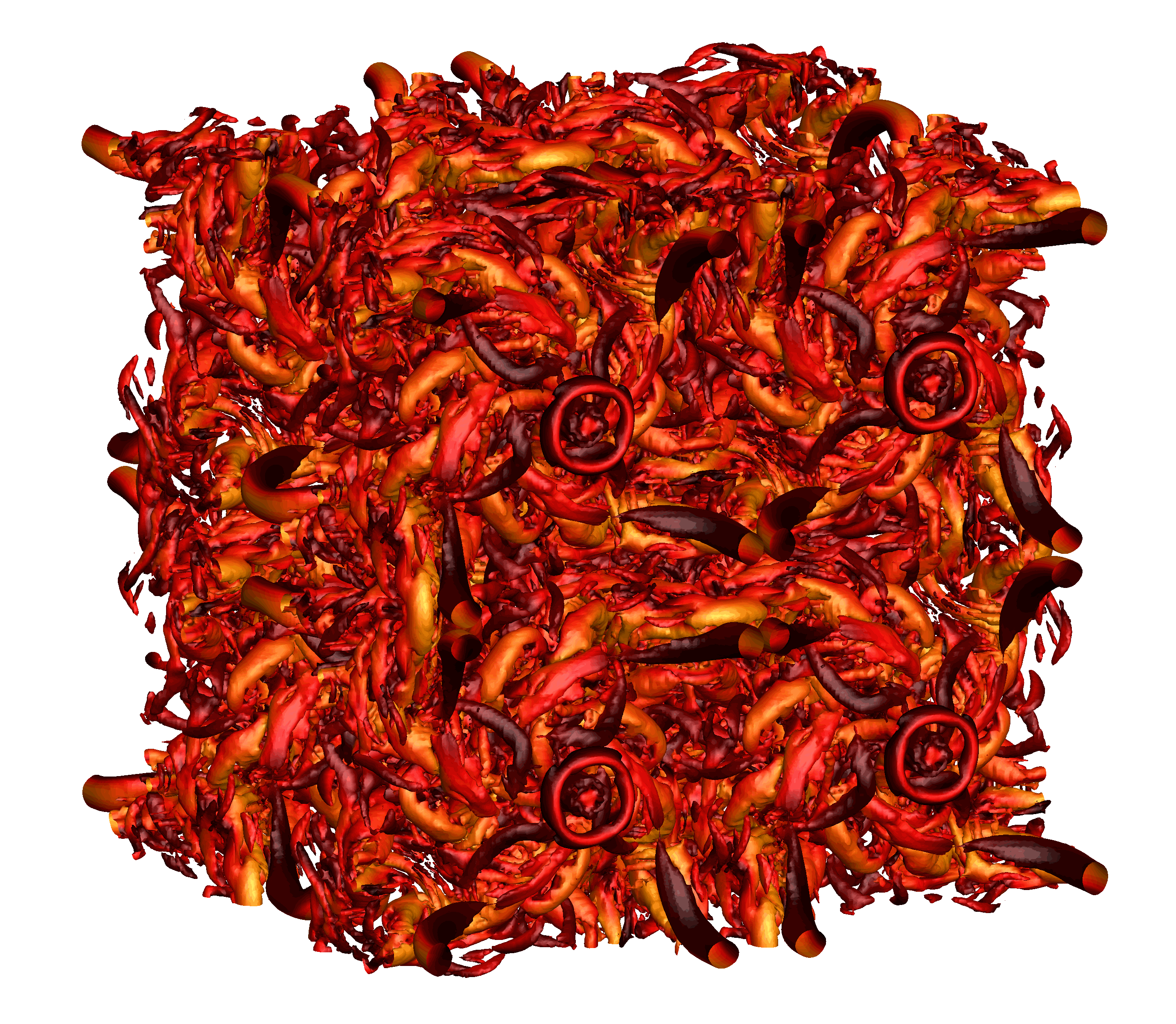}
        \label{f:tgv:Q:t11}
    }
    \subfloat[]{
        \hspace{-0.35in}
        \includegraphics[height=\fsize,keepaspectratio]
        {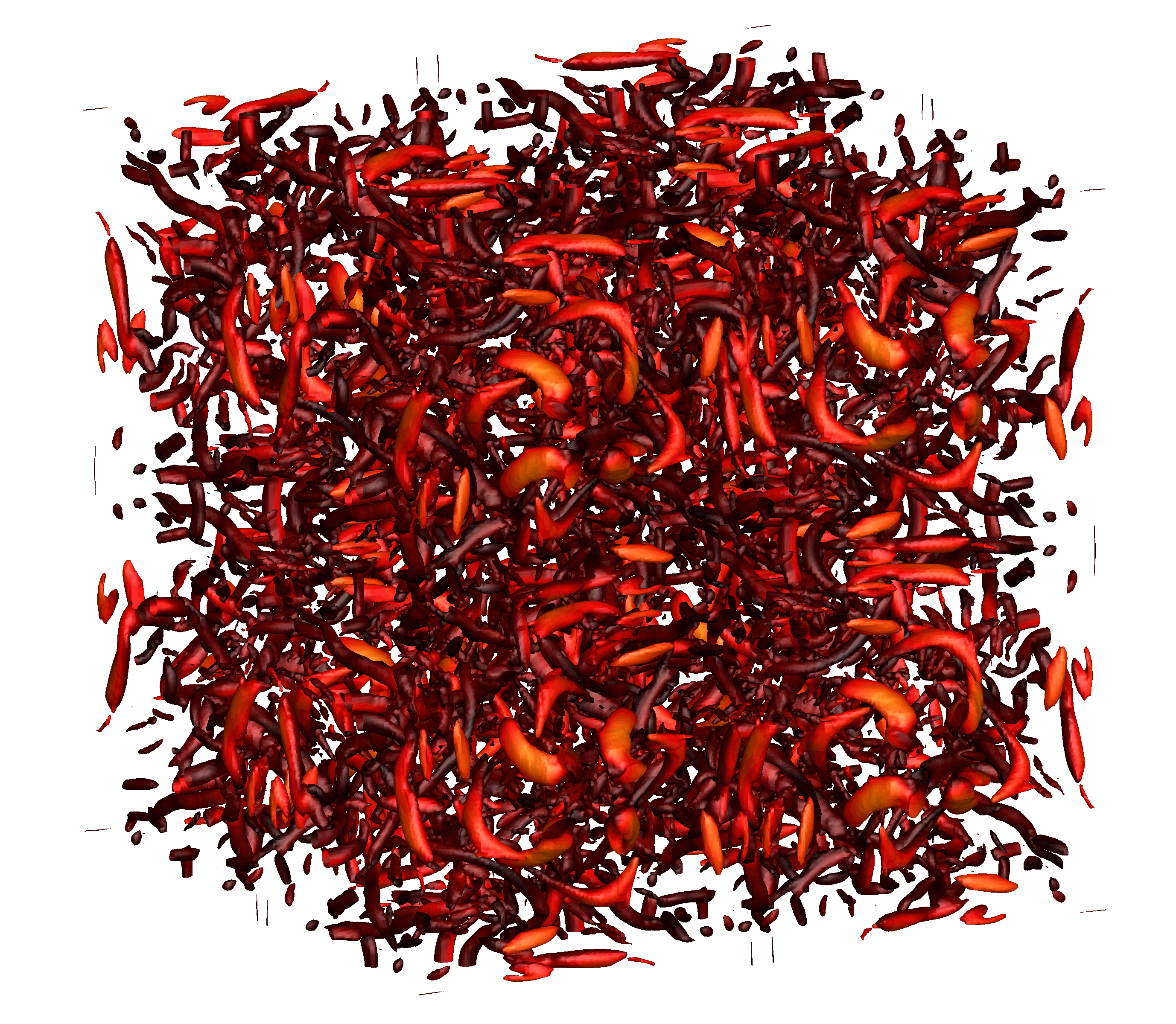}
        \label{f:tgv:Q:t20}
    }
    
    \caption{TGV: SD result of \emph{Q}-criterion ($QL^2/u^2_{o}=1.5$) colored by velocity magnitude at (a) $t^\star=5$, (b) $t^\star=11$ and (c) $t^\star=20$ on a $64^3$ grid using $\p=3$ ($256^3$ DoF).}
    \label{f:tgv:Q}
\end{figure}

\subsubsection{Well-resolved}\label{s:tgv:wellresolved}

\begin{figure}[H]
    \renewcommand{\fsize}{63mm}
    \centering
    \subfloat[]{
        \hspace{-0.24in}
        \includegraphics[height=\fsize+4mm,keepaspectratio]
        {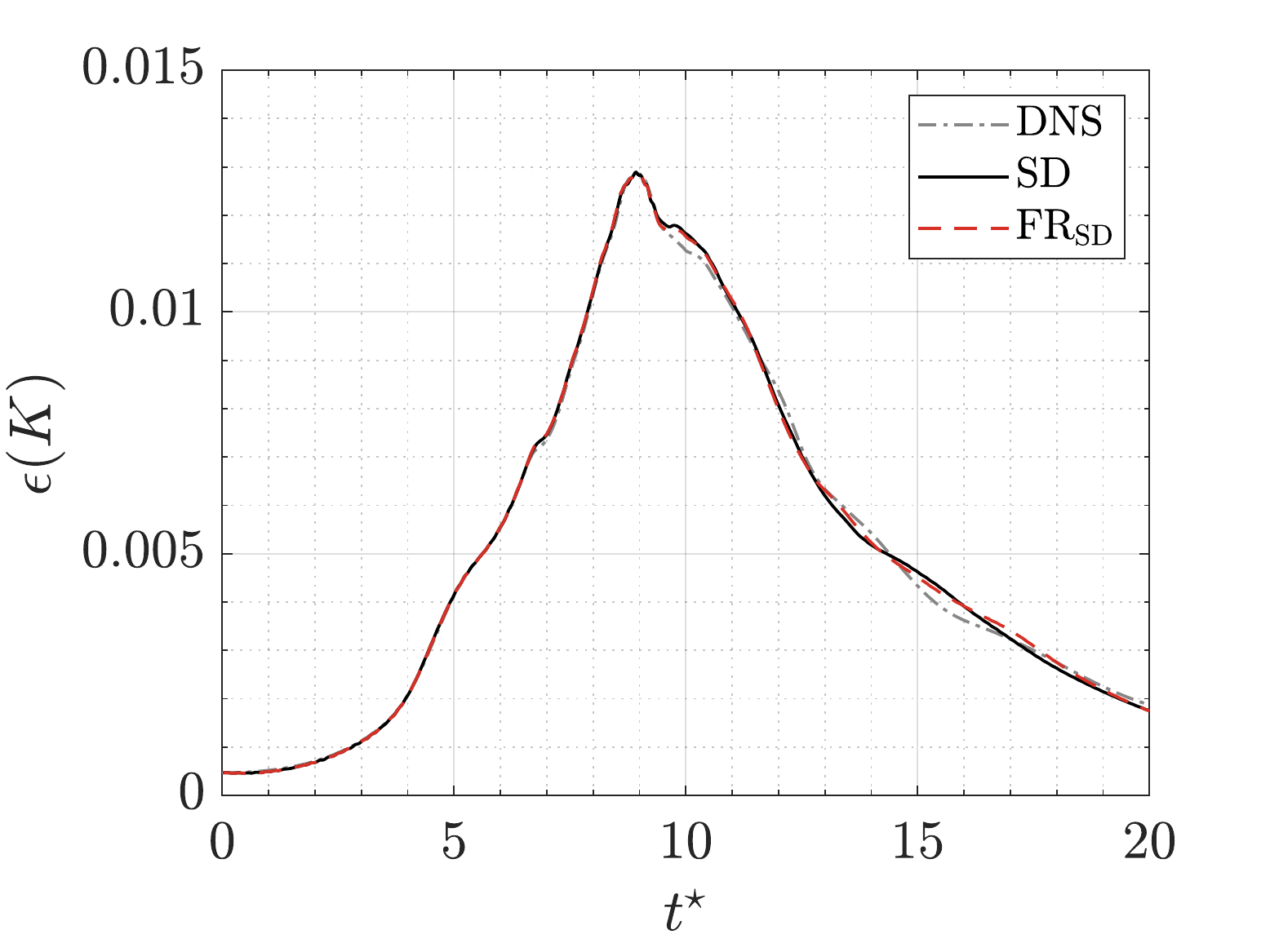}
        \label{f:tgv:64:dKdt}
    }
    \subfloat[]{
        \hspace{-0.24in}
        \includegraphics[height=\fsize+4mm,keepaspectratio]
        {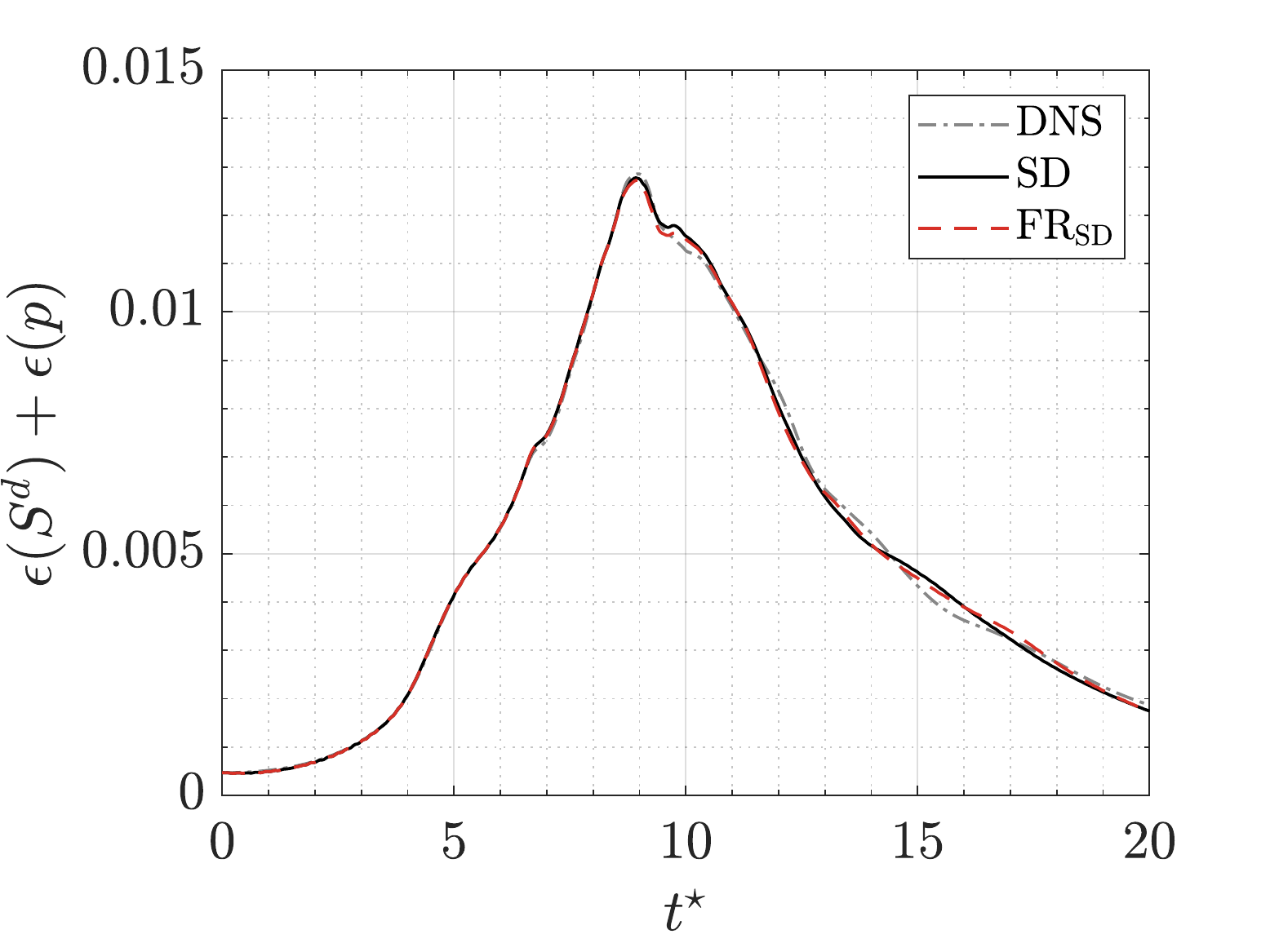}
        \label{f:tgv:64:dissStrainPdila}
    }
    \\
    \vspace{-0.1in}
    \subfloat[]{
        \includegraphics[height=\fsize,keepaspectratio]
        {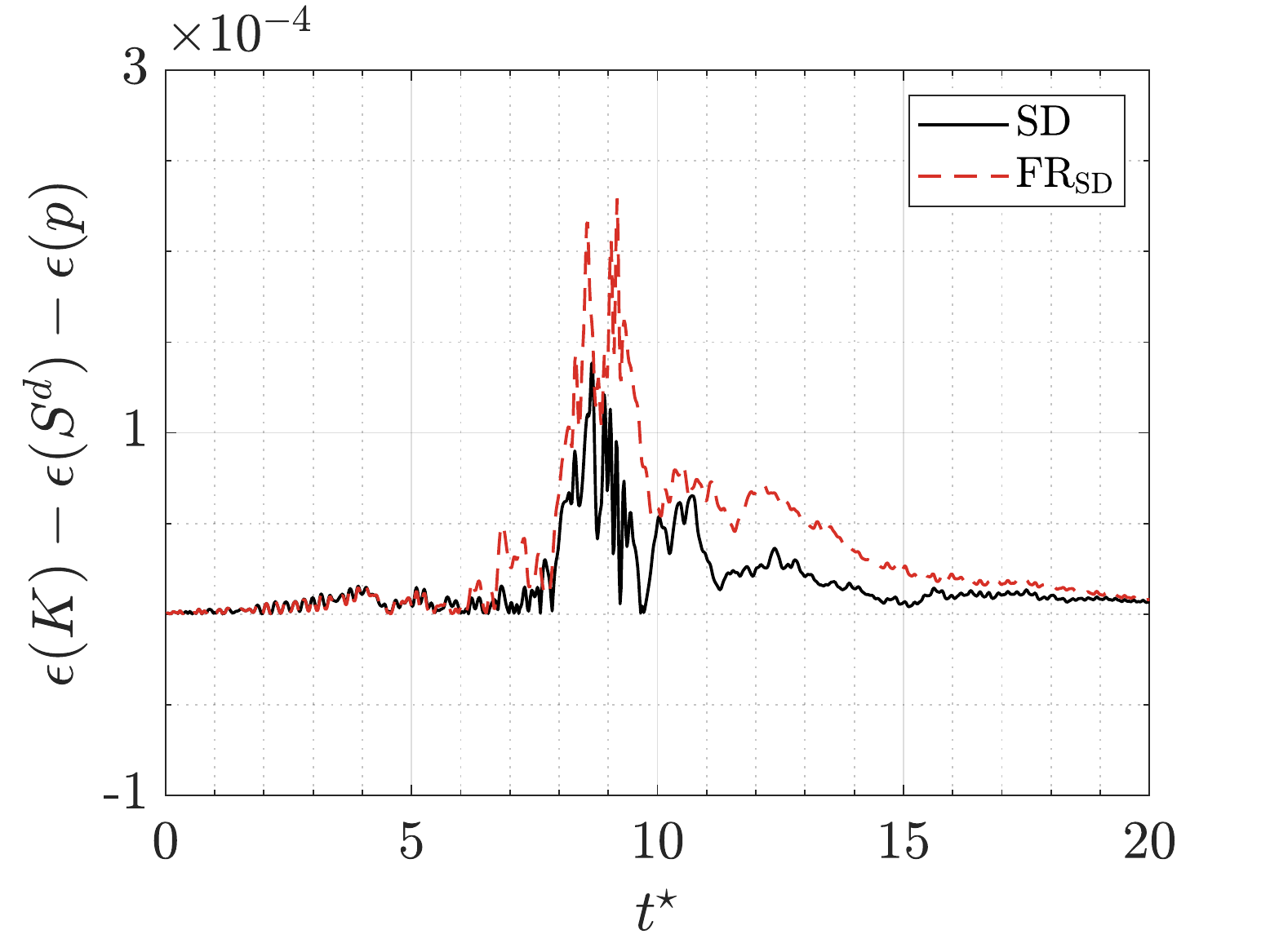}
        \label{f:tgv:64:dkdt-strain-pdila}
    }
    \subfloat[]{
        \includegraphics[height=\fsize,keepaspectratio]
        {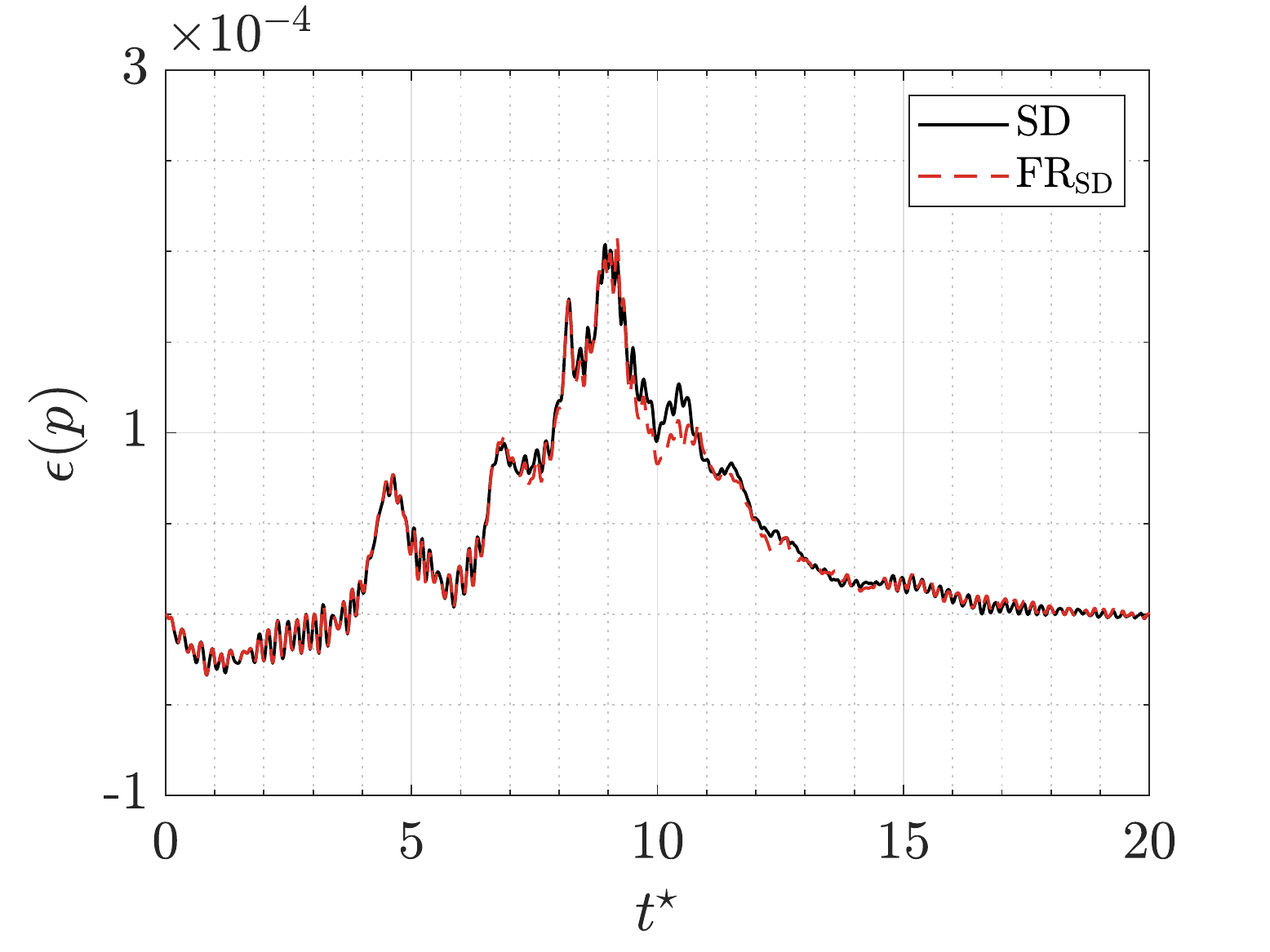}
        \label{f:tgv:64:pdila}
    }
    
    \caption{TGV: (a) measured dissipation rate based on kinetic energy $\epsilon(K)$, (b) theoretical dissipation based on strain-rate $\epsilon(\bm{S}^d)$ and pressure dilatation $\epsilon(p)$, (c) difference between (a) and (b) $\epsilon(K) - \epsilon(\bm{S}^d) - \epsilon(p)$, (d) pressure dilatation $\epsilon(p)$. Results are from a $64^3$ grid using $\p=3$ ($256^3$ DoF). DNS results have been provided by van Rees~\etal~\protect\cite{vanrees:2011}.}
    \label{f:wellresolved}
\end{figure}

First, we perform well-resolved simulations of the TGV using a $64^3$ grid and $\p=3$, giving a total of $256^3$ DoF, to show the ability of both $\SD$ and $\FRSD$ to accurately capture the flow physics of the TGV and its transition to and subsequent decaying of turbulence. All simulations for this test case are run using Davis' form of the Rusanov approximate Riemann solver such that a close comparison can be made to the results from Vermeire~\etal~\cite{vermeire-witherden-vincent:2017} who used $\FRDG$ with similar initial conditions. Figure~\ref{f:tgv:Q} demonstrates the roll-up of the vortex sheets at $t^\star=5$, the transition to turbulence leading to the production of small-scale vortical structures at $t^\star=11$, and the subsequent decaying of these structures depicted at $t^\star=20$. Results of $\epsilon(K)$ and $\epsilon(\bm{S}^d)+\epsilon(p)$ in Fig.~\ref{f:tgv:64:dKdt} and Fig.~\ref{f:tgv:64:dissStrainPdila} indicate little discrepancy between the measured and theoretical dissipation rates, with the peak dissipation rate occurring near $t^\star=9$. The actual difference between $\epsilon(K)$ and $\epsilon(\bm{S}^d)+\epsilon(p)$ is plotted in Fig.~\ref{f:tgv:64:dkdt-strain-pdila} and can be attributed to numerical dissipation and dispersion, non-conservation in evaluating the derivative of the conservative variables since the scheme is only guaranteed to be $C^0$ continuous~\cite{vermeire:2016}, and numerical errors aliased from the higher modes to the lower ones. We can observe that the maximum difference under $\SD$ is approximately 60\% of that exhibited under $\FRSD$. The pressure dilatation-based dissipation rate---which measures compressibility effects on the dissipation of turbulent energy---among the two schemes is essentially identical and shown in Fig.~\ref{f:tgv:64:pdila}. Maximum values of $\epsilon(p)$ are approximately \num{2e-4}.

\begin{figure}[H]
    \renewcommand{\fsize}{80mm}
    \centering
    \includegraphics[height=\fsize,keepaspectratio]
    {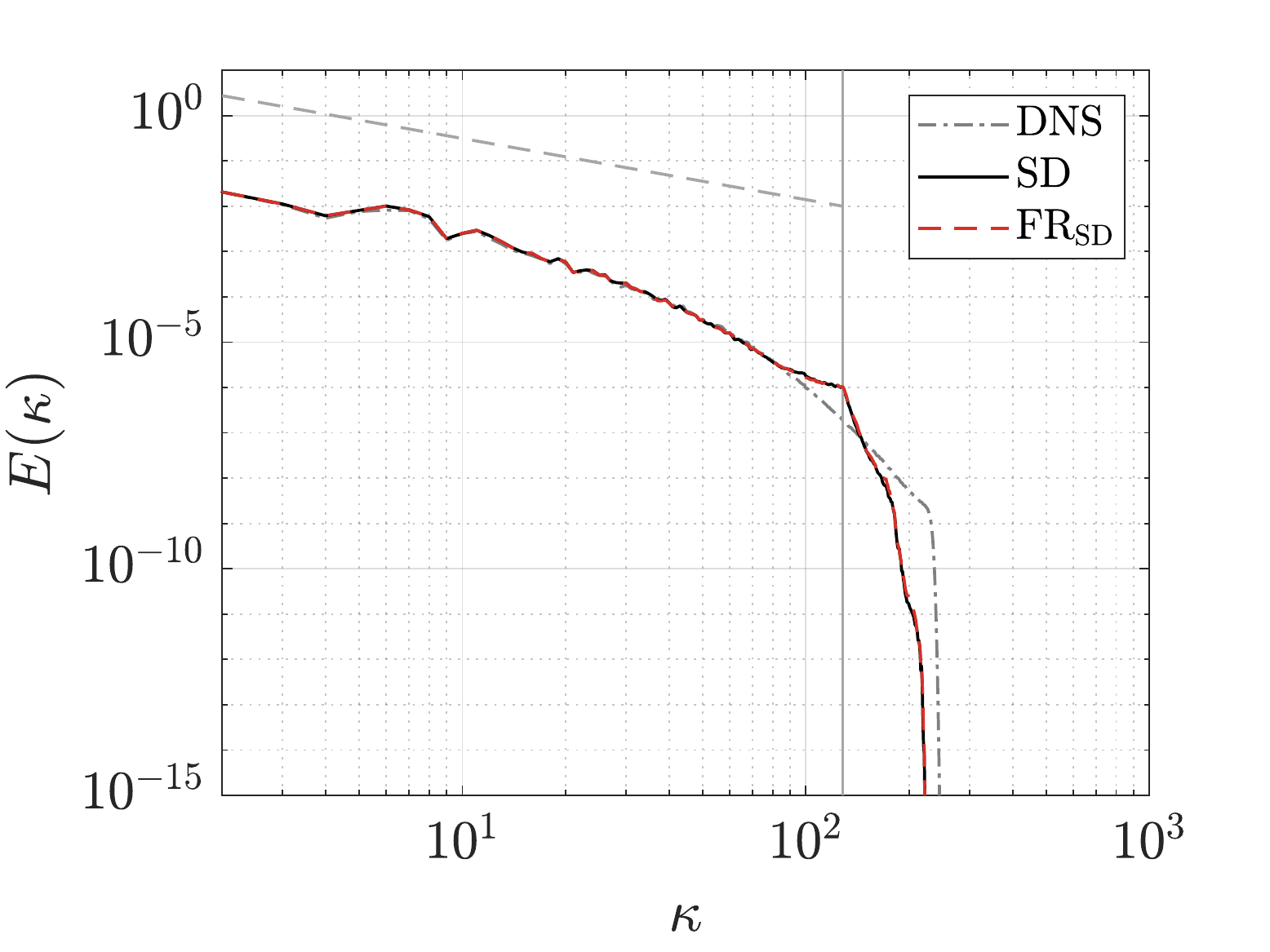}
    \label{f:tgv:64:energy_spectra}
    
    \caption{TGV: energy spectra at $t^\star=9$ on a $64^3$ grid using $\p=3$ ($256^3$ DoF). The cutoff wavenumber ($\color{gray} -$) and the -5/3 slope ($\color{gray} --$) are plotted in gray. DNS results have been provided by van Rees~\etal~\protect\cite{vanrees:2011}.}
    \label{f:energy_spectra}
\end{figure}

Following the procedure laid out in Brachet~\etal~\cite{brachet:1983}, we compute the spherically-averaged energy spectra $E(\kappa)$ at the peak dissipation rate ($t^\star=9$). Results are plotted in Fig.~\ref{f:energy_spectra} for both schemes against the reference DNS result. Both $\SD$ and $\FRSD$ exhibit an accumulation of energy near the cutoff wavenumber $\kappa=128$ due to the dissipation inherent to the Riemann solver~\cite{moura:2017}. Sharp dissipation is known to promote this pile-up of energy prior to the dissipation range and induce a more pronounced bottleneck effect~\cite{falkovich:1994}. This build-up of energy at the smallest captured scales is related to contamination of the true physics by numerical errors such as dispersion.

\subsubsection{Under-resolved}\label{s:tgv:underresolved}

\begin{figure}[H]
    \renewcommand{\fsize}{61mm}
    \centering
    \subfloat[]{
        \includegraphics[height=\fsize,keepaspectratio]
        {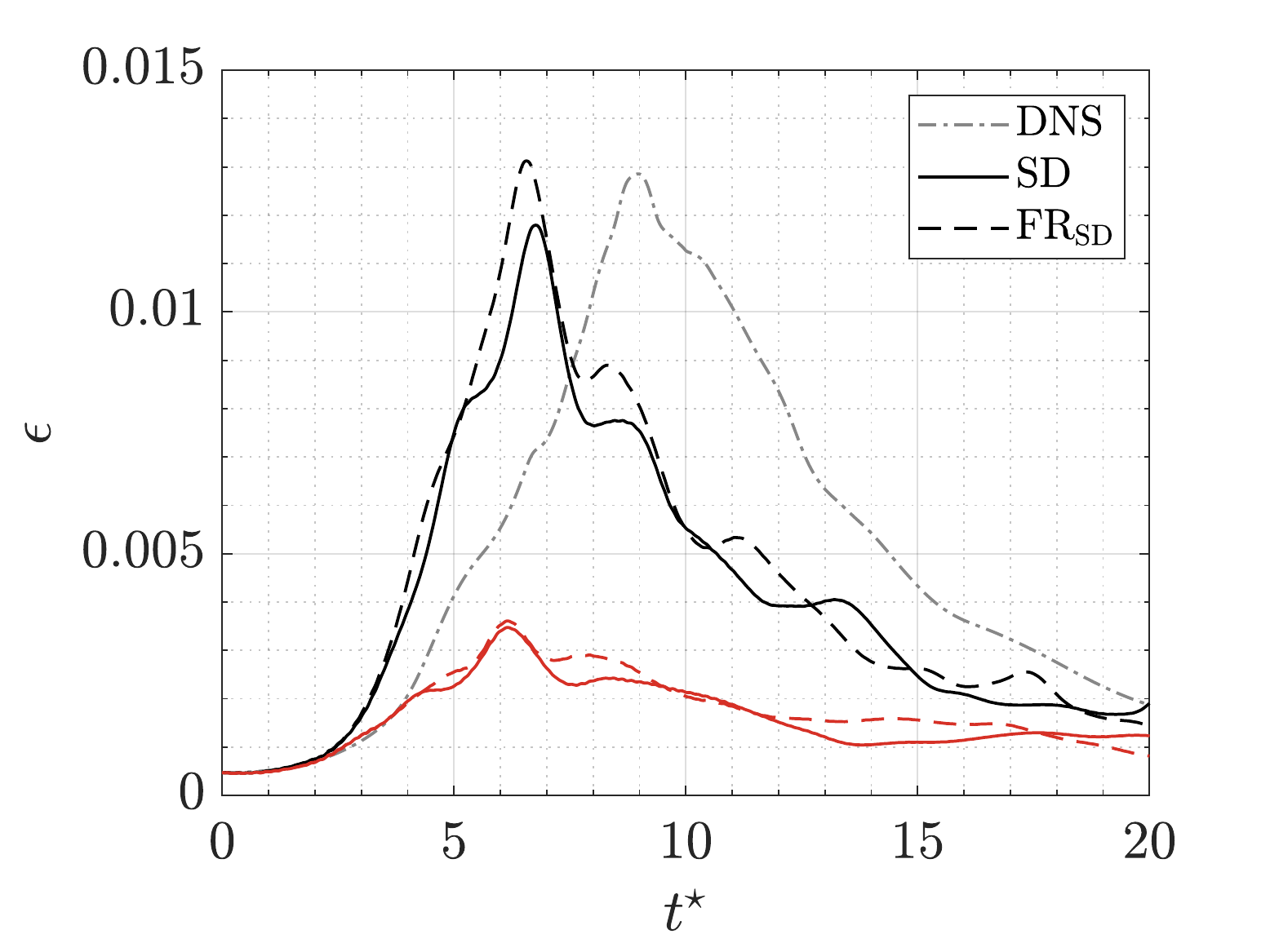}
        \label{f:tgv:8:p3:diss}
    }
    \subfloat[]{
        \includegraphics[height=\fsize,keepaspectratio]
        {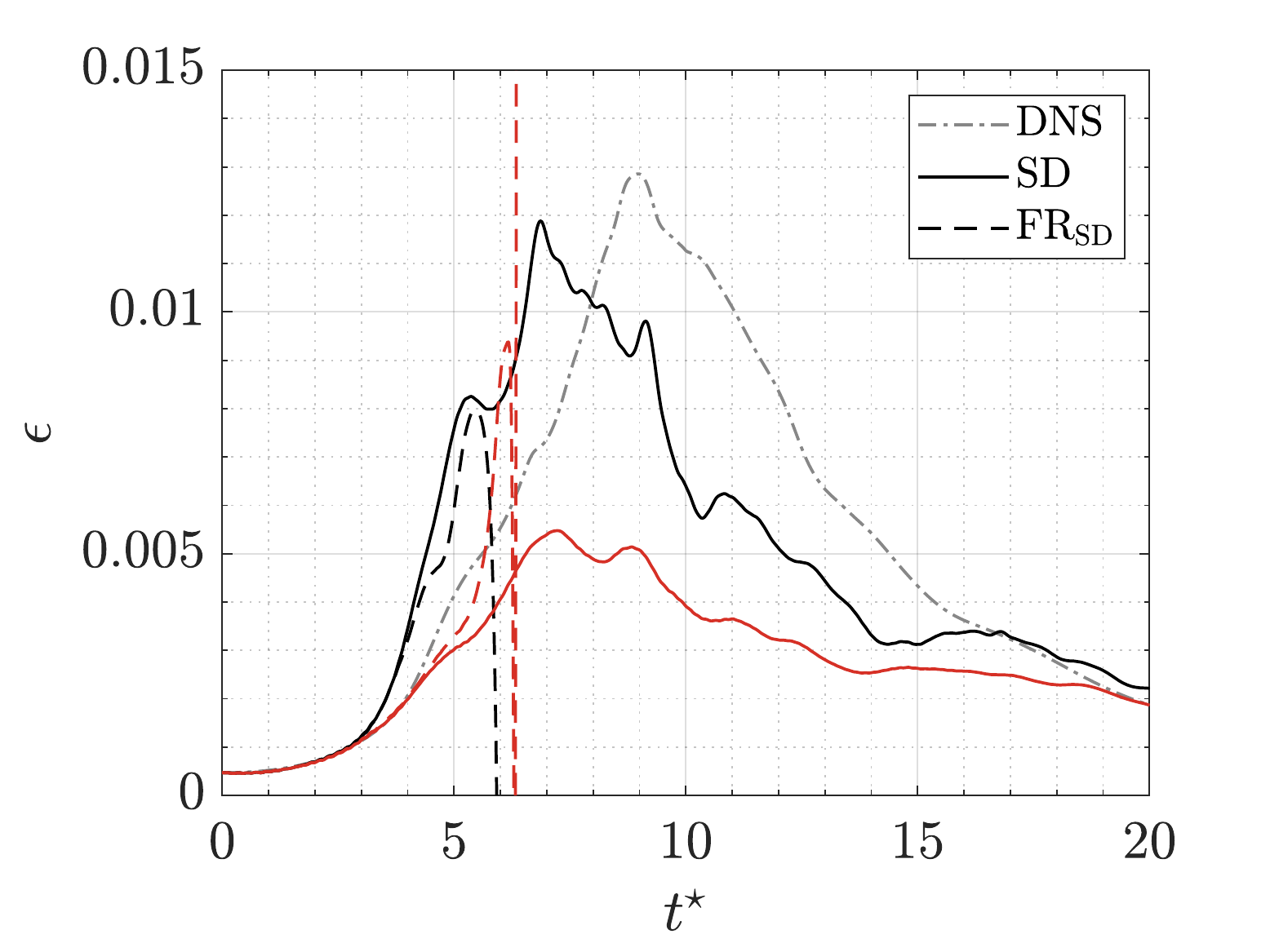}
        \label{f:tgv:8:p4:diss}
    }
    \\
    \vspace{-0.2in}
    \subfloat[]{
        \includegraphics[height=\fsize,keepaspectratio]
        {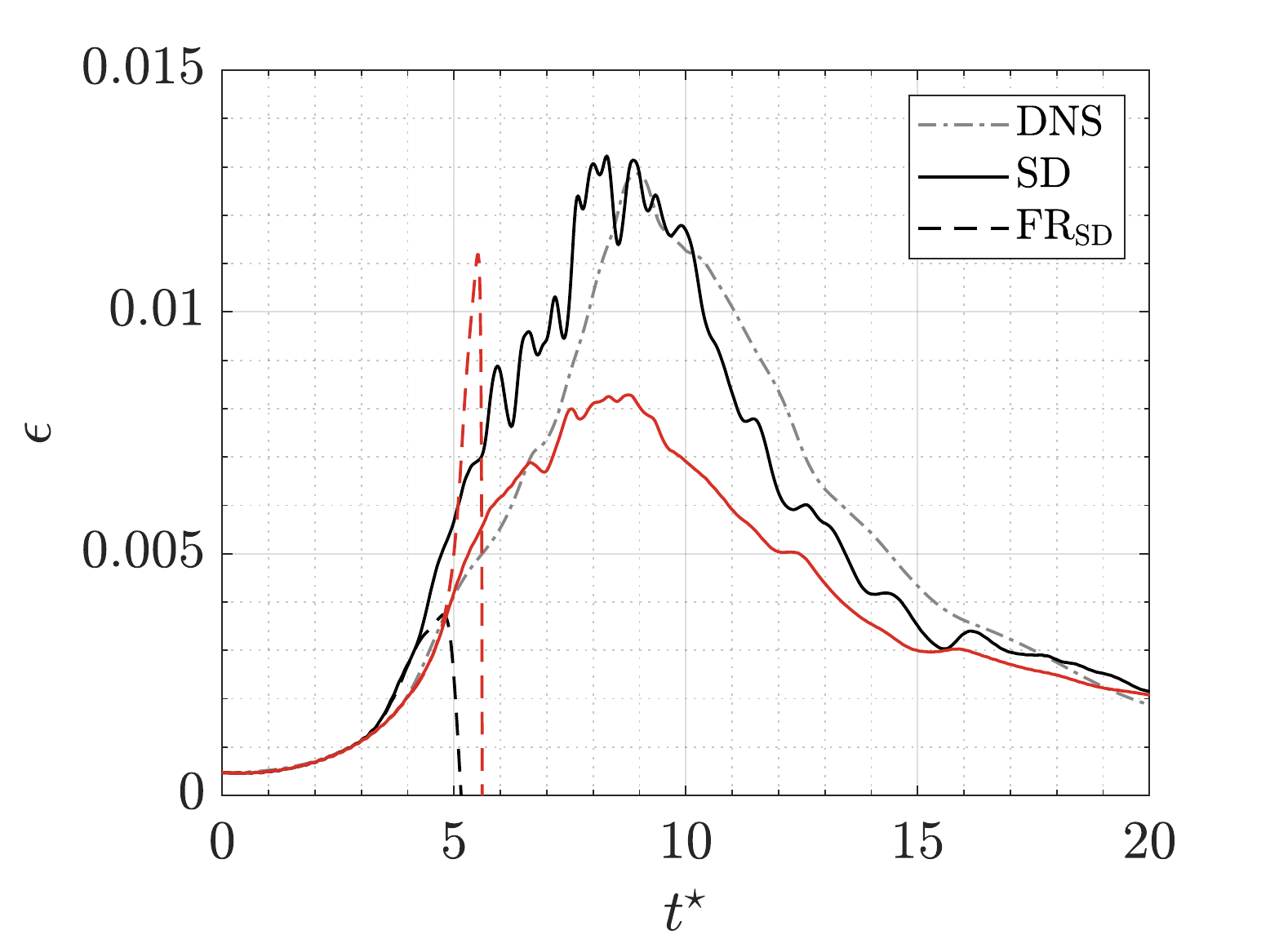}
        \label{f:tgv:8:p5:diss}
    }
    \subfloat[]{
        \includegraphics[height=\fsize,keepaspectratio]
        {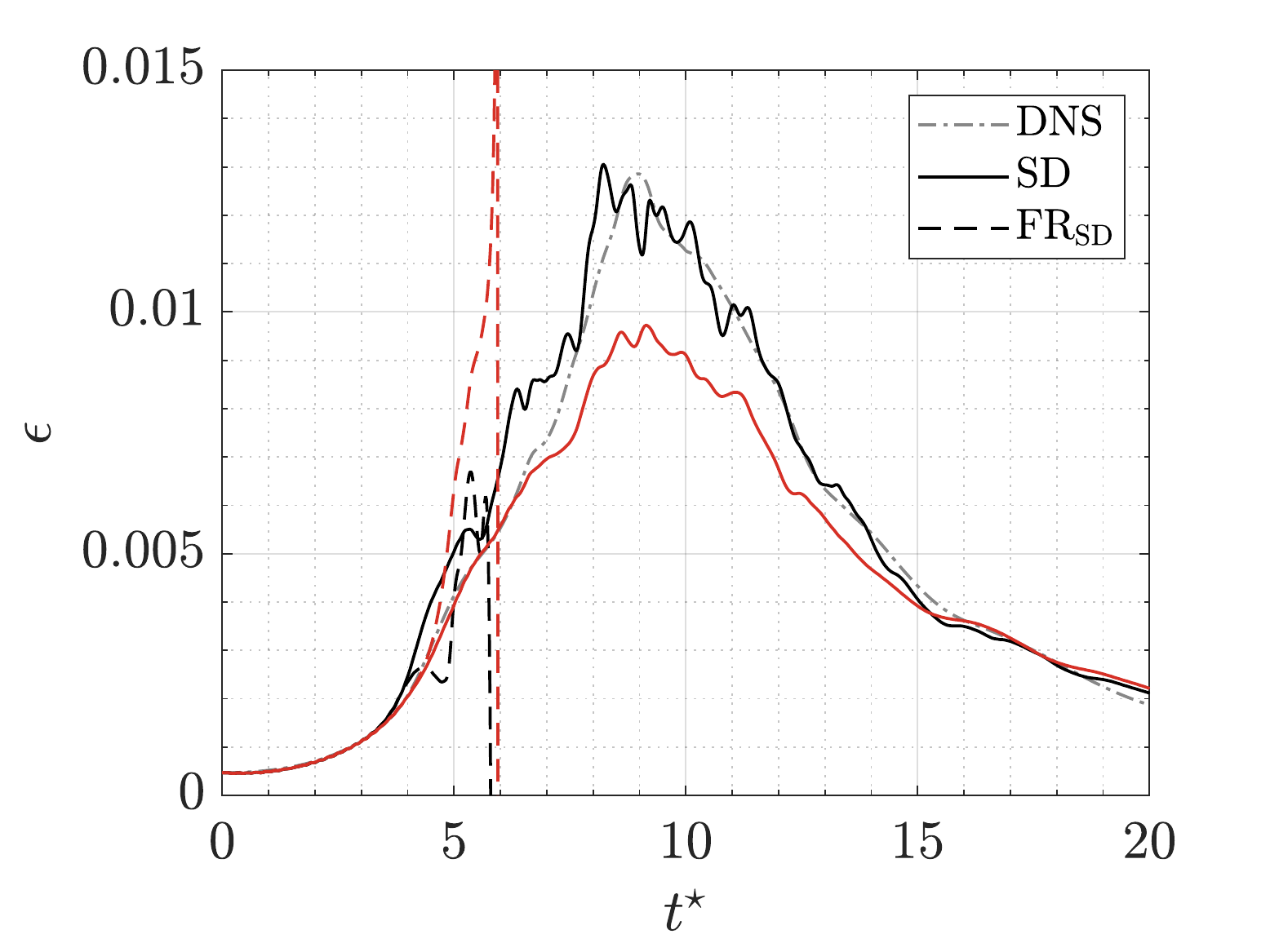}
        \label{f:tgv:8:p6:diss}
    }
    \\
    \vspace{-0.2in}
    \subfloat[]{
        \includegraphics[height=\fsize,keepaspectratio]
        {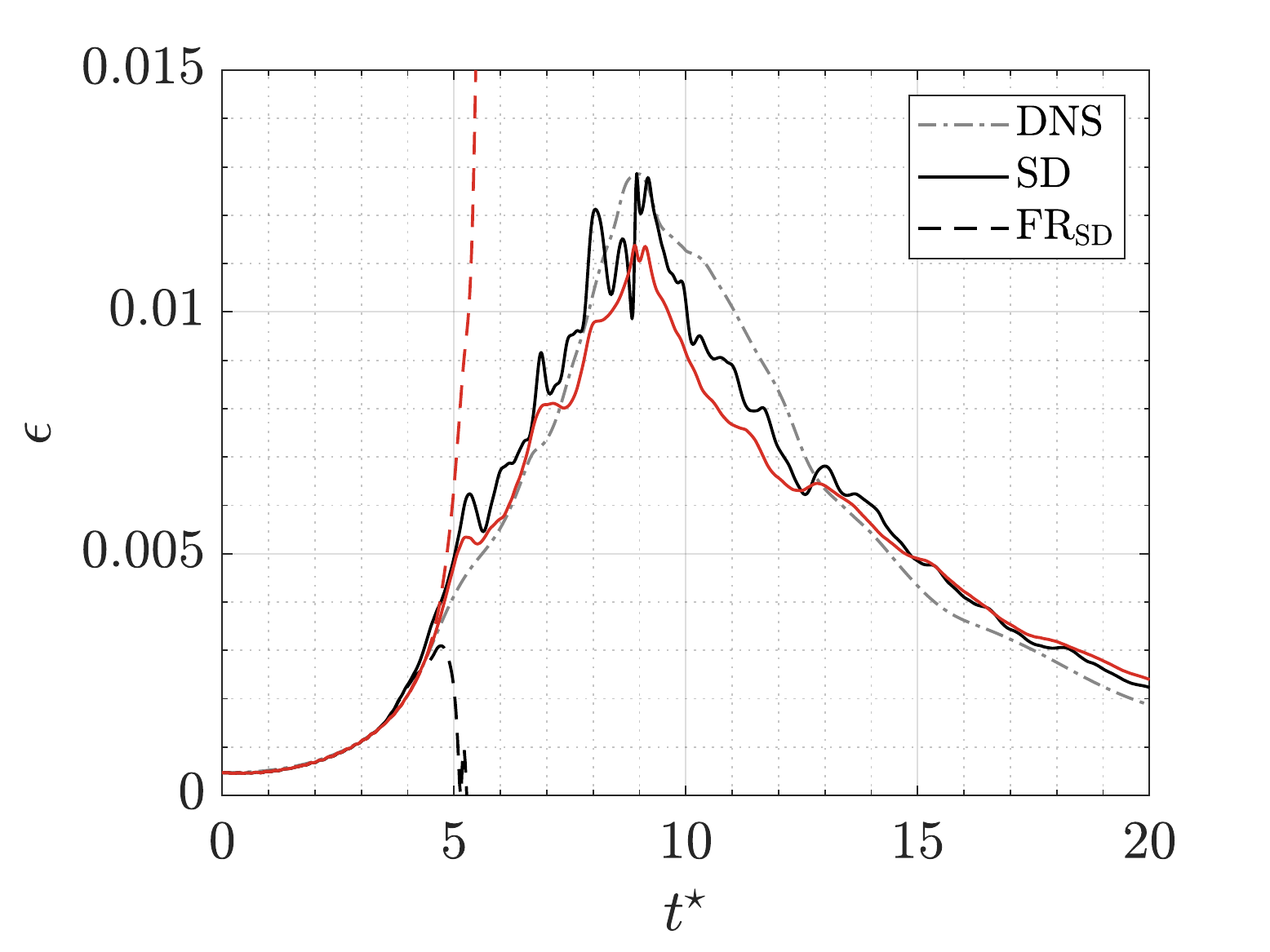}
        \label{f:tgv:8:p7:diss}
    }
    \subfloat[]{
        \includegraphics[height=\fsize,keepaspectratio]
        {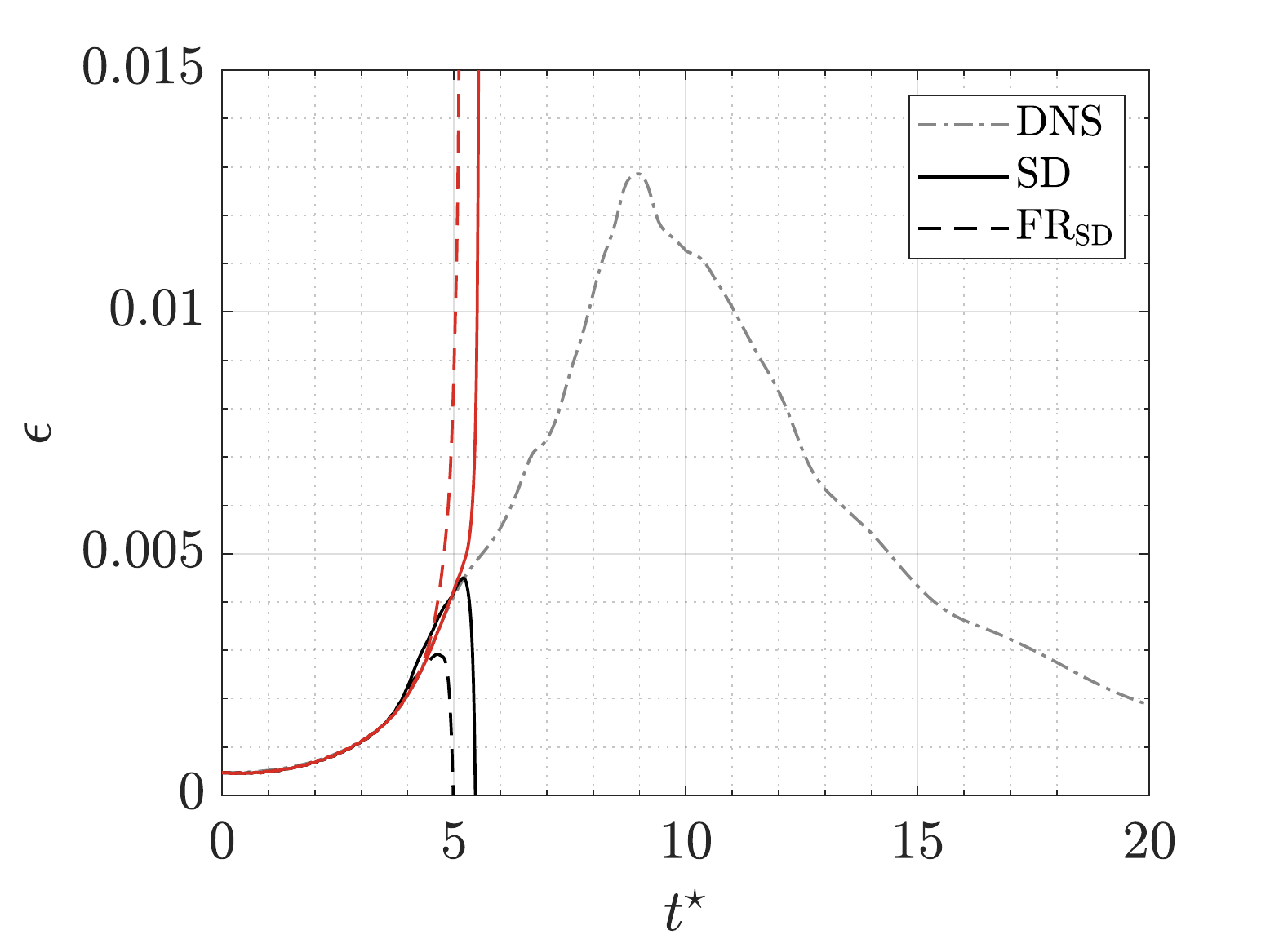}
        \label{f:tgv:8:p8:diss}
    }
    
    \caption{TGV: measured dissipation rate based on kinetic energy $\epsilon(K)$ ({\it black}) and theoretical dissipation rate based on strain-rate and pressure dilatation $\epsilon(\bm{S}^d) + \epsilon(p)$ ({\it red}) on a $8^3$ grid; (a) $\p=3$, (b) $\p=4$, (c) $\p=5$, (d) $\p=6$, (e) $\p=7$, (f) $\p=8$. DNS results have been provided by van Rees~\etal~\protect\cite{vanrees:2011}.}
    \label{f:underresolved:1}
\end{figure}

We perform under-resolved simulations of the TGV using an $8^3$ grid while increasing $\p$ to see the effect of higher polynomial orders on stability for true spectral difference and the modified spectral difference recovered via the flux reconstruction formulation. We start the simulations at $\p=3$ and increment the polynomial order by 1 until both schemes produce unstable solutions, which occurs at $\p=8$. Therefore, we are considering seven different levels of resolution: $24^3$, $32^3$, $40^3$, $48^3$, $56^3$, $64^3$ and $72^3$ DoF. To reduce the amount of numerical dissipation, we run all simulations for this test case using Roe's scheme~\cite{toro:1999} for the approximate Riemann solver. Results of $\epsilon(K)$ and $\epsilon(\bm{S}^d)+\epsilon(p)$ are plotted in Fig.~\ref{f:underresolved:1}. In Fig.~\ref{f:tgv:8:p3:diss}, we observe a large amount of numerical dissipation in the results computed using $\p=3$, whereby the rate of kinetic energy loss is overestimated at earlier times in the simulation, where the flow is restricted to a smaller range of scales. The simulation from $\FRSD$ is quickly rendered unstable at $\p=4$, largely due to aliasing errors produced at the higher wavenumbers when substantial roll-up of the vortex sheets occurs near $t^\star=5$---this blowup in the solution occurs at similar times for all higher values of $\p$. The simulations from the $\SD$ scheme, on the other hand, demonstrate that as $\p$ is increased further, the solution is stable and the difference between the measured dissipation rate due to kinetic energy and the theoretical dissipation rate becomes smaller, and the result from $\epsilon(\bm{S}^d)+\epsilon(p)$ approaches the DNS result up to $\p=7$. However, the $\SD$ solution does become unstable at $\p=8$ near $t^\star=5$. Overall, these results indicate suppressed aliasing errors in and enhanced stability of the $\SD$ scheme on coarse grids with higher polynomial orders when performing under-resolved turbulence simulations without any filtering, subgrid-scale modeling, or de-aliasing.

\subsection{SD7003 at $Re=60\,000$, $\alpha=8^\circ$}\label{s:sd7003}

We perform implicit large eddy simulations of the transitional flow of a Selig--Donovan (SD) 7003 airfoil~\cite{selig-donovan:1989,selig:1995} at $Re=60\,000$, Mach number $M=0.2$ and angle-of-attack $\alpha=8^\circ$. This test case is commonly used to assess a numerical scheme's ability to predict separation and transition in a turbulent flow~\cite{visbal:2009,galbraith:2010,uranga:2011,garmann:2013,beck:2014}, and we compare results from the flux reconstruction and spectral difference schemes without any filtering, subgrid-scale modeling, or de-aliasing. Laminar flow separation and reattachment occurs on the upper surface of the airfoil, forming a laminar separation bubble (LSB) near the leading edge. Lift and drag on an airfoil can be significantly affected by an LSB, which can cause stability and control issues. The flow experiences transition near reattachment in the unsteady solution, which causes a region of turbulence over a large portion of the airfoil's upper surface and a turbulent wake downstream of the airfoil. 

\begin{figure}[H]
    \renewcommand{\fsize}{88mm}
    \centering\setcounter{subfigure}{0}
    \subfloat[]{
        \includegraphics[width=\fsize,keepaspectratio]
        {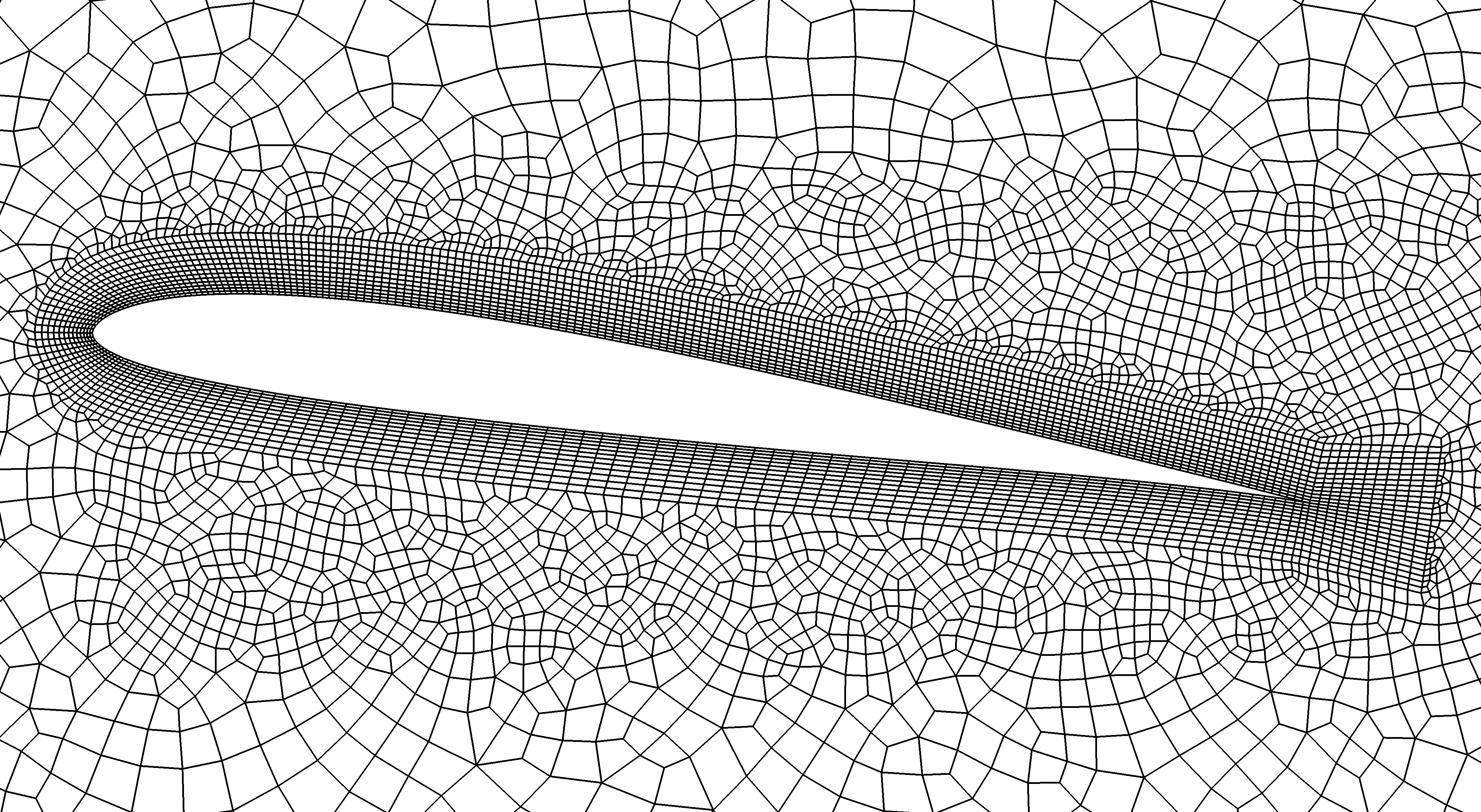}
    }
    \subfloat[]{
        \includegraphics[width=\fsize,keepaspectratio]
        {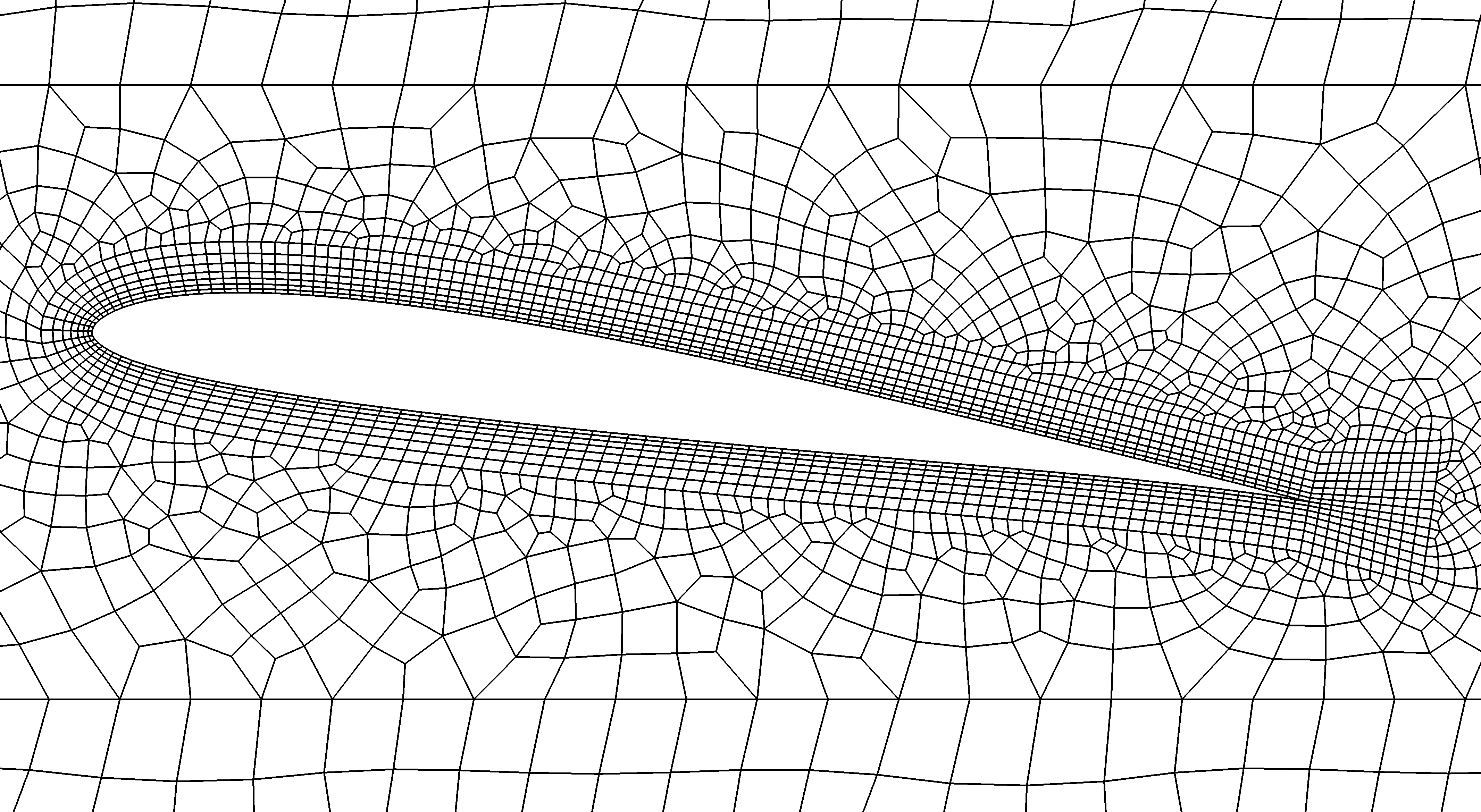}
    }
    
    \caption{SD7003 at $Re=60\,000$, $\alpha=8^\circ$: (a) near wall region of mesh A, provided by Vermeire~\etal~\protect\cite{vermeire-witherden-vincent:2017} (b) near wall region of mesh B.}
    \label{f:sd7003:mesh}
\end{figure}

\begin{figure}[H]
    \renewcommand{\fsize}{80mm}
    \centering
    \includegraphics[height=\fsize,keepaspectratio]
    {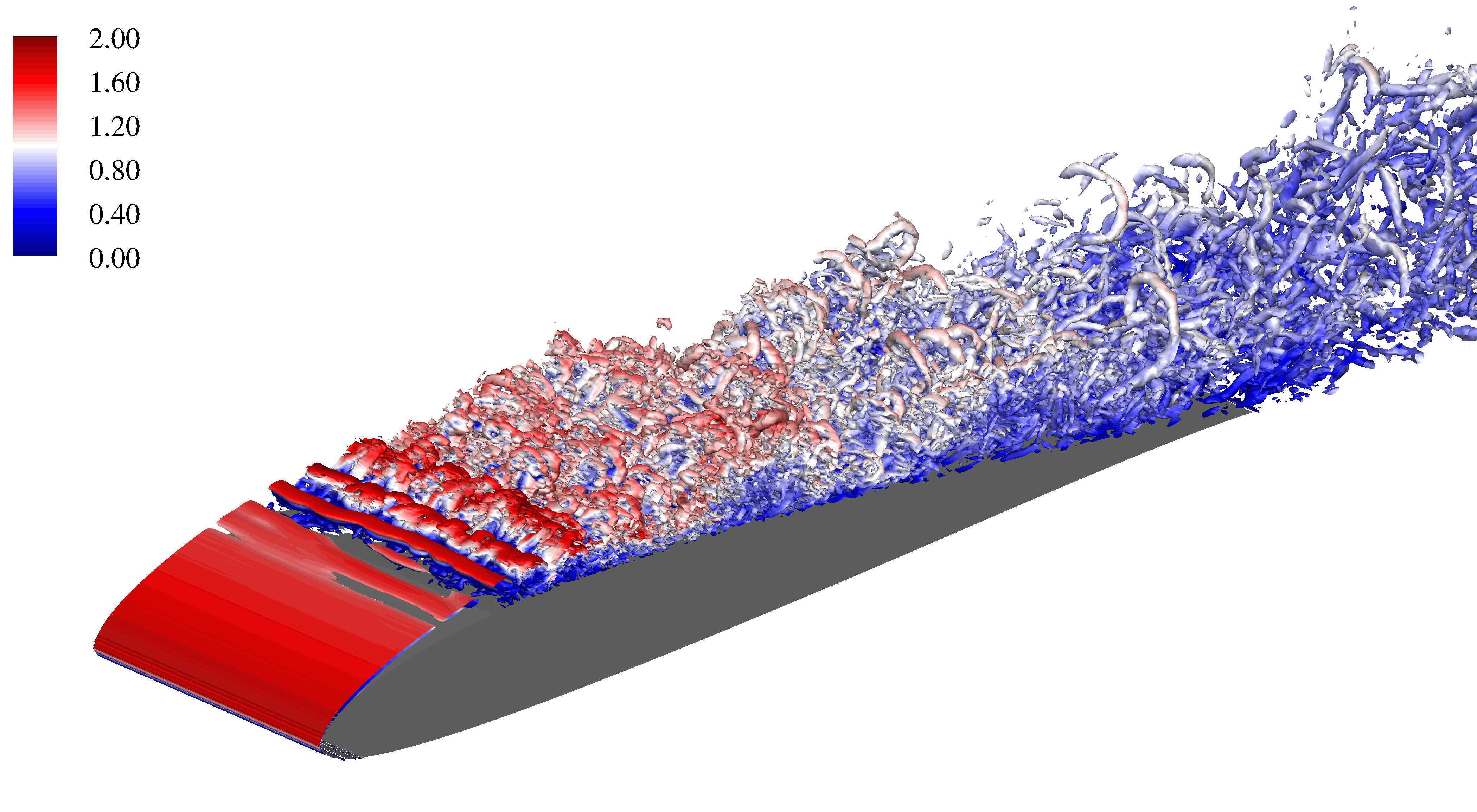}
    
    \caption{SD7003 at $Re=60\,000$, $\alpha=8^\circ$: isosurface of \emph{Q}-criterion ($Qc^2/u^2_{\infty}=500$) colored by velocity magnitude obtained using the $\SD$ scheme at $\p=7$.}
    \label{f:sd7003:Q}
\end{figure}

To perform these simulations, we use two meshes of different resolution as shown in Fig.~\ref{f:sd7003:mesh}, the first (mesh A) of which was provided by Vermeire~\etal~\cite{vermeire-witherden-vincent:2017}. We use these two different meshes to study the ability of each scheme to simulate under-resolved transitional and turbulent flow at varying levels of $\p$. Mesh A contains a total of $137\,916$ hexahedral elements with 12 elements in the spanwise direction. The domain extends $10c$ upstream and $20c$ downstream of the airfoil and extends in the spanwise direction by $0.2c$, where $c$ is the chord length. This spanwise length is deemed sufficient for capturing spanwise structures~\cite{uranga:2011}. We use this mesh to verify our implementation and directly compare results to those from a well-established FR implementation in PyFR~\cite{vermeire-witherden-vincent:2017}. For this mesh, we set $\p=4$ to make a direct comparison to these results which gives approximately \num{1.723e7} 
DoF. The second mesh constructed (mesh B) is a coarser mesh that contains a total of $33\,264$ elements with 8 elements in the spanwise direction, which provides roughly the same number of degrees of freedom (\num{1.703e7}) 
using $\p=7$. The upper surface of the airfoil in mesh A and mesh B is represented with 173 and 110 elements along the chord, respectively. This gives a total of \num{5.19e4} DoF on the upper surface in mesh A and \num{5.63e4} DoF in mesh B. To better capture the solid boundary curvature, the airfoil surface is represented by quartic elements. A no-slip adiabatic boundary condition is used for the airfoil surface, Riemann invariant boundary conditions are applied to the far field, and periodic conditions are applied in the spanwise direction. We use the low-storage, four-stage, third-order embedded pair time integration scheme (RK[4,3(2)]-2N) with adaptive time-stepping to integrate in time. We march forward in time for $30t_c$, where $t_c=c/u_{\infty}$ is one convective time period. At $20t_c$ the flow is considered fully developed, and we collect time and spanwise-average statistics between $20t_c$ and $30t_c$.

\begin{figure}[H]
    \renewcommand{\fsize}{70mm}
    \centering
    \subfloat[]{
        \includegraphics[height=\fsize,keepaspectratio]
        {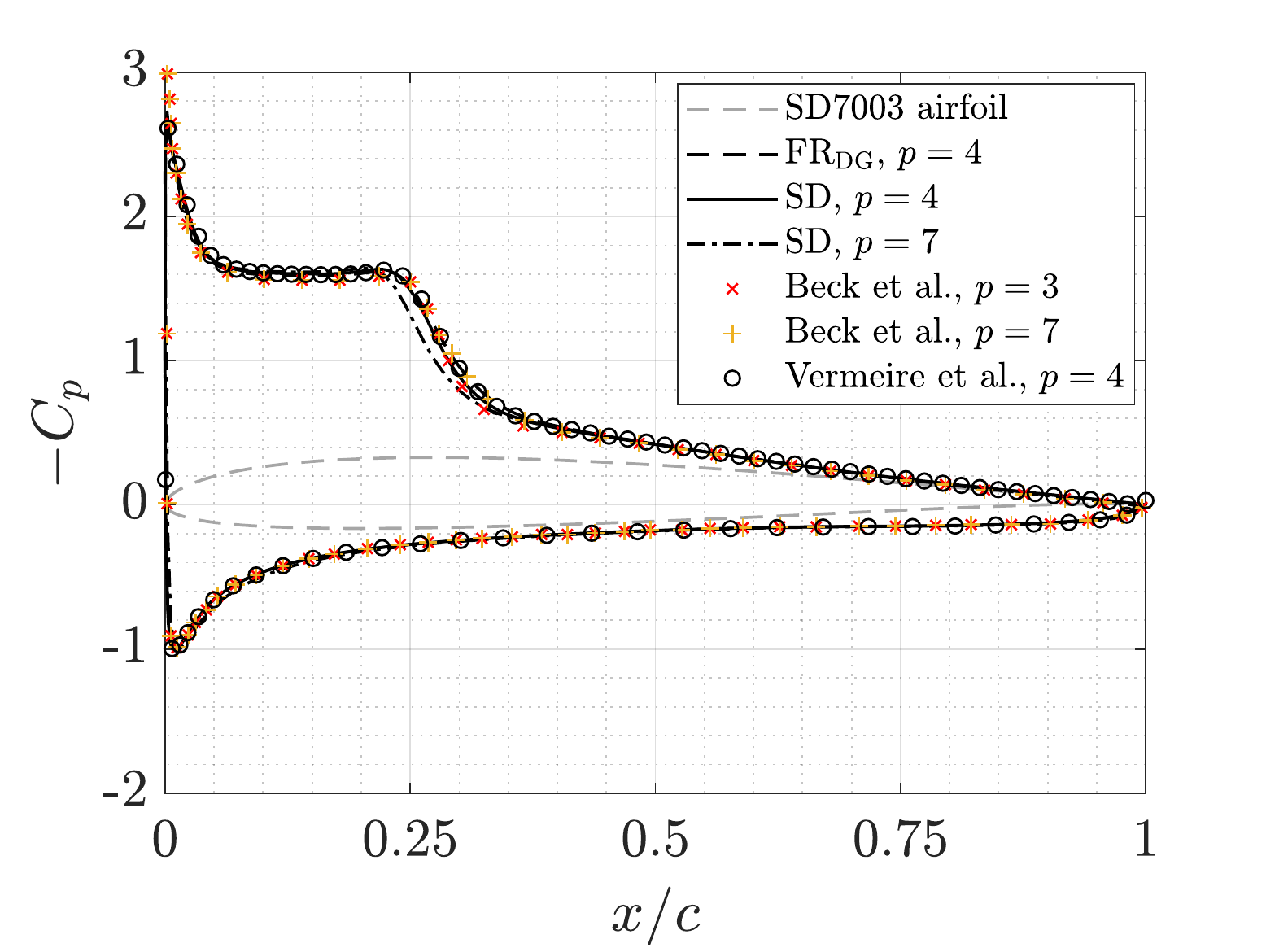}
        \label{f:sd7003:cp}
    }
    \subfloat[]{
        \hspace{-0.25in}
        \includegraphics[height=\fsize,keepaspectratio]
        {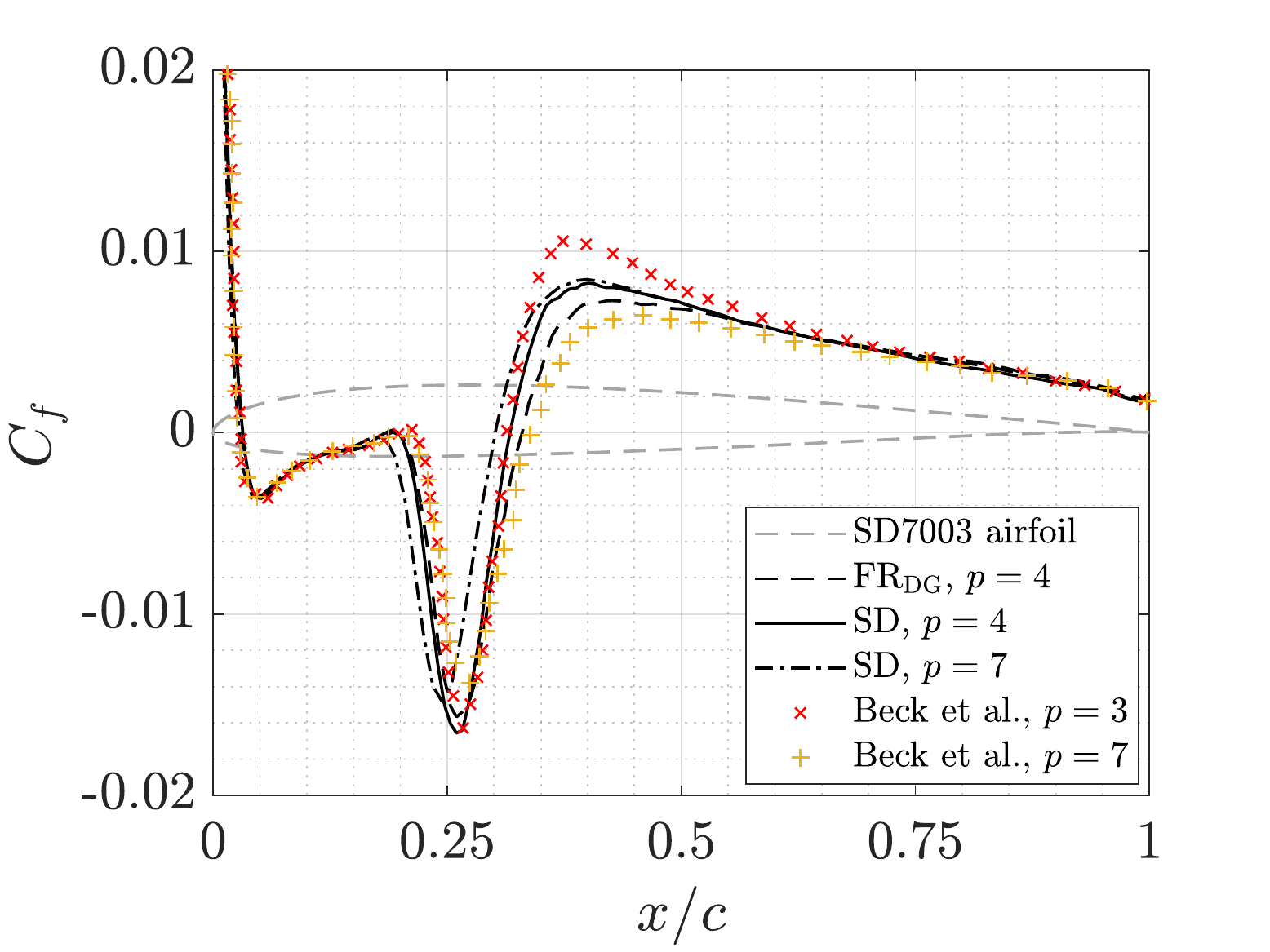}
        \label{f:sd7003:cf}
    }
    
    \caption{SD7003 at $Re=60\,000$, $\alpha=8^\circ$: (a) pressure coefficient $C_p$, (b) upper surface skin friction coefficient $C_f$. Results corresponding to $\p=4$ and $\p=7$ are obtained on mesh A and B, respectively. Results from Beck~\etal~\protect\cite{beck:2014} and Vermeire~\etal~\protect\cite{vermeire-witherden-vincent:2017} are provided for reference.}
    \label{f:sd7003:cp:cf}
\end{figure}

Figure~\ref{f:sd7003:Q} displays an isosurface of the \emph{Q}-criterion ($Qc^2/u^2_{\infty}=500$) colored by velocity magnitude from the $\SD$ scheme with $\p=7$. Time and spanwise-averaged plots of the pressure and skin friction coefficients are shown in Fig.~\ref{f:sd7003:cp:cf}. We report maximum skin friction values in the turbulent region above the airfoil using $\SD$ of \num{8.3e-3} (mesh A, $\p=4$) and \num{8.5e-3} (mesh B, $\p=7$) and $\FRDG$ of \num{7.3e-3} (mesh A, $\p=4$). This gives $y^+$ values of 8.95, 12.54 and 8.40, respectively. However, the corresponding $y^+$ values of the first solution point nearest the airfoil surface, $y^+|_{sp}$, are 0.42, 0.25 and 0.39. Table~\ref{t:sd7003} demonstrates that averaged values of the lift coefficient $\overbar{C}_L$ and drag coefficient $\overbar{C}_D$ as well as time and spanwise-averaged values of flow separation $x_s/c$ and reattachment $x_r/c$ locations of the laminar separation bubble are in agreement with various discontinuous spectral element results of implicit large eddy simulation found in the literature. The ILES results from Garmann~\etal~\cite{garmann:2013}, who used a 6th order finite difference scheme, are also provided in the table. We report here that under $\SD$, the simulation is stable on both the coarse mesh ($\p=7$) and fine mesh ($\p=4$). Under $\FRDG$, the simulation is rendered unstable only on the coarse mesh, and under $\FRSD$, the simulation is unstable on both meshes. These findings demonstrate the extra stability afforded by the staggered arrangement of flux points inherent to the $\SD$ scheme for achieving a stable under-resolved implicit large eddy simulation of transitional flow using a higher polynomial order on a coarse grid. Furthermore, in light of the FR results for this test case, we recommend the use of $\FRDG$ instead of $\FRSD$ when filtering or de-aliasing is not applied for these under-resolved simulations of turbulent flows.

\begin{table}[H]
    \centering
    \setlength{\tabcolsep}{10pt}
    {\footnotesize
    \begin{tabular}{lllllllll} \toprule
        Author & Scheme & Mesh & Elements & $\p$ & $\overbar{C}_L$  & $\overbar{C}_D$  & $x_s/c$ & $x_r/c$ \\ \midrule
        {\it Current}
            & $\SD$   &  A   & 137\,916 &  4  & 0.938  & 0.049  & 0.032  & 0.317   \\
            & $\FRSD$ &  A   & 137\,916 &  4  & \xmark & \xmark & \xmark & \xmark  \\
            & $\FRDG$ &  A   & 137\,916 &  4  & 0.942  & 0.051  & 0.031  & 0.330   \\
            & $\SD$   &  B   &  33\,264 &  7  & 0.940  & 0.048  & 0.028  & 0.301   \\
            & $\FRSD$ &  B   &  33\,264 &  7  & \xmark & \xmark & \xmark & \xmark  \\
            & $\FRDG$ &  B   &  33\,264 &  7  & \xmark & \xmark & \xmark & \xmark  \\
        Vermeire~\etal~\cite{vermeire-witherden-vincent:2017}
            & $\FRDG$ &  A   & 137\,916 &  4  & 0.941  & 0.049  & 0.045  & 0.315 \\
        Romero~\cite{romero-thesis:2017}
            & $\DFR$  &  -   & 202\,500 &  4  & 0.950  & 0.045  & 0.035  & - \\
        Beck~\etal~\cite{beck:2014}
            & DGSEM   &  -   &  66\,500 &  3  & 0.923  & 0.045  & 0.027  & 0.310  \\
        Beck~\etal~\cite{beck:2014}
            & DGSEM   &  -   &   8\,900 &  7  & 0.932  & 0.050  & 0.030  & 0.336  \\
        Garmann~\etal~\cite{garmann:2013}
            & FD (6th order) &  -   & 12\,549\,120 & - & 0.969 & 0.039 & 0.023 & 0.259 \\ \bottomrule
    \end{tabular}
    }
    
    \caption{SD7003 at $Re=60\,000$, $\alpha=8^\circ$: averaged lift coefficient $\overbar{C}_L$, drag coefficient $\overbar{C}_D$, separation location $x_s/c$, and reattachment location $x_r/c$. Unstable simulations are indicated by the symbol~\xmark. Results from various authors are provided for reference.}
    \label{t:sd7003}
\end{table}

\subsubsection{Computational Cost}\label{s:performance}

Performance of the spectral difference and the flux reconstruction schemes was measured using the simulations on mesh A in terms of wall-clock time taken to compute the divergence of the flux $\nabla \cdot \bm{F} = \pxvar{\bm{f}}{x} + \pxvar{\bm{g}}{y} + \pxvar{\bm{h}}{z}$, normalized by the total degrees of freedom, number of equations to solve, and number of stages $k$ in the time stepping scheme, such that $t^{\star}_{wall}=t_{wall}/DoF/N_{eq}/k$. All simulations have been done using double precision. The results shown in Tab.~\ref{t:sd7003_performance} demonstrate that, with the current high-order framework of the solver, the performance of the spectral difference and flux reconstruction schemes is approximately identical on mesh A using $\p=4$ in computing transitional flow past the SD7003 airfoil. In addition to previous computational performance assessments~\cite{liang-cox-plesniak:2013}, these results offer a complimentary and more supportive view on the efficiency of the spectral difference scheme.

\begin{table}[H]
    \centering
    \setlength{\tabcolsep}{24pt}
    {\footnotesize
    \begin{tabular}{ll} \toprule
        Scheme & $t^{\star}_{wall}$ (\SI{1e-9}{\second}) \\ \midrule
        $\SD$  & 0.5920 \\
        $\FR$  & 0.5924 \\ \bottomrule
    \end{tabular}
    }
    
    \captionsetup{width=120mm}
    \caption{Wall-clock time to compute $\nabla \cdot \bm{F}$ in mesh A using 48 Intel Xeon E5-2680 v4 processors, normalized by total degrees of freedom, number of equations, and number of RK stages. All calculations are done using double precision.}
    \label{t:sd7003_performance}
\end{table}

\section{Conclusions}\label{s:conclusions}

We reported the development of various discontinuous spectral element methods within a single high-order coding framework such that a fair and impartial comparison among several numerical schemes may be performed---most notably the true spectral difference and flux reconstruction methods. With this construct, we were able to assess the accuracy, stability, and performance of these two schemes. Furthermore, we provided a novel nonlinear stability analysis of the spectral difference scheme and demonstrated that the error bound for this scheme can be smaller than the flux reconstruction scheme due to the staggered nature of the flux points. We performed a number of numerical experiments to support this analysis, such as heterogeneous linear advection, isentropic Euler vortex, inviscid, subsonic flow over a cylinder, Taylor--Green vortex at $Re=1\,600$, and transitional flow past the SD7003 at $Re=60\,000$. These results highlighted the advantages of using the baseline $\SD$ scheme on coarse grids with higher polynomial orders and demonstrated the potential for extra stability afforded by the $\SD$ scheme in achieving stable under-resolved implicit large eddy simulations of turbulent flow. Based on both numerical analysis and experiments, we can conclude that the pure spectral difference method can be more robust for nonlinear problems than its flux reconstruction analog, incurring less of a need for de-aliasing.


\section*{Acknowledgments}\label{s:acknowledgments}
We would like to thank the support received under the Texas A\&M Chancellor's Research Initiative for partially funding this work. We would also like thank Guido Lodato for helpful discussions and for sharing results from his spectral difference flow solver. Lastly, we thank the Texas Advanced Computing Center and Texas A\&M University's High Performance Research Computing facility for providing the resources to perform these simulations.

%
%

\bibliographystyle{elsarticle-num}
\bibliography{./library}

\end{document}